\documentstyle[11pt,aaspp4]{article}
\newcommand{\ee}[1]{\mbox{${} \times 10^{#1}$}}
\newcommand{\eten}[1]{\mbox{$10^{#1}$}}

\newcommand{\degree}{\mbox{$^{\circ}$}}
\newcommand{\am}{\mbox{\arcmin}}
\newcommand{\as}{\mbox{\arcsec}}

\newcommand\cmc{\mbox{cm$^{-2}$}}

\def\lsim {$\rlap{\raise.4ex\hbox{$<$}}\lower.55ex\hbox{$\sim$}\,$}

\newcommand{\mm}{millimeter}
\newcommand\submm{submillimeter}
\newcommand\smm{submillimeter}
\newcommand\fir{far-infrared}
\newcommand\mir{mid-infrared}
\newcommand\nir{near-infrared}

\newcommand{\lsun}{\mbox{L$_\odot$}}
\newcommand{\msun}{\mbox{M$_\odot$}}

\newcommand{\td}{\mbox{$T_d$}}
\newcommand{\lbol}{\mbox{$L_{bol}$}} 
\newcommand{\tbol}{\mbox{$T_{bol}$}} 

\newcommand{\mv}{\mbox{$M_V$}} 
\newcommand{\mean}[1]{\mbox{$\langle#1\rangle$}} 
\newcommand{\fsmm}{\mbox{$L_{smm}/L_{bol}$}} 
\newcommand{\lsmm}{\mbox{$L_{smm}$}} 
\newcommand{\alphanir}{\mbox{$\alpha_{NIR}$}} 
 
\newcommand{\form}{H$_2$CO}

\newcommand{\ammonia}{\mbox{{\rm NH}$_3$}}

\newcommand{\hcop}{HCO$^+$}

\newcommand{\nthp}{N$_2$H$^+$}
 
\input{epsf}

\newcommand{\Inu}{\mbox{$I_{\nu}(b)$}}
\newcommand{\Snu}{\mbox{$S_{\nu}$}}
\newcommand{\Bnu}{\mbox{$B_{\nu}(\Td)$}}
\newcommand{\nInu}{\mbox{$I_{\nu}(b)/I_{\nu}(0)$}}
\newcommand{\kappanu}{\mbox{$\kappa_{\nu}$}}
\newcommand{\Td}{\mbox{$T_{d}$}}
\newcommand{\Tdr}{\mbox{$T_{d}(r)$}}
\newcommand{\rhor}{\mbox{$\rho (r)$}}
\newcommand{\ppc}{pre-protostellar core}
\newcommand{\m}{\mbox{$m$}}
\newcommand{\p}{\mbox{$p$}}
\newcommand{\beam}{\mbox{$\theta_{mb}$}}

\begin{document}


\title {\bf Tracing the Mass during Low-Mass Star Formation. \
 I. Submillimeter Continuum Observations}
\author {Yancy L. Shirley and Neal J. Evans II\altaffilmark{1}\altaffilmark{2}}
\affil{Department of Astronomy, The University of Texas at Austin,
       Austin, Texas 78712--1083}
\and
\author{Jonathan M. C. Rawlings}
\affil{Department of Physics and Astronomy, University College London,
        Gower Street, London WC1E 6BT}
\and
\author {Erik M. Gregersen}
\affil{Department of Physics and Astronomy, McMaster University,\
 Hamilton, ON L8S 4M1, Canada}

\altaffiltext{1}{Sterrewacht Leiden, Postbus 9513, 2300 RA Leiden, The Netherlands}
\altaffiltext{2}{Department of Physics and Astronomy, University College London,
        Gower Street, London WC1E 6BT}

 
\begin{abstract}
 
We have obtained 850 and 450
\micron\ continuum maps of 21 low mass cores with SED's ranging from 
pre-protostellar to Class I (18K $<$ \tbol\ $<$ 370K), using 
SCUBA at the JCMT. 
In this paper we present the maps, radial intensity profiles, and
photometry. 
Pre-protostellar cores do not
have power-law intensity profiles, whereas the intensity
profiles of Class 0 and Class I sources
can be fitted with power laws over a large range of radii.
A substantial number of sources have companion sources within a few
arcminutes (2 out of 5 pre-protostellar cores, 9 out of 16 Class 0/I 
sources). The mean separation between sources is 10800 AU. The median
separation is 18000 AU including sources without companions as a
lower limit.
The mean value of the spectral index
between 450 and 850 \micron\ is $2.8 \pm 0.4$, with \ppc s having slightly
lower spectral indices ($2.5 \pm 0.4$). The mean mass of the sample, based on 
the dust emission in a 120\arcsec\ aperture, is $1.1 \pm 0.9$ \msun.
For the sources fitted by power-law intensity distributions
($\nInu = (b/b_0)^m$), the mean value of $m$ is 1.52 $\pm$ 0.45 for
Class 0 and I sources at 850 \micron\ and 1.44 $\pm$ 0.25 at 450 \micron.
Based on a simple analysis, assuming the emission is in the Rayleigh-Jeans
limit and that $\Tdr \propto r^{-0.4}$, 
these values of $m$ translate into power-law density distributions
($ n \propto r^{-p}$) with $p \sim 2.1$.
However, we show that this result may be changed by more careful
consideration of effects such as beam size and shape, finite outer radii, more
realistic \Tdr, and failure of the Rayleigh-Jeans approximation.

\end{abstract}


\section{Introduction}

Theories of isolated, low-mass star formation predict the distribution 
of density, \rhor,
and dust temperature, \Tdr, before and during the star formation process.
These can be used to predict the spectral energy distribution (SED) and the
spatial intensity distribution, \Inu, of dust continuum emission.
Up to now, the primary tool for determining the evolutionary state
of a particular core has been the SED, but the relationship between the
SED and the distribution of matter is not unique (Butner et al. 1991,
Men'shchikov \& Henning 1997).
Observing the spatial intensity distribution of dust continuum emission at 
long wavelengths, where it becomes optically thin, provides a powerful 
tool for constraining the actual distribution of matter (Adams 1991, Ladd
et al. 1991).
New instruments have recently been developed at \submm\ wavelengths
that greatly enhance our capability in this area (Hunter et al. 1996,
Cunningham et al. 1994).
In this paper, we present maps of dust emission around 21 cores in various
evolutionary states using the Submillimetre Common User Bolometer Arrray
(SCUBA) (Holland et al. 1999) at 
wavelengths of 1.3 mm, 850\micron, and 450 \micron.
By mapping the extended dust emission, we can probe the density structure
from 7\as\ to 100\as\ (for the nearest sources in our sample, at 125 pc,
these angles correspond to 870 AU to 12500 AU).

Our conception of the evolution of a dense core, first to a  protostar, an
object whose luminosity is dominated by accretion, then to a pre-main sequence
star has been guided by an empirical evolutionary sequence (Lada 1987).
Theoretical modelling of the SED (Adams, Lada, \& Shu
1987) shows a good correspondence with this classification. In this
scheme, sources are classified by the shape of their SED.
Thus, an infrared spectral index is defined for the wavelength range
$\lambda$=2--20$\mu$m;
\begin{equation}
\alphanir = \frac{d(log\lambda S_{\lambda})}{d(log\lambda)},
\end{equation}
where $S_{\lambda}$ is spectral flux density per wavelength interval.
Class I sources were identified as the youngest protostars, deriving most of
their luminosity from accretion. They are embedded in an envelope and have
SEDs that rise ($\alphanir \geq 0$) to a peak in the \fir. 
Class II and III sources are progressively less embedded than Class I sources.
Class II SEDs peak in the \nir\ but possess a \mir\ excess 
($-1.5<\alphanir<0$), and they
are normally associated with star/disk systems without a significant envelope.
Class III SEDs are reddened blackbodies ($-1.5>\alphanir$) and are associated 
with stars without optically thick disks.
Class II and Class III sources typically correspond to classical and weak-line
T-Tauri stars respectively.
 
Within the last decade, this classification scheme, which was defined in the
context of infrared SEDs, has been modified to include more deeply embedded
sources, which presumably represent a phase earlier than Class I.
Andr\'{e} et al. (1993) proposed the name Class 0 for sources that are
so highly enshrouded that their SEDs peak longward of 100 \micron\ and their
\nir\ emission is very faint. Class 0 sources are defined to be cores which
possess a central source, but which have $\fsmm \geq 0.005$, where
\lsmm\ is the luminosity at $\lambda >$350$\mu$m. This criterion corresponds
approximately to the mass of the centrally condensed protostellar core being
less than that of the collapsing envelope. \fsmm\ should 
decrease with time (Andr\'e et al. 1993).

Starless cores provide plausible candidates for a still earlier stage.
These starless cores are associated with dense gas cores 
(Myers \& Benson 1983; Benson \& Myers 1989) for which no source
was detected by IRAS.
Ward-Thompson et al. (1994) detected \submm\ emission from a sample
of these objects, which they labelled ``pre-protostellar cores'' (PPCs,
sometimes called Class $-1$ sources). 
Pre-protostellar cores may be in
hydrostatic equilibrium, or they may be gravitationally bound, magnetically
sub-critical cores that are undergoing
quasi-static contraction as a result of ambipolar diffusion.
 
In an alternative approach,
Myers \& Ladd (1993) defined a continuous variable, \tbol , which is
the temperature of a blackbody with the same mean
frequency as the observed SED.  Class 0 sources have \tbol\ $<$ 70K
while Class I sources have 70K $\leq$ \tbol\ $<$ 650K (Chen et al. 1995).  
\tbol\ has been
hard to determine for pre-protostellar cores
because of an absence of data at $\lambda <
450$ \micron, but a few values are available from
space-based observations in the \fir\
(Ward-Thompson, Andr\'e, \& Motte 1998, Ward-Thompson \& Andr\'e 1999).  
 
The empirical classification scheme has been compared to theoretical models
of star formation.  Shu and co-workers
(e.g. Shu, Adams, \& Lizano 1987, Shu et al. 1993) developed a detailed
theory of low mass star formation from the stage of cloud core formation to
an emergent pre-main-sequence star.  The simplest form of the theory
(not including rotation or magnetic fields) begins with collapse inside
a centrally condensed isothermal sphere ($n(r) \propto r^{-2}$ , Shu 1977).
The inside-out collapse model predicts that a wave of infall propagates
outward at the sound speed of the gas.  The density inside the infall
radius approaches $r^{-1.5}$ toward the center as appropriate for
free fall.  Core formation would
then correspond to the \ppc s. Collapse would begin with the
Class 0 stage and continue into the Class I stage. 

This model is somewhat simplistic, and alternatives have been proposed.  
For instance, 
observational evidence exists for sharp density contrasts
near the edges of cloud cores (Abergel et al. 1998).
Submillimeter continuum emission
from pre-protostellar cores indicates that the density
distributions in the core flatten, rather than continuing to follow
a single power law to small radii (Ward-Thompson et al. 1994), and
line profiles consistent with infall motions have been detected in 
a substantial fraction of \ppc s (Lee, Myers, \& Tafalla 1999,
Gregersen \& Evans 2000).
If dynamical collapse begins before the
core is fully relaxed to the isothermal sphere, there is an early
stage of fast mass accretion (Basu \& Mouschovias 1995,
Henriksen et al. 1997), that may distinguish Class 0 from Class I sources.

Determination of the density distribution of dust in a sample including
PPCs, Class 0, and Class I sources can answer some questions.
Do the PPCs have density distributions predicted by theoretical models
of core formation leading to inside-out collapse?
Are there differences between the distributions
in those with evidence for infall and those without? Are there qualitative
differences in the distribution of matter around Class I sources and Class
0 sources, or is the difference only quantitative?

\section{Observations}

\subsection{The Sample}

The sample of sources (Table 1) was chosen to span a range 
of evolutionary states, from pre-protostellar cores to Class I
sources, and to include sources with (11) and without (10) evidence of infall,  
based on line profile shapes in \hcop, CS, or \form\
(Gregersen et al.\ 1997, 2000; Gregersen and Evans 2000; Mardones et al.\ 1997).
The coordinates were taken from a variety of references, including 
Ward-Thompson et al. (1994) (PPC), Gregersen et al. (1997) and 
Mardones et al. (1997), and corrected or updated in a few cases.
Distances were obtained from a literature search, making extensive use of the
compendia of Hilton \& Lahulla (1995) and Lee \& Myers (1999), but going back to 
the original references. We chose the newer, closer distances to the Ophiuchus 
complex ($125\pm25$ pc) 
favored by de Geus et al. (1990) over the traditional choice of 160 pc, and to 
the Perseus clouds ($220\pm20$ pc) according to \v{C}ernis (1990) over the 
usual 350 pc. 

Our original sample was selected to contain roughly equal numbers of
PPCs, Class 0, and  Class I sources. In the end, more Class 0 sources
were observed, and
several sources moved from Class I to Class 0 when we recalculated their
properties, including our data.
Changes in estimates of \lbol, \tbol, and \fsmm\ are discussed in \S 3.2. 
We use $\tbol = 70$ K as the dividing line between Class I and 
Class 0 (Chen et al. 1995), giving us 3 Class I sources in our final sample.
Andr\'e et al.  (1993) required $\fsmm > 0.005$ for Class 0 status; with this
criterion, SSV13 would be the only Class I source remaining in our sample.
Consequently, we often discuss Class 0 and I sources together, referring
to them as Class 0/I.

\subsection{Observing and Calibration}

The cores were mapped simultaneously at 850 and 450 $\mu$m using SCUBA during 
parts of 12 nights in 1998 January, April, and May with the James Clerk Maxwell
Telecsope (JCMT) on Mauna Kea, Hawaii.  Nine cores
were also mapped at 1.3 mm using the single bolometer detector on SCUBA. 
A 120\as\ chopper throw in azimuth was used for all the cores.  
Using the 64-point jiggle map mode, each SCUBA map fully samples a 2\farcm3 
region simultaneously at 850 and 450 $\mu$m.  
The telescope was nodded during each map.  Each jiggle map produces 
4.2 minutes of integration time on the source.  Because
we are interested in mapping extended low brightness emission, we made
5-point maps (each a 64-point jiggle map) with a spacing of 30\as.  
Such maps also mitigate the effects of bad bolometers.
The inner
2\am\ of the final image was mapped in each of the 5-point maps with a 
total integration time of 21 minutes.  The signal-to-noise ratio varied between 
5 and 97 for our images (see Table 9); these estimates are conservative
because the main part of the image had 5 times as much integration as
the outer parts, where the noise was determined. 
In some cases, extra positions were
observed to cover additional sources in the field.

        Pointings and skydips were performed between 5-point maps. 
The pointing varied by less than 2\as\  between objects.  We measured
$\tau$$_{850}$ and $\tau$$_{450}$ during each skydip. The CSO radiometer
was monitored simultaneously to obtain $\tau$$_{cso}$ (measured at
225 GHz).  Our observations
confirm the correlation between this opacity and those obtained from the JCMT
skydips (Chapin 1998).  We used the skydips at 850
and 450 \micron\ immediately preceding and following a 5-point map to
interpolate the extinction correction.  The average and standard deviation
over the night in the
opacities at 850 and 450 \micron\ for each night are listed in Table 2.
Opacities derived from the peak fluxes of Uranus before correction 
for extinction generally agreed with the opacities derived from skydips.  
The uncertainties in the opacity dominate the uncertainties in the
total flux calibration, but they have little effect on our primary objective of
imaging the sources, because the maps were obtained with a 2-D array in 
a short time.
   
To assess the effects of image smearing when we average the components
of our 5-point maps taken with different pointings, 
we averaged all the Uranus maps on all the runs with 1\arcsec\ pixels
to produce an average Uranus map.  From this map, the average FWHM 
($\beam$) at 850 and 450 \micron\ were 15\farcs 2 and 7\farcs 9 
respectively (roughly 1\as\ larger than values derived from a single map).  
Because even this worst-case experiment produced only marginal broadening, 
the maps are not significantly distorted by averaging data with different
offsets. 
The average radial profiles for Uranus observed during April
are shown in Figure 1 for both wavelengths;
the sidelobe structure is clear and consistent with other measurements
(W. Holland, personal communication).
Night to night variations in the sidelobes can be seen, but amplitudes of the 
sidelobes vary by less than a few dB and positions of the sidelobes are roughly
constant.  April and August observations were made during second shift
(01:30 -- 09:30 HST) when the dish shape and focus had stabilized.
Significant variations in sidelobe structure are seen for observations
taken during first shift (17:30 -- 01:30 HST) (C. Chandler personal
communication).
Our average 850 \micron\ beam is characterized by sidelobes at 
24\arcsec ($-17$dB) and at
47\arcsec ($-23$dB).  The average 450 \micron\ beam is characterized by sidelobes
at 13\arcsec ($-12$dB), at 24\arcsec ($-16.5$dB), and at 40\arcsec ($-23.5$dB). 
Removing Uranus from the data makes a small change in the central Gaussian
width ($\le 10$\%) and negligible difference in the sidelobe structure
(The diameter of Uranus during April 1998 was 3\farcs 2).
We use the actual beam profile in the the modeling described in \S 4.2.

The total flux was calibrated using 120\as\ and 40\as\ diameter apertures in 
extinction corrected maps of Uranus, AFGL 618, and Mars.  The observed
flux densities for an aperture of diameter $\theta$ were computed from
$\Snu(\lambda,\theta) = C^{\lambda}_{\theta} V(\lambda,\theta)$, where
$V(\lambda,\theta)$ was the voltage measured at wavelength $\lambda$ in 
an aperture of diameter $\theta$.
The calibration factors, $C^{\lambda}_{\theta}$, 
were calculated from the fluxes of Uranus and Mars.
The total observed flux from previous SCUBA measurements was 
used for AFGL618 (SCUBA secondary flux calibrator webpage):  
$4.56 \pm 0.17$ Jy/beam at 850 \micron; 
$11.2 \pm 1.4$  Jy/beam at 450 \micron.
The flux calibration did not vary substantially from night to night within 
an observing run but did vary from run to run.  Therefore,
we use an average flux calibration for each run (January, April, and August),
listed in Table 2.

\subsection{Image Reduction}

        The initial reduction of each image was performed using SURF,
the SCUBA User Reduction Facility software package (Jenness \& Lightfoot, 
1997). The raw images of 64-point jiggle maps already have removed the
effects of the chopping.  The raw images are further reduced by 
removing the telescope nod and correcting for different bolometer 
gains (flat-fielding). Sky variations
were subtracted by averaging the response of multiple bolometers off the
source.  Because some of our sources are very extended,
care was taken to choose bolometers that were free of significant 
low level emission.  
After sky noise subtraction, each image was rebinned to 
$0.5\beam$ per pixel on a B1950.0 coordinate system. 
SCUBA's bolometers are subject to microphonic and 1/f noise.  
Excessively noisy bolometers (RMS voltage $\geq$ 60 nV in noise tests) 
were removed.  Also, noisy sections
of the integration were removed.  These are most likely caused by imperfect 
subtraction of sky noise and are usually observed across several bolometers
at the same time.  The noisiest bolometers were typically found near the edge
of the array, resulting in increased noise near the edge of the
maps.

        The final 5-point maps were rebinned by shifting the individual
images by their centroid.  Corners of the 5-point map occasionally
chopped onto source emission, causing the negative beam to become visible.
Bolometers in the negative beam were removed from the final image,
resulting in irregularly shaped edges on many maps.

\section{Results}

\subsection{Images}
        
        Contour plots of our SCUBA images are shown in Figures 2-8.  
Contour levels
are indicated in each plot caption with the lowest contour $\geq 3\sigma$.  
Outflow axes are marked with a solid line.
The (0,0) positions are given in Table 1.  Typical rms noise near the 
edge of the maps was 20 mJy/beam at 850\micron\ and 100 mJy/beam at
450\micron .  Assuming a dust temperature of 10K near the edge of the
cloud, we are sensitive to $A_{V} = 3$mag (1$\sigma$ rms) at
850\micron\ and $A_{V} = 14$mag at 450\micron .  These correspond to
column densities $N$(H$_2$) of $2.7\times 10^{21}$ \cmc\ and $1.3\times
10^{22}$ \cmc\ respectively.

Pre-protostellar cores are clearly more diffuse than the Class 0/I
sources.  L1512 is the most extreme in this sense, showing no evidence 
for a centrally peaked source.  
L1544 is an elongated structure with a central peak. L1689B
has an elongated peak at high contour levels. 
However, their intensities are not as strongly peaked as the Class 0/I sources.
Two out of five \ppc s have companions within 2\arcmin.
L1689A has two sources visible at both wavelengths, separated by
about 0.03 pc, with roughly comparable intensity. 
B133 has a weaker companion to the southeast.
L1689B may also have multiple peaks along the east-west ridge. 
The maps of L1544, B133, and L1689B are consistent with maps 
of 1.3 mm emission
toward those sources (Ward-Thompson et al. 1999; Andr\'e et al. 1996).

        While the pre-protostellar cores are diffuse, the
Class 0/I sources are strongly  centrally peaked. All the Class 0
sources except L1172 and L1455 have one well-defined centroid.  
B335 and B228 appear to be quite circularly symmetric, but B335 has a 
slight extension to the south-east visible in the 450 \micron\ map 
(cf. Huard et al. 1999) that may be associated with the outflow 
(Bontemps et al. 1996).  The other Class 0
sources all have non-spherical extensions.  These extensions correspond
roughly to the outflow directions in L1527, L1157, and L1455.
L1157 is a particularly good example with a sizable extension to 
the south along the outflow axis.  
Other sources (L483 and IRAS03282+3035) are extended perpendicular to the
outflow direction. SSV13 is elongated perpendicular to the outflow
direction at high contours, but the lower contours lie along the
outflow axis.
It is possible that extensions along outflow directions are caused by 
heating of dust by short wavelength radiation escaping along the outflow 
cavity, while extensions perpendicular to the outflow direction reflect
the distribution of maximum column density, which most models predict
to be perpendicular to the outflows. This subject will be analyzed in
later papers, where two-dimensional dust radiative transport can be modeled.
        
Nine of our sixteen Class 0/I sources have
a secondary source within 2\arcmin.  The L1448 cloud (L1448NW, N, and C)
and SSV13 complex were known to contain multiple sources. In addition, the 
eastern source in L43 is seen in both the 450 \micron\ map of Bence 
et al. (1998) and the 1.3 mm map of Ward-Thompson et al. (1999).
We are not aware of previous detections of the additional sources
toward CB244 and L1455 (see Table 3). 
L1455 is an extreme case with 5 sources within the SCUBA map, only one
of which corresponds to a known source in this region.
Continuum emission is observed to bridge between sources, making it 
difficult to disentangle the envelope density structure.
L1172 has at least 2 peaks within a 20\as\ region.
For the seven Class 0/I sources without secondary sources, 
the lack of contamination 
coupled with high signal-to-noise (Table 9) will allow radial intensity
profiles extending up to
11 beams (at 450 \micron ) from the central source (see section 3.3).

For cores with more than one source, the mean separation in the plane 
of the sky is 10800 AU, less than twice the fragmentation scale of 6000 AU
found in the $\rho$ Ophiuchi cores by Motte et al. (1998).
A mean separation of \eten{4} AU is also close to the break in the
distribution of optical binaries in Taurus, which Larson (1995)
associates with the Jeans length, and to the length scale
inferred for dynamical collapse based on specific angular momentum 
arguments (Ohashi et al. 1997). Looney, Mundy, \& Welch (2000) find
evidence for multiplicity on still smaller scales in L1448 and SSV13.
They describe sources separated by $>6000$ AU as ``separate envelope"
multiplicity, and our data are consistent with this picture.  The 
median separation, including the map size as a lower limit for
sources with no detected companion, is 18000 AU.

\subsection{Photometry, Classification, Spectral Index and Masses}

The photometry is presented in Table 3, including
calibration uncertainties.  Fluxes are reported in 120\as\ and 40\as\ apertures
at 1.3 mm, 850 \micron, and 450 \micron.  
Uncertainties reported include statistical
uncertainties and calibration uncertainties calculated from
\begin{equation}
\sigma_{\Snu}^{2} = \Snu^{2} \left[ \left( \frac{\sigma_{C}^{2}}{C^{2}} 
\right)_{Run \: Avg} + \ \left( \sec^{2}z \: \sigma_{\tau}^{2} \right)_{Source}
\right]
\end{equation}
where \Snu\ is the flux density (Jy) and $C$
is the calibration factor (Jy/V).
The uncertainties are $\sigma_C$ and $\sigma_{\tau}$, 
the standard deviation over the run in $C$ and over the night 
in $\tau$ (Table 2) and $z$ is the
mean zenith angle of the observations of the source.
Photometry is presented for each source in the map that is sufficiently
strong and well-defined. The offsets of the centroids from previous
infrared/submillimeter positions (Table 1) are reported in Table 3.

The spectral index $\alpha_{450/850}$, defined by 
\begin{equation}
\alpha_{450/850} = \frac{\log \left( \frac{S_{450}}{S_{850}} \right) }{
\log \left( \frac{850}{450} \right) },
\end{equation}
is given in Table 4 for both 40\arcsec\ and 120\arcsec\ photometry, when
available. Note that this definition differs from that in equation 1.
The uncertainties include calibration uncertainty, which
usually dominates. The values for the two aperture sizes do not
differ significantly; we use the values for a 40\arcsec\ aperture
in the following discussion because more data are available. 
The mean spectral index for the collapse candidates is indistinguishable
from that of sources with no evidence of collapse. 
The mean for the \ppc s ($\mean{\alpha_{450/850}} = 2.5 \pm 0.4$) 
is slightly less than that for Class 0 and I sources 
($\mean{\alpha_{450/850}} = 2.9 \pm 0.4$).  A lower spectral index may be an
indication of lower \Td\ in \ppc s, resulting in failure of the Rayleigh-Jeans
approximation (see \S 4.2). 
Because the difference is not statistically
significant, we also calculated the mean and standard deviation 
of all the measurements  ($\mean{\alpha_{450/850}} = 2.8 \pm 0.4$).
If the emission were in the Rayleigh-Jeans limit, this result
would imply that that $\kappa_\nu \propto \nu^{0.8}$ between 450 and
850 \micron, but this exponent should be interpreted as
a lower limit if the Rayleigh-Jeans approximation fails.  
The average spectral index between 850\micron\ and 1.3mm,
defined in the same manner as in equation 3, is
$\mean{\alpha_{850/1.3}} = 3.4 \pm 0.3$.  
A higher $\alpha_{850/1.3}$ would be expected if the Rayleigh-Jeans approximation were
failing at 450 \micron.

Values of \lbol, \tbol, and \fsmm, where $L_{smm}$ includes all flux 
at $\lambda > 350$ \micron, were newly computed (Table 8), including archival 
data and the results of our photometry. We include the archival data
and references in Tables 5-7.
For isolated sources, we chose 
data in the largest available apertures at long wavelengths, 
because our data show that much of the flux density comes from very extended
regions. Two different methods were used to integrate the data. 
The uncertainties reflect uncertainties in the photometry and 
differences in the
method of integration, but the uncertainties in \lbol\ do not include 
uncertainties in distance, since these are unavailable for many sources. 
Among the \ppc s, only L1544 and L1689B have the requisite \fir\ data.
The \lbol\ of L1544 is comparable to some Class 0 sources, but the origin
of the relatively strong \fir\ emission is unclear.

Some sources changed classification as a result of our data or
analysis. The position
of IRAS04166+2706 is close to the IRAS position, but clearly displaced from
the position in Mardones et al. (1997). We calculate a lower \tbol, but the
source remains a Class I source. L1448N is clearly a Class 0 source
with our photometry, whereas it was borderline for Mardones et al.
(who referred to it as $03225+3034$). The situation is similar for
L1455. We find a very low \tbol\ for
IRAS03282+3035, consistent with the upper limit of Mardones et al. (1997).
For L1527 and CB244, we excluded the \nir\ data, which is clearly displaced from
the \smm\ source, leading to a lower \tbol\ than previous estimates 
(Chen et al. 1995, Mardones et al. 1997).
Our value of \tbol\ for B335 is 28 K, rather than
the 37 K of Mardones et al. (1997), presumably because we include the larger
flux densities that we find. Adding data at \smm\ wavelengths also decreased
\tbol\ for L1157 and L1172, moving the latter source to Class 0.
The values of \tbol\ for SSV13 and L43 include \nir\ data, because the
emission peaks on the \smm\ position. However, SSV13 has varied substantially
and
\tbol\ depends on the epoch; we used pre-flare \nir\ data (Harvey et al. 1998). 
The emission in L43 (RNO91) is polarized, hence scattered. If we exclude
the \nir\ data, \tbol\ would be $83\pm5$, still a Class I source.

We have estimated the masses (gas and dust) 
from the dust emission and the equation

\begin{equation}
M_D = {\Snu D^2 \over \Bnu \kappanu} = 3.69\ee{-6} \msun \Snu (Jy) D^2(pc) (e^{16.9K/\Td} -1),
\end{equation}
where \Snu\ is the flux density at 850 \micron\ in a 120\arcsec\ beam
(Table 8), \Bnu\ is the Planck function, \kappanu\ is the opacity per
gm of gas and dust at 850 \micron, and we have assumed optically thin
emission in a constant density sphere. The values in Table 8 were computed
assuming $\kappanu = 2\ee{-2}$ cm$^2$ gm$^{-1}$ and $\Td = 20$ K. 
Masses
computed with $\Td = 10$ K are a factor of 3.3 higher. 
Estimates of \kappanu\
vary by at least a factor of 3.  We have used the value for agglomerated
grains with thin ice mantles (col. 5 of Table 1 of Ossenkopf \& 
Henning 1994, hereafter OH5 dust), a model which has reproduced other
data well (van der Tak et al. 1999, 2000).  
For comparison, we also computed the virial mass in the same (60\arcsec) 
radius using the width of the line
most likely to be optically thin (see references in Table 8). 
Both calculations
assumed a uniform density cloud (no temperature or density gradients).  The
virial mass estimate would be decreased by a factor of 0.6 in a cloud with
$n(r) \propto r^{-2}$ for example. Given the uncertainties in each calculation,
the agreement is good. The mean and standard deviation of
the ratio are $M_D(20K)/\mv = 0.5 \pm 0.3$. 
This result is consistent (within reasonable uncertainties in \Td\ and distance)
with the assumption that the sources are gravitationally bound, but
the uncertainties make this assumption hard to test conclusively.
The mean mass is $M_D(20K) = 1.1\pm 0.9$ \msun.

\subsection{Radial Profiles}

To compute the average intensity distribution, we assume azimuthal
symmetry. While many images are not circular, experiments with taking
cuts along different axes, cutting out sectors with elongated emission,
etc. indicate that the overall results are not significantly affected by
deviations from azimuthal symmetry. Chandler \& Richer (2000) modeled
the effects of outflow cavities and found that the resulting
intensity profile has only a slightly lower value of $m$.

Normalized, azimuthally averaged radial profiles were made for each 
SCUBA image.  Each image was rebinned to $0.5 \theta_{mb}$ spacing. The mean 
\Inu\ in an annulus about impact parameter $b$ was computed from the 
intensity map, weighted by $A_i/\sigma^{2}_{i}$, where $A_i$ represents the area
of the $i$th pixel intercepted by the annulus and $\sigma_{i}$ is
the uncertainty in
the map intensity in the  $i$th pixel.
The error bars were calculated by propagating the uncertainties 
from the map. The radial profiles were normalized to the peak emission, and we
plot \nInu.
The image centroids in Table 3 were used for the center of the 
radial profile. To avoid effects of chopping, we terminated the
radial profiles at
98\as\  from the centroid; points are binned at $0.5 \theta_{mb}$ spacing 
(7\as\  at 850 $\mu$m and 3\farcs 5  at 450 $\mu$m). We have used the distances
in Table 1 to convert angles to impact parameters ($b$) in units of AU. 
In Figs. 9--12, the normalized radial profiles (\nInu) are plotted versus $b$.
The inflections in some profiles at large radii are due to
contamination by secondary sources. 
The point-to-point fluctuations in 
the profile are substantially less than the errorbars because
half-beam sampling was used.
The radial profiles agree well with those of 1.3 mm emission
presented by Andr\'e et al. (1996) and Ward-Thompson et al. (1999).
The radial intensity profiles of L1448-C agree well with those of
Chandler \& Richer (2000), but our profile of L1527 is steeper than
those found by either Chandler \& Richer (2000) or Hogerheijde \& Sandell (2000).
        
Previous observations of \ppc s showed that the radial intensity
profile did not follow a 
single power law, but a broken power law was able to fit the
limited data (Ward-Thompson et al. 1994). Maps of \mm\ emission
(Ward-Thompson, Motte, \& Andr\'e 1999) have clearly shown the flattening
of \nInu\ in the inner regions. While the outer regions can be approximated
by a power law, Andr\'e et al. (1996) commented that the north-south
cut through L1689B would be described better by a Gaussian than by a power
law.  
It is clear from Figure 9
that a power-law does not fit any portion of
the \ppc\ radial intensity profiles in the \submm, confirming the result
of Ward-Thompson et al. (1999).
Our results for \ppc s are consistent with observations of a 
larger sample (Andr\'e  et al.  2000). 

Models of core formation leading to inside-out
collapse predict a flat inner core approaching $p \sim 2$ at larger radii,
and the flat core should shrink toward the center with time. While
our observations
do not appear to be consistent with these models, we caution that
detailed modeling is still needed. If the evidence for large-scale infall
in all of these cores but L1512 (Lee, Myers, \& Tafalla 1999; Gregersen \& Evans
2000) is correct, then infall may begin
before the core has fully relaxed to a singular isothermal sphere with
$n(r) \propto r^{-2}$ (e.g., Shu 1977).

In contrast, the intensity profiles of most Class 0/I sources (Figures 10--12)
can be fitted with
power laws, if the inner three points, which are affected by the finite
beam size, and the outermost points (noisy, and possibly affected by the
finite chop size or other sources) are ignored.
Power law fits ($\nInu = (b/b_0)^{-m}$, with $b_0$ corresponding to 
$0.25 \theta_{mb}$) were made to 8 cores at 850 
$\mu$m and 10 cores at 450 $\mu$m.  
The uncertainty in the value of $m$ (Table 9) includes
the deviations from a straight line and the standard deviation from the
radial profiles, as described above. 
Fits used only points in the profile where the signal was greater than 
the noise in each bin. 
Fits for images with multiple sources are 
terminated at the intensity minimum between the sources.    
The slopes ($m$) from these fits are tabulated in Table 9.
The average slopes are $\mean\m = 1.52 \pm 0.45$ at 850 \micron\ 
and $\mean\m = 1.44 \pm 0.25$ at 450 \micron\ for Class 0/I sources.
Since the values of \mean\m\ determined at different wavelengths agree well,
we average all values to obtain $\mean\m = 1.48 \pm 0.35$.
The average slope for Class I sources does not differ significantly 
from that for Class 0 sources, but a larger sample of Class I sources
is needed.
We are unable to distinguish statistically significant 
differences in the average slope between
sources with ($\mean\m = 1.60 \pm 0.38$) and without
($\mean\m = 1.29 \pm 0.15$)  evidence for collapse.

\section{Analysis}

\subsection{Density Distributions: A Simple Analysis}

Combining the value of the slopes in the radial intensity 
profiles with a knowledge 
of the temperature distribution of dust grains, \Tdr,
constrains the density distribution of dust grains, \rhor.  If the 
emission is optically thin and the 
opacity ($\kappa_{\nu}$) 
of the dust grains does not vary with radius, the observed
intensity at an impact parameter, $b$, is given by
\begin{equation}
I_{\nu}(b) = 2 \int_{b}^{r_o} B_{\nu}(T_{d}(r)) \kappa _ {\nu} 
\rho (r) \frac{r}{\sqrt{r^{2} - b^{2}}} dr 
\end{equation}
(Adams 1991), where $r_o$ is the outer radius.   
If we assume power law distributions for the density and temperature,
\begin{equation}
\rho (r) = \rho (r_f) \left( \frac{r}{r_f} \right) ^{-p}
\end{equation}

\begin{equation}
T_{d}(r) = T_{d}(r_f) \left( \frac{r}{r_f} \right) ^{-q} 
\end{equation})
where $r_f$ is a fiducial radius, then, if the emission is in the Rayleigh-Jeans
limit and if $r_0$ $\gg$ $b$, equation (5) simplifies to
\begin{equation}
\nInu =  (b/b_0)^{-m} \; \; \;  , \; m = p + q -1.
\end{equation}

        The dust opacity for grains in the \submm\ portion of the
spectrum roughly follows a power law 
$ Q_{\nu} \propto \nu ^ {\beta} $
(see Ossenkopf \& Henning 1994).  Using this assumption 
the temperature distribution around a centrally
heated source follows a power law of the form
\begin{equation} 
\Tdr \propto L^{q/2}r^{-q}\; \; \; , \; q = 2/(4 + \beta )
\end{equation}
(cf. Doty \& Leung 1994).
Estimates of $\beta$ typically vary between 0 to 2 in the submillimeter.
For $\beta = 1$, consistent with our data (\S 3.2), $\Tdr \propto r^{-0.4}$.  
In this case, $p = m - q +1 = m + 0.6$, and the \mean\m\ found above
translates into $\mean\p = 2.08 \pm 0.35$ for Class 0/I sources.
For $\beta = 2$ the slope of the density  power law changes slightly
to $p = m + 0.67$.
These values are consistent with those found by Chandler \& Richer (2000), 
with the exception of L1527 (see also Hogerheijde \& Sandell 2000).
Within the uncertainties, these values are consistent with the density
distribution expected from an isothermal sphere or the outer parts
of a Bonnor-Ebert sphere (Shu 1977). 
However, there are quite a few caveats.

\subsection{Some Caveats}

For sources without confusing secondary sources, the data often fall
below the fit at large radii.
This behavior could be attributed to an outer radius where the profile 
becomes steeper (e.g., Abergel et al. 1998). We are wary of this conclusion
for several reasons.  Some of the turn-down 
near the edge may be caused by using a finite chop.  Since our sources have
very extended envelopes, we may have chopped onto low level emission,
decreasing the observed emission.  
While we tried to avoid this effect by only carrying
the radial profiles out to 98\arcsec, 
it still may be a problem. 

To investigate the importance of various effects 
on the radial profiles, we constructed 5 spherically symmetric 
models (Fig. 13).  All 5 models calculate the observed intensity, generalizing
equation 5 to allow finite optical depth, and convolve the result with a beam
profile; they use a code generously supplied by L. Mundy.  We assumed power
laws for \Tdr\ and $ \rho (r) $ ($q = 0.4$ and $p = 2$) and 
use OH5 opacities for coagulated dust grains with icy mantles 
(Ossenkopf \& Henning 1994).  The source was placed at a distance of 200 pc
with a total mass of 1.1 \msun,
equal to the mean distance ($200\pm 60$ pc) and mean mass of the 
sample.  
An inner radius for the dust shell of 60 AU and an outer radius
of 30000 AU (corresponding to 0\farcs24 and 120\as) were used.
Model 1 assumes 
gaussian beams with the FWHMs of 15\farcs 2 and 7\farcs 9 at
850\micron\ and 450\micron\ respectively.  Models 2 -- 5
have been convolved with our observed beam profiles.
The sidelobes clearly increase the normalized intensity throughout the profile
and spread the effects of a finite outer radius back to smaller impact
parameters. If very small and very large impact parameters are excluded from the
fit, the effect on the fitted slope is small ($\Delta m < 0.2$).

Another important issue is  Rayleigh-Jeans
failure in equation 5, which  occurs when the dust 
temperature falls below $h\nu / k$   
($h\nu/k$ = 32 K and 17 K at 450 and 850 \micron, respectively).
This is a problem for the emission at large distances from the low luminosity,
highly embedded
sources that we are observing because the dust temperature does drop 
below $h\nu / k$.  
Model 2 in Fig. 13
shows the observed intensity profile for a source in which 
$T_d(r) > 2h\nu / k$ over the entire profile.  Models 3 and 4
use the same $q = 0.4$ but were normalized to lower temperatures, such  that 
\Tdr\ dropped below $h\nu / k$ at 450 \micron\ (model 3) and 850 \micron\
(model 4) at 8000 AU.  Rayleigh-Jeans failure results in an increase in
$m$ by as much as 0.5, leading to an overestimate of \p\ by the same amount.     

The temperature is only approximated
by a power-law (eq. 7). The actual \Tdr\ is probably 
steeper in the inner regions, where the radiative transfer of heating 
radiation is optically thick (e.g., Doty \& Leung 1994). 
More importantly for this analysis, if the core
is exposed to the interstellar radiation field, \Tdr\ can flatten out or even
rise again in the outer regions.
Figure 13 also shows model 5 in which the temperature distribution
is isothermal ($\Tdr = 10K$) from 3000 AU outward.
A flattening of \Tdr\ causes a decrease in $m$ by as much as 0.5.   
The typical interstellar radiation field is
capable of heating dust to about 14 K at the extinctions probed by
\submm\ emission.  This slight rise would be expected to cause a further
decrease in $m$. 
Consequently, uncertainties of about $\pm 0.5$ are present in deducing
the value of \p\ from simple fits assuming power-laws for temperature and
the Rayleigh-Jeans approximation. To correct for these effects,
careful modeling of each source is needed.  These models will be the subject
of a later paper.

Although we have assumed spherical symmetry in the
previous analysis, the actual geometry of these cores is certainly
much more complex. Elongated extensions at low contour levels are seen 
in many sources.  The slope
of the intensity profile is not greatly affected by these extensions.
The $m$ at 850 and 450 \micron\ was modified by 0.1 when sectors
containing the extensions were eliminated from the azimuthal average.
This appears to be a small correction compared to other 
uncertainties in the analysis.

We have also assumed that $\beta$ is constant throughout the radial profile.
Ambipolar diffusion can cause a relative drift between the gas and the dust and
between the different dust grain populations. The former may lead to spatial
variations in the dust-to-gas ratio, and the latter will also cause $\beta$
to vary across the cores.
Substantial molecular depletions and grain aggregations 
in the central regions of some cores can
also lead to changes in $\beta$.
Visser et al. (1998) interpreted decreases in spectral index toward
column density peaks in NGC 2024 in terms of a decrease in $\beta$ in
dense cores, possibly caused by grain growth.
The correction factor to $\beta$ for Rayleigh-Jeans failure is
\begin{equation} 
\gamma(\td) = 1 + \frac{\log\left[\frac{\exp(h\nu_{850}/k\td) - 1}{\exp(h\nu_{450}/k\td) - 1}\right]}{\log\left(\frac{850}{450}\right)}
\end{equation}
where the spectral index is given by $\alpha = 2 + \beta + \gamma(\td)$ (Visser et al. 1998).
If we assume $\td = 20K$, then $\gamma(20K) \approx -0.7$.  Applying this
correction increases the
estimate of $\beta$ from 0.8 found in \S 3.2 to 1.5 for the data in a 40\as\ aperture.
The Rayleigh-Jeans correction factor varies quickly at low temperatures.
For example, $\gamma(\td)$
changes by 0.5 between 10 and 14 K, plausible
changes in \Tdr\ from the center to edge of an externally  heated \ppc.
Chandler \& Richer (2000) saw little evidence for changes in $\beta$
with radius, but Hogerheijde \& Sandell (2000) did observe changes in a
few cases.

\section{Conclusions}

The main conclusions of our study are as follows.

Pre-protostellar cores are clearly more diffuse that Class 0/I sources.
Pre-protostellar cores do not have central peaks that are as well defined as
those in Class 0/I sources.
Many sources had companion sources within
2\arcmin\ (2/5 \ppc, 9/16 Class 0/I sources). 
The presence of several
sources in an IRAS beam means that previous studies of SEDs may have
been distorted. Observations with higher spatial resolution in the \fir\
are needed. 
For the sources with companions, the mean projected separation
is 10,800 AU, more than the  mean separation of 6000 AU in
the $\rho$ Ophuichi cluster (Motte et al. 1998), and close to radii
previously suggested to be significant for setting the scale for
star formation (Larson 1995, Ohashi et al. 1997).  The median
separation is 18000 AU including lower limits for sources with no
detected companions. 
These results suggests that truly isolated star formation is uncommon.

Some Class 0/I sources show extensions, sometimes along the outflow
axis and sometimes perpendicular to it. Both heating and column
density effects may play a role in defining the shapes at low
contour levels.

The mean spectral index for all sources between 450 and 850 \micron\ is
$\mean{\alpha_{450/850}} = 2.8\pm 0.4$. This value would imply
an exponent in the opacity law, $\beta \sim 1$, but Rayleigh-Jeans 
failure could increase this value. 
Pre-protostellar cores have a slightly
lower average spectral index ($\mean{\alpha_{450/850}} = 2.5\pm 0.4$).
The average spectral index measured at 850\micron\ and 1.3mm is higher
($\mean{\alpha_{850/1.3}} = 3.4\pm 0.3$).

The mean mass in the sample is $ \mean{M_d} = 1.1 \pm 0.9 \msun $. 
The masses computed
from the dust emission agree reasonably with those computed from the
virial theorem, supporting the hypothesis that the cores are gravitationally
bound and that the values used for the dust opacity are reasonable.

The radial intensity profiles of \ppc s cannot be fitted with power laws
over a significant range of radii. In contrast, most Class 0/I sources
can be fitted with power laws if the inner and outer points are excluded.
For some sources, the fit must be truncated before emission from secondary
sources affects the profile.
The mean slope is $\mean{m} = 1.48 \pm 0.35$ for Class 0/I sources. 
We are unable to distinguish between Class 0 and Class I radial profiles
with our limited sample. A simple analysis
suggests that a density power law $\rhor \propto r^{-p}$, with $p \sim 2.1$
would fit the data. 

Models that include more accurate \Tdr, 
account for Rayleigh-Jeans failure, and
include the actual beam shape show
that the simple analysis can be misleading.  These models still 
can be fitted by power laws in the normalized intensity, but the fitted
slope may vary by $\pm 0.5$ compared to the simple analysis.

Given the likely complexities in real cores, it is somewhat surprising
that simple power-law models fit the Class 0/I sources as well as they
do.  
Neither the mean spectral indices nor the slopes in the intensity
profiles distinguish between Class 0 and Class I sources nor between
candidates and non-candidates for collapse in the present sample.
Firmer conclusions await a larger sample of Class I sources and
detailed, source-by-source modeling.

\section*{Acknowledgments}

We are grateful to T. Jenness for assistance with SURF, 
to L. Mundy for providing the computer code used to
generate the models in Figure 13, and to the staff of the JCMT for crucial 
support while observing. We thank Claire Chandler and the
referee for comments that improved the paper.
The JCMT is operated by the Joint Astronomy Centre on behalf of the Particle 
Physics and
Astronomy Research Council of the United Kingdom, The Netherlands Organization
for Scientific Research and the National Research Council of Canada.
We thank the State of Texas and NASA (Grant NAG5-7203) for support.
NJE thanks the Fulbright Program and PPARC for support while at University
College London and NWO and NOVA for support in Leiden.



\begin{figure}
\figurenum{1}
\plotone{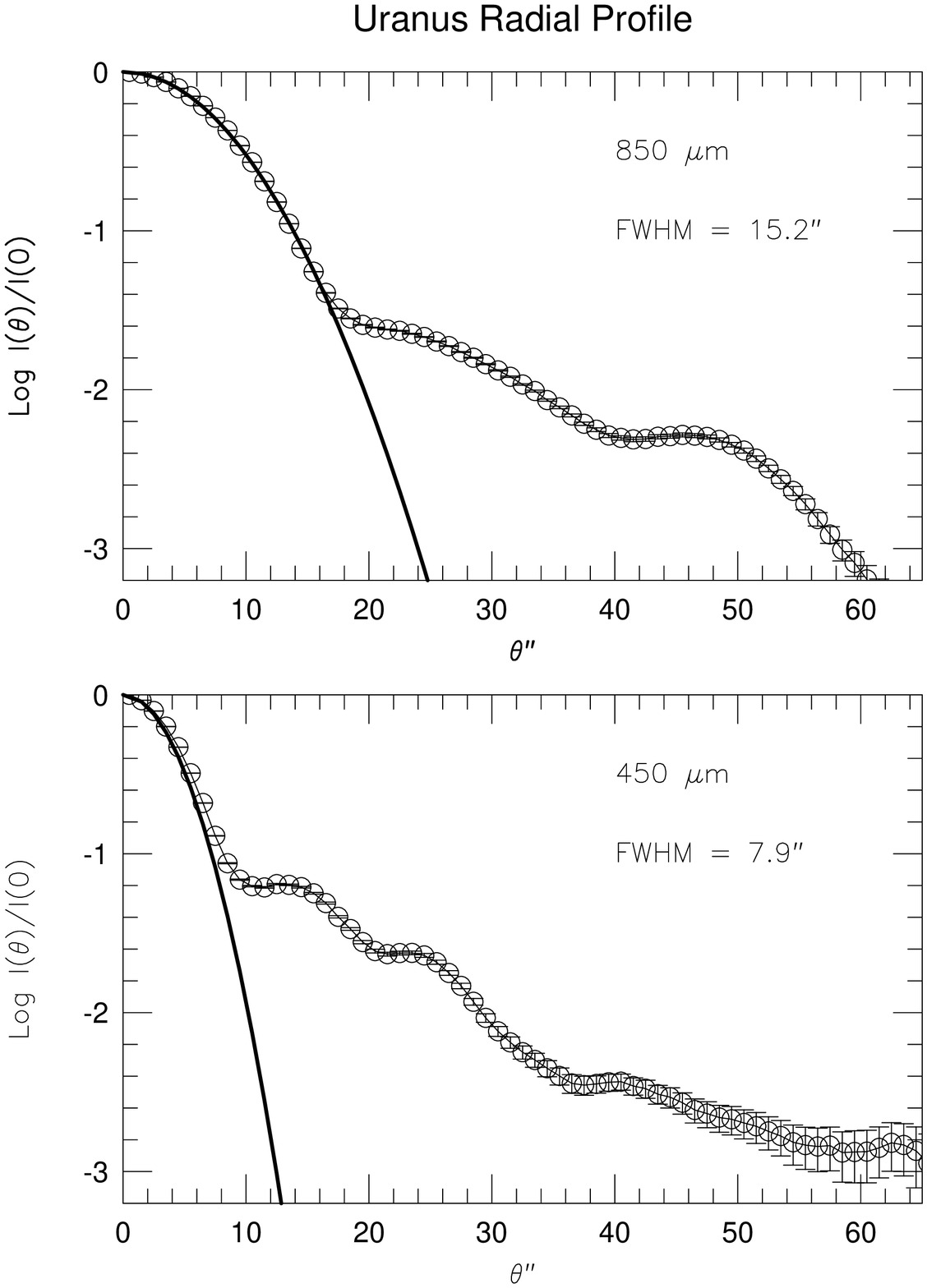}
\figcaption{
The average Uranus radial profile from the April 1998 run, based
on an image with 1\arcsec\ pixels.   
Nine maps were averaged for the 850 \micron\ profile, and eight maps were
averaged for the 450 \micron\ profile.
The normalized intensity is
plotted as a function of angle from the center.  The solid line shows a 
gaussian profile with the FWHM of the main beam.
}
\end{figure}

\begin{figure}
\figurenum{2}
\centering
 \vspace*{7.8cm}
   \leavevmode
   \includegraphics{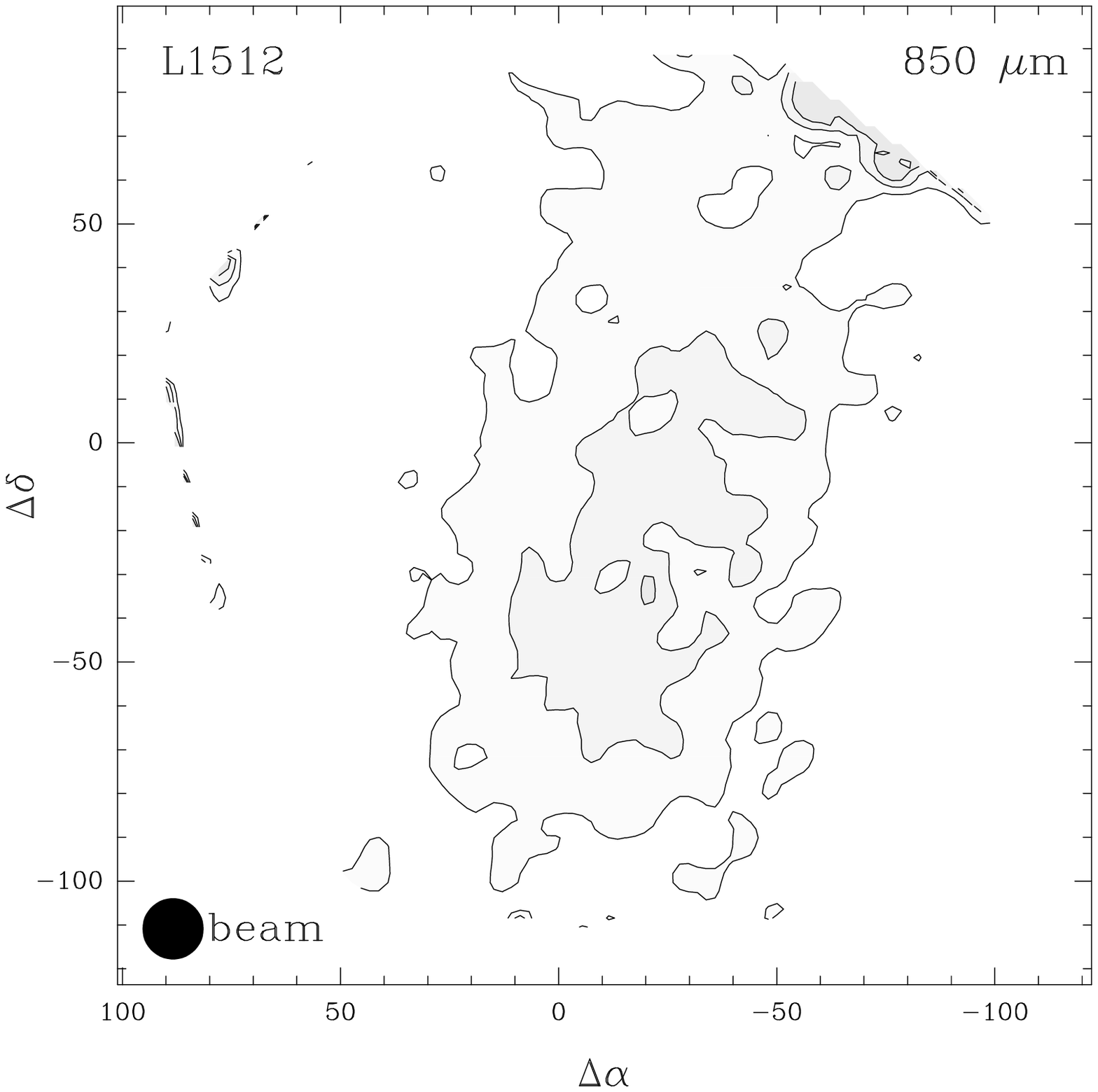}
   \includegraphics{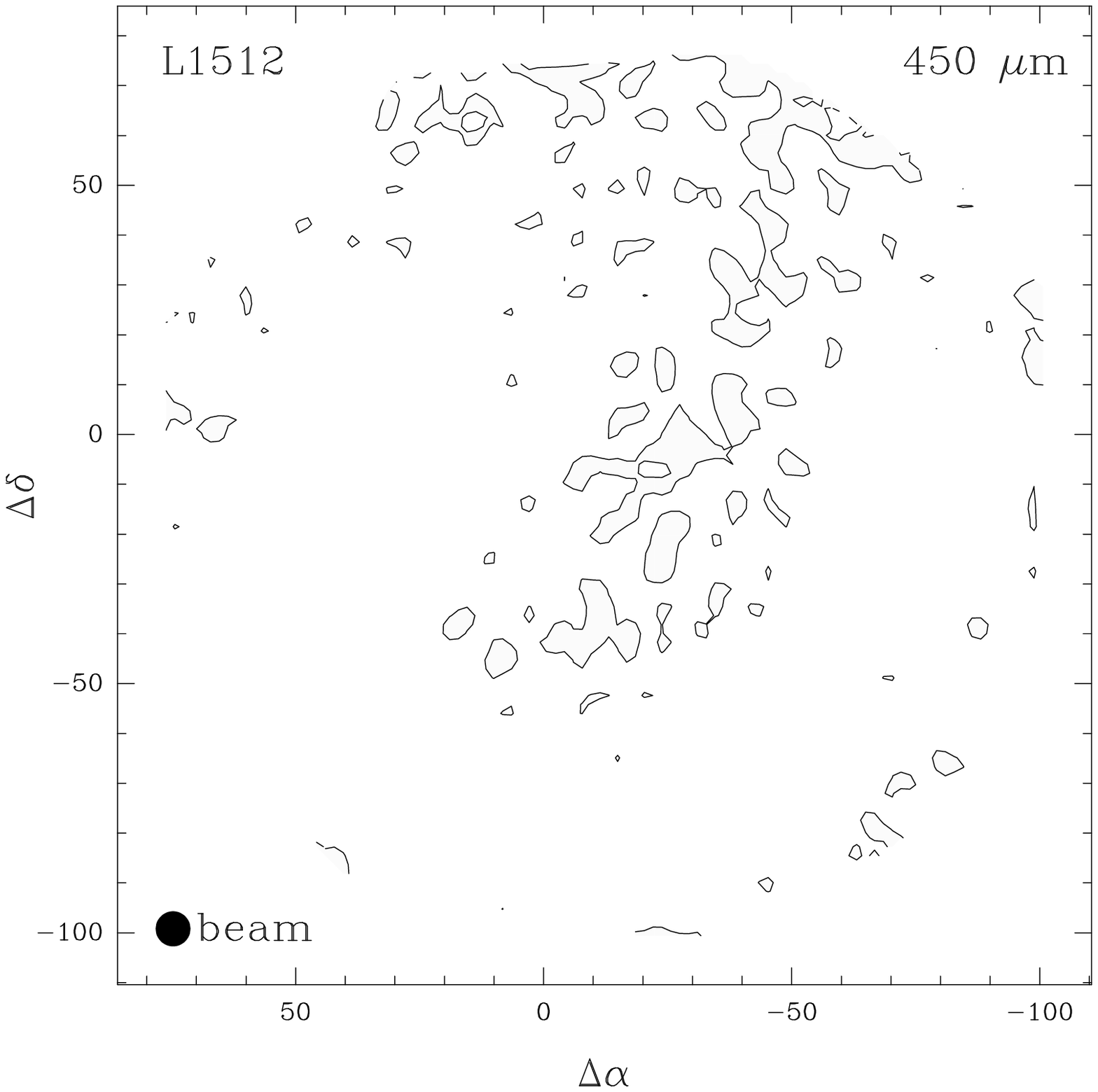}
   \includegraphics{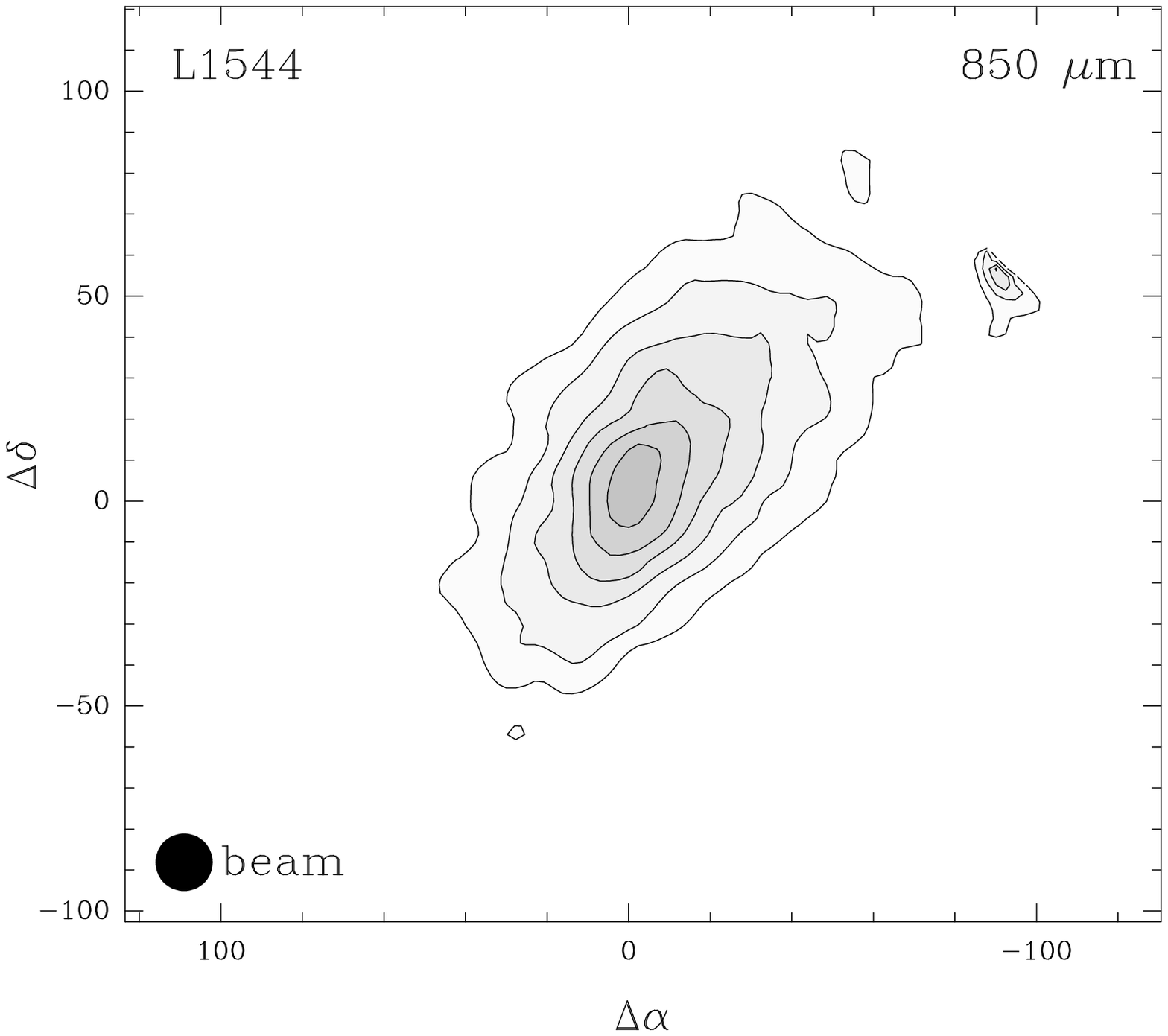}
   \includegraphics{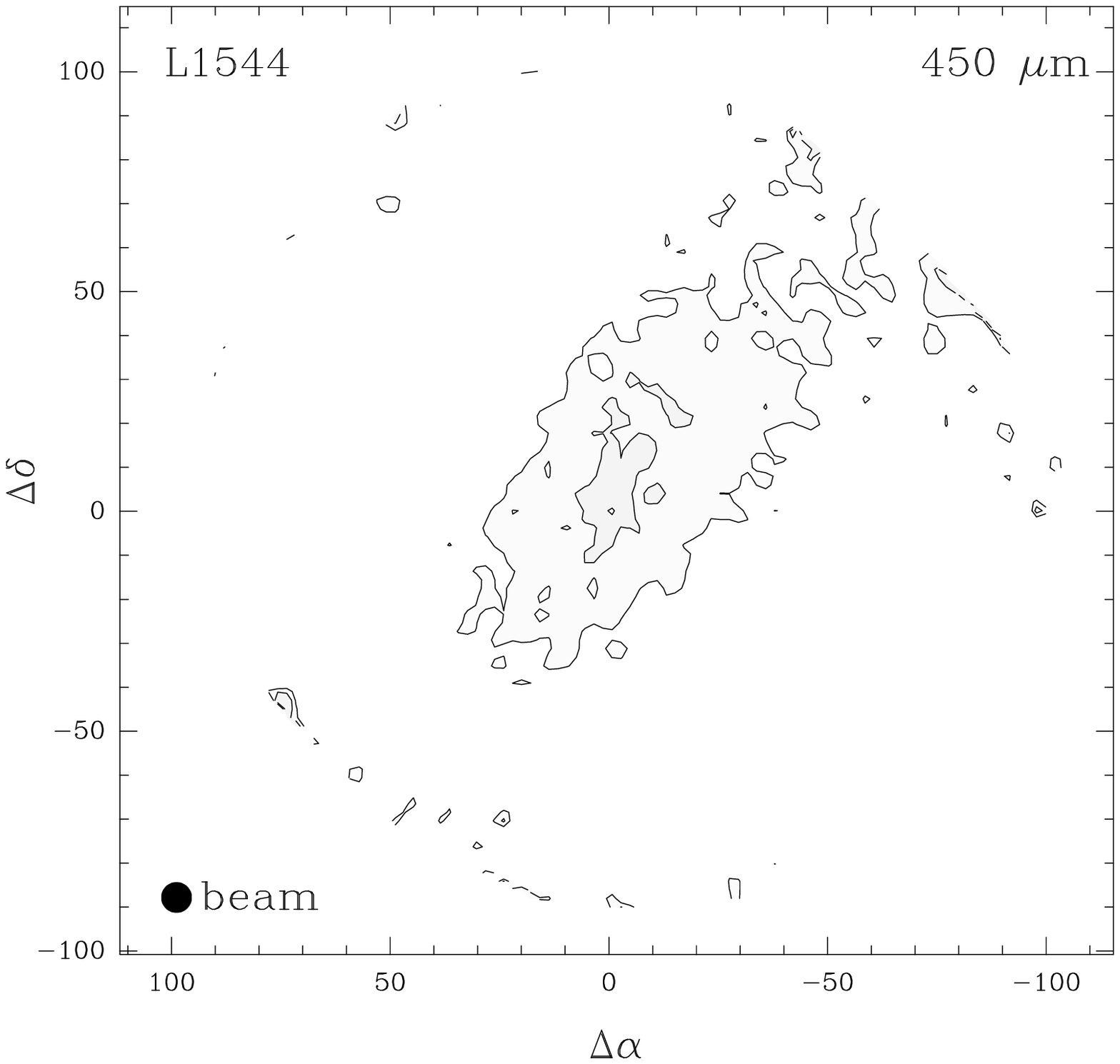}
   \includegraphics{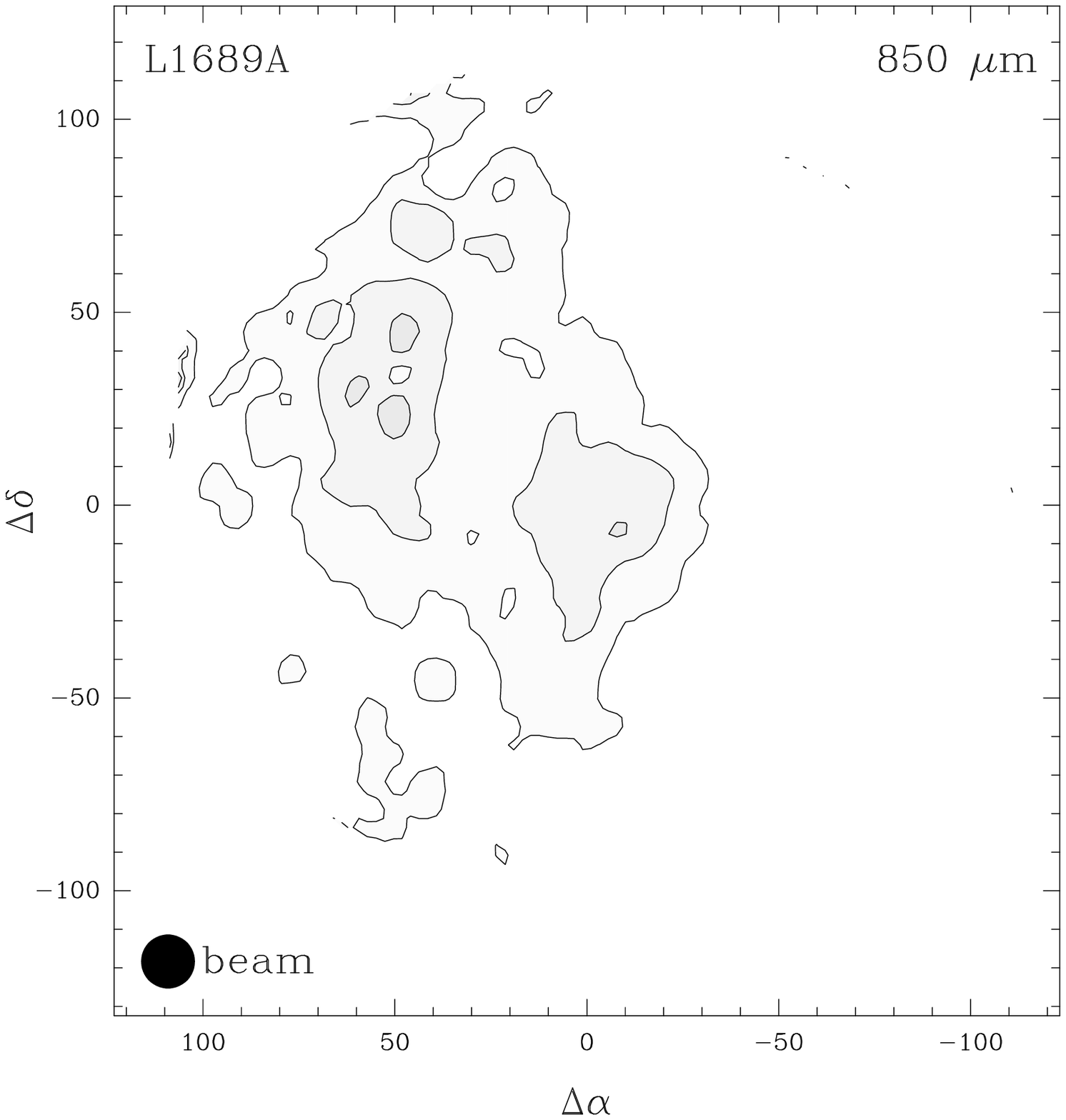}
   \includegraphics{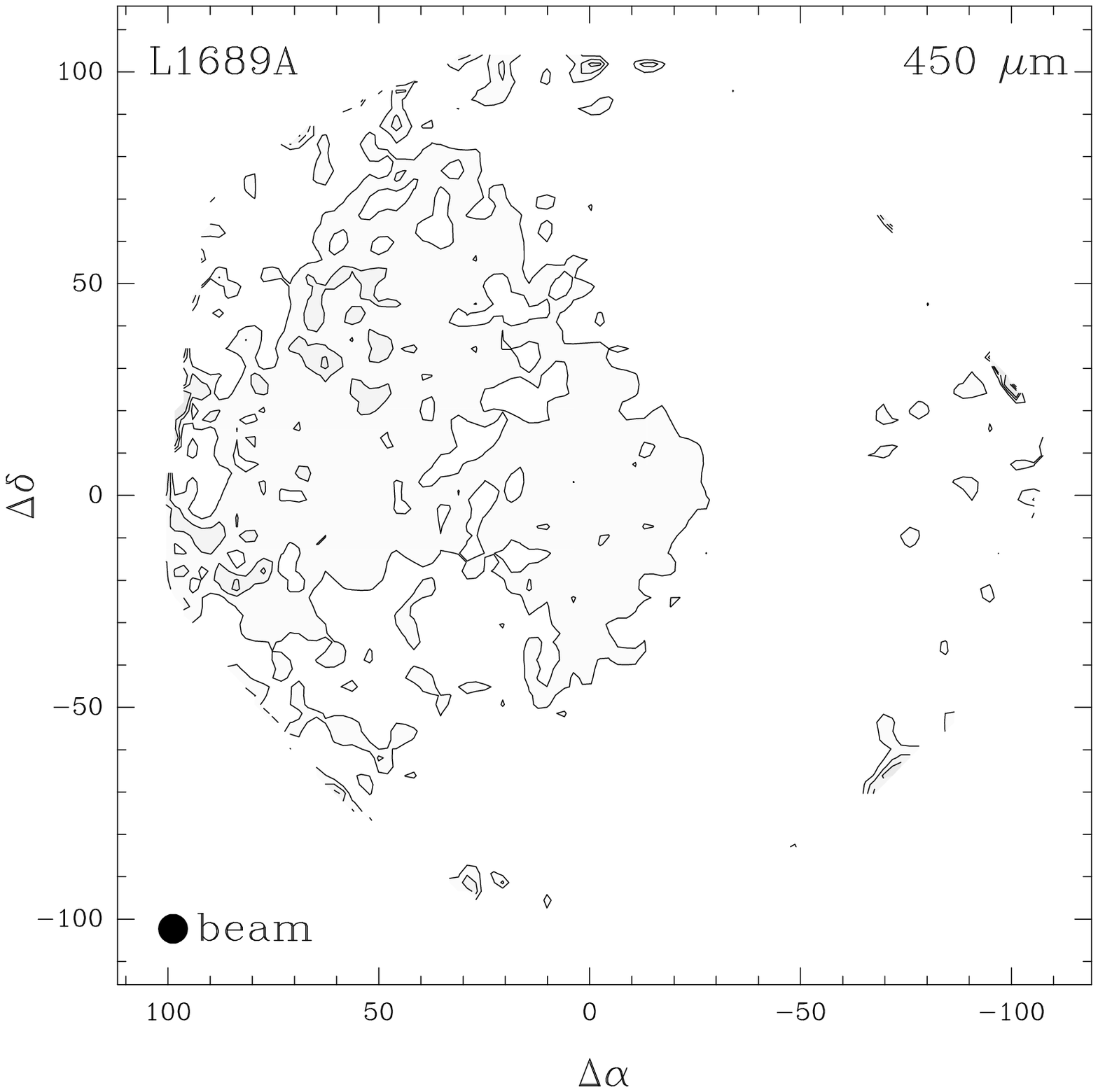}
\vskip 4.25in
\figcaption{
Contour maps of \ppc s.  
The left column is 850 \micron\ and the right column is 450 \micron.
The contour levels are as follows (lowest contour and contour increment
in percentage of the peak flux). 
L1512 (850\micron) 40\%(3$\sigma$) increasing by 29\%(2$\sigma$). 
L1512 (450\micron) 68\%(3$\sigma$) increasing by 45\%(2$\sigma$). 
L1544 (850\micron) 20\%(3$\sigma$) increasing by 13\%(2$\sigma$). 
L1544 (450\micron) 50\%(3$\sigma$) increasing by 33\%(2$\sigma$). 
L1689A (850\micron) 40\%(3$\sigma$) increasing by 26\%(2$\sigma$). 
L1689A (450\micron) 41\%(3$\sigma$) increasing by 33\%(2$\sigma$).
Contours near the edge of the maps should be ignored due to 
noisy pixels, less integration time, and inability of the plotting
package to handle irregular edges.
}
\end{figure}

\begin{figure}
\figurenum{3}
\centering
 \vspace*{7.8cm}
   \leavevmode
   \includegraphics{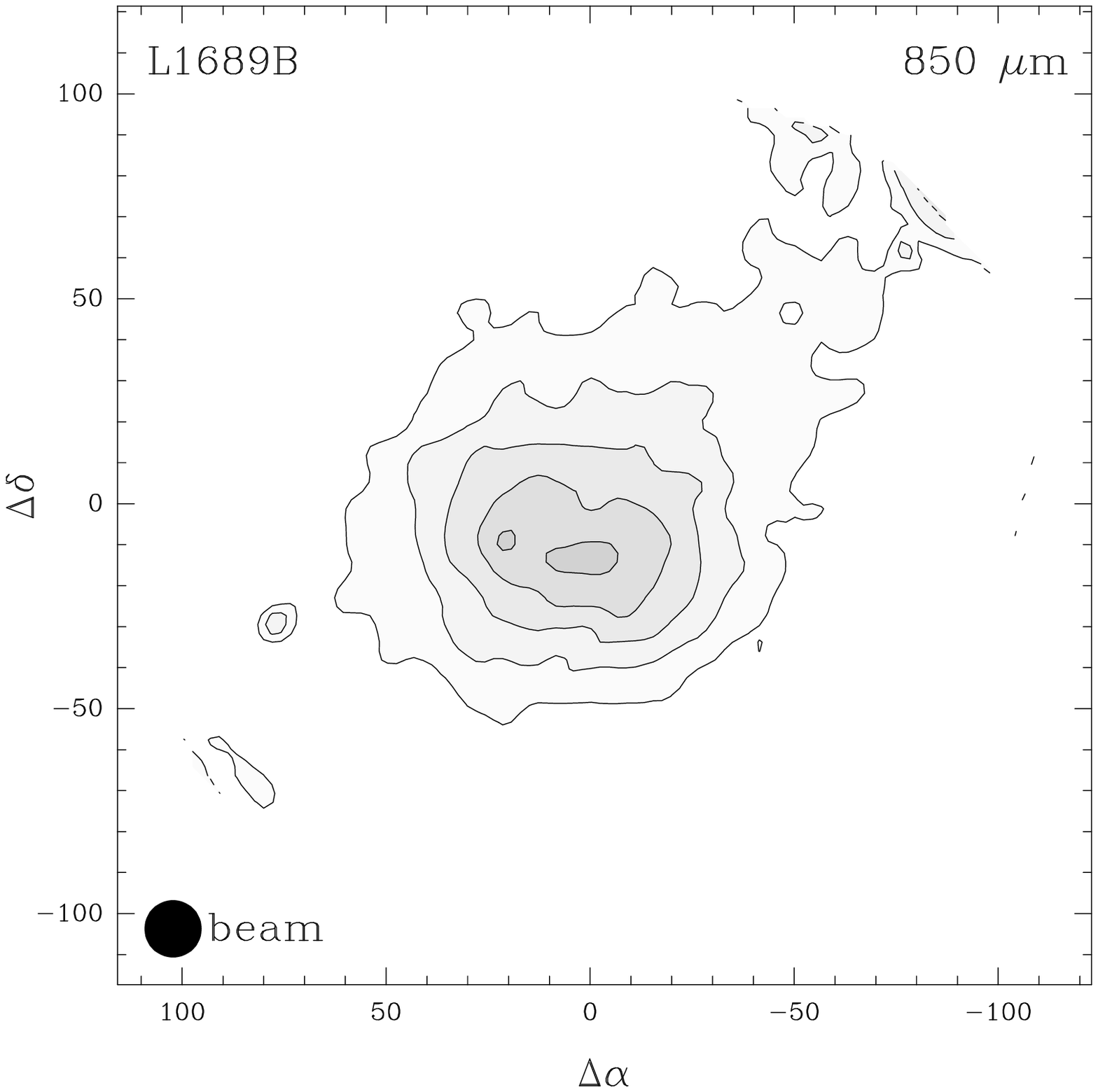}
   \includegraphics{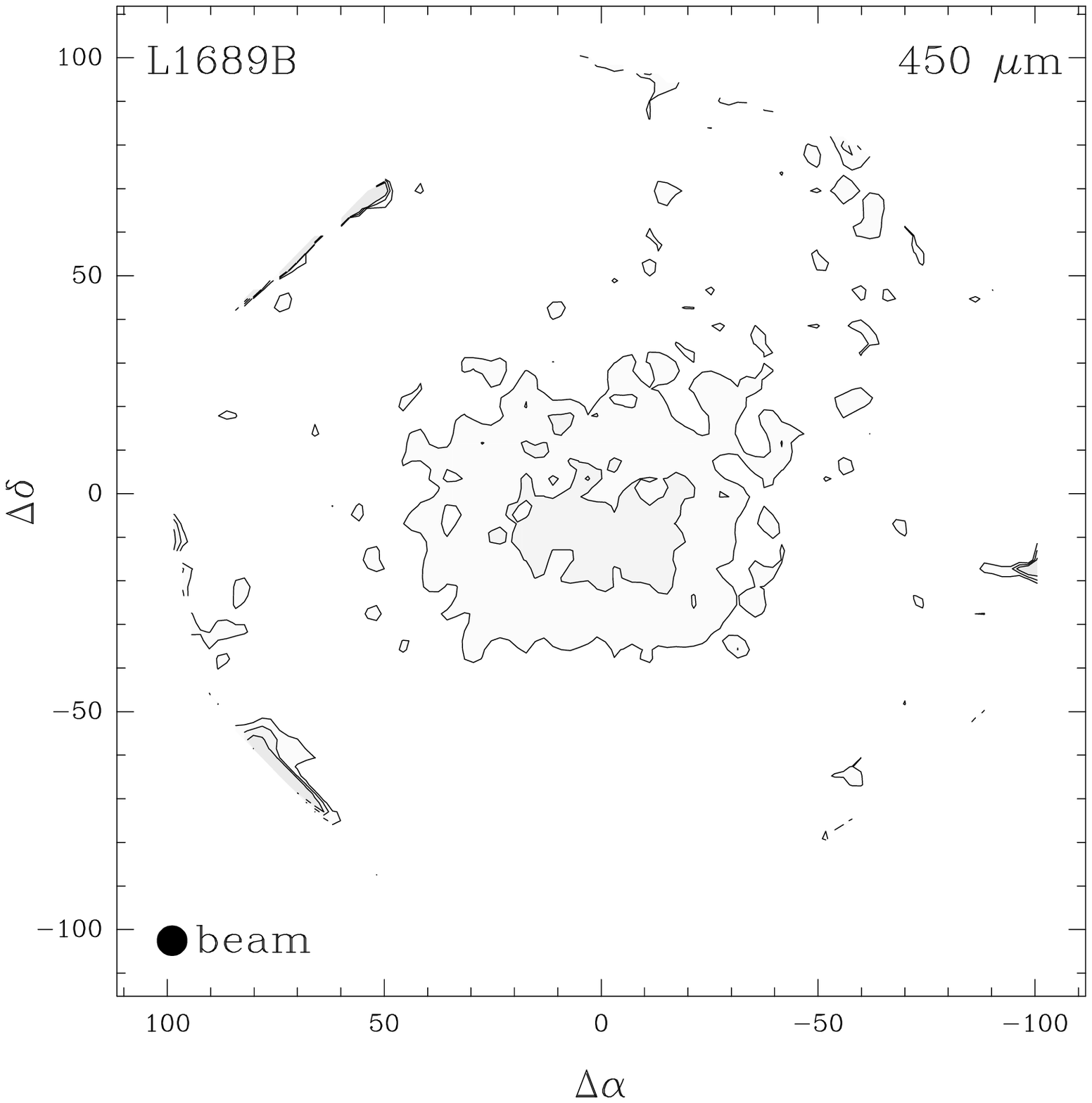}
   \includegraphics{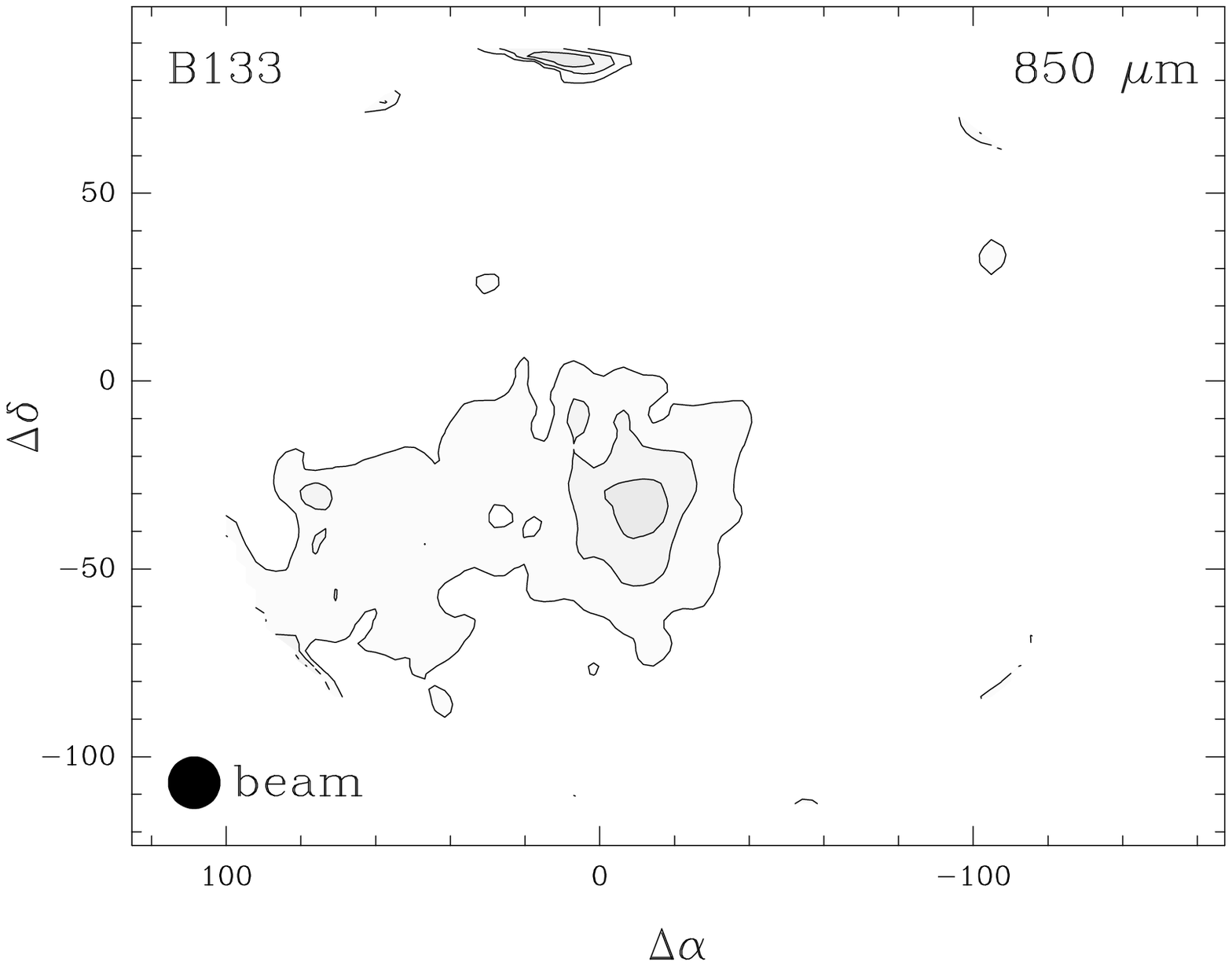}
   \includegraphics{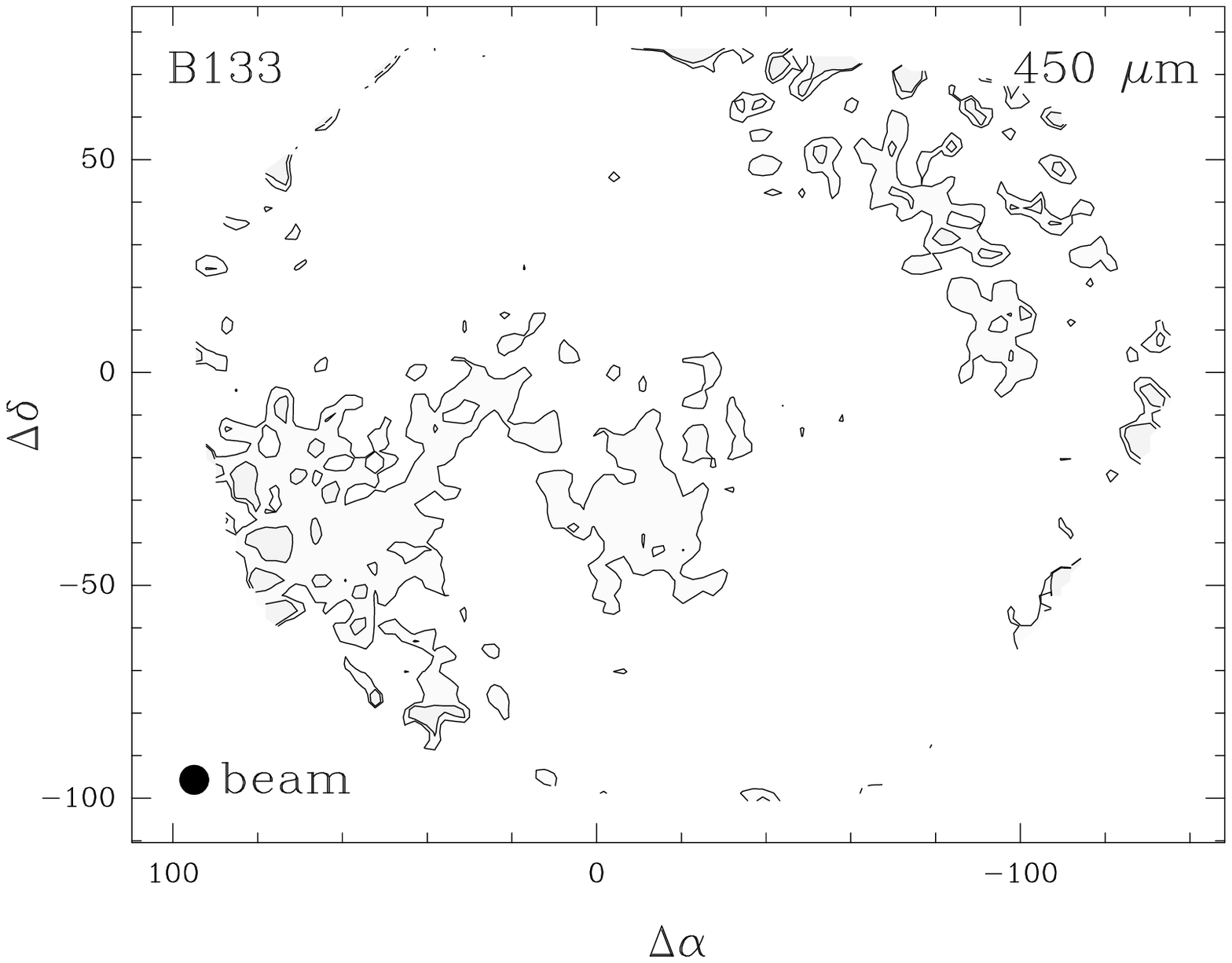}
\vskip 4.25in
\figcaption{
Contour maps of \ppc s.  
The left column is 850 \micron\ and the right column is 450 \micron.
The contour levels are as follows (lowest contour and contour increment
in percentage of the peak flux).  
L1689B (850\micron) 27\%(3$\sigma$) increasing by 18\%(2$\sigma$). 
L1689B (450\micron) 41\%(3$\sigma$) increasing by 27\%(2$\sigma$).
B133 (850\micron) 36\%(3$\sigma$) increasing by 24\%(2$\sigma$).
B133 (450\micron) 59\%(3$\sigma$) increasing by 39\%(2$\sigma$). 
Contours near the edge of the maps should be ignored due to 
noisy pixels, less integration time, and inability of the plotting
package to handle irregular edges.
}
\end{figure}

\begin{figure}
\figurenum{4}
\centering
 \vspace*{7.8cm}
   \leavevmode
   \includegraphics{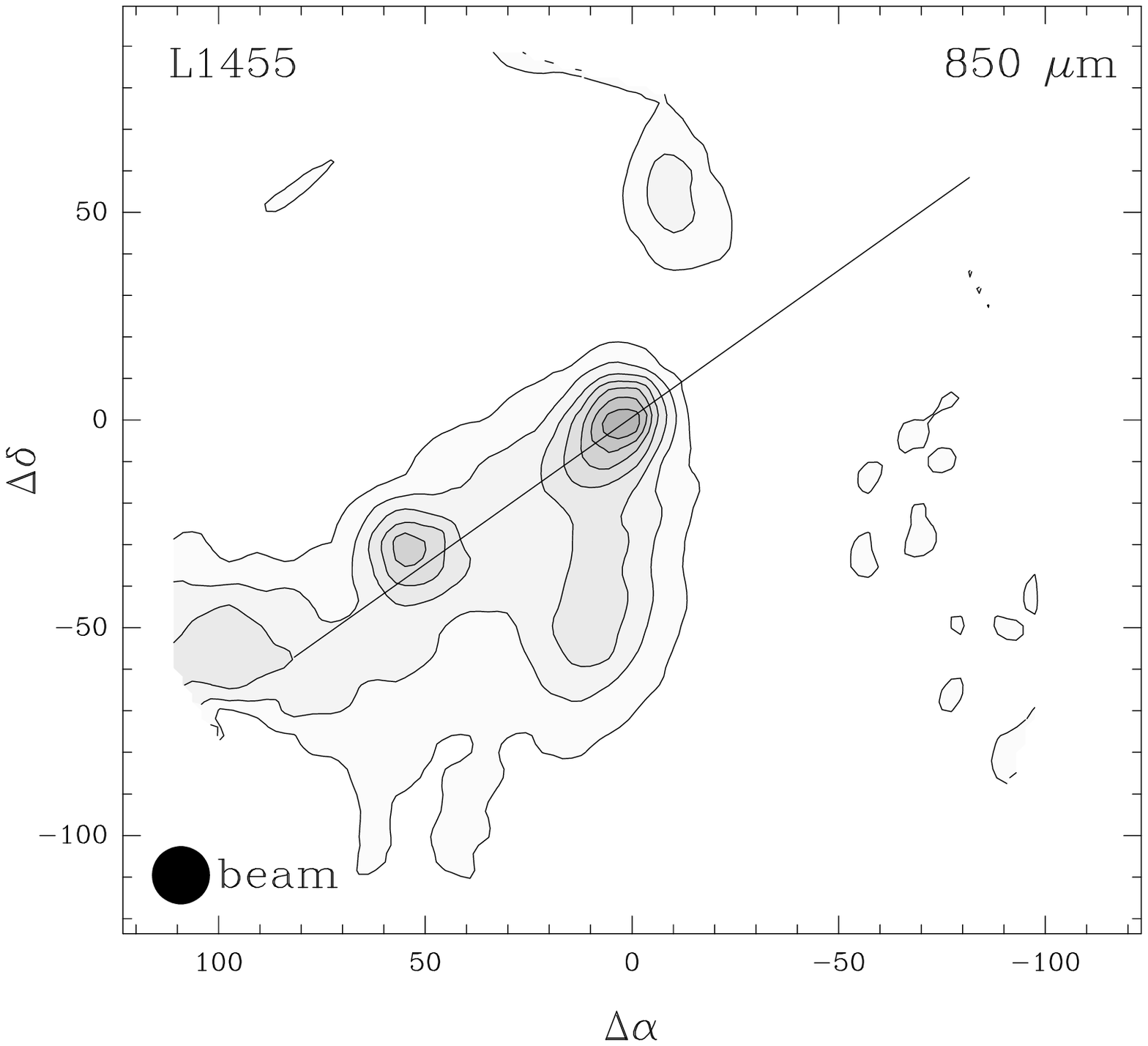}
   \includegraphics{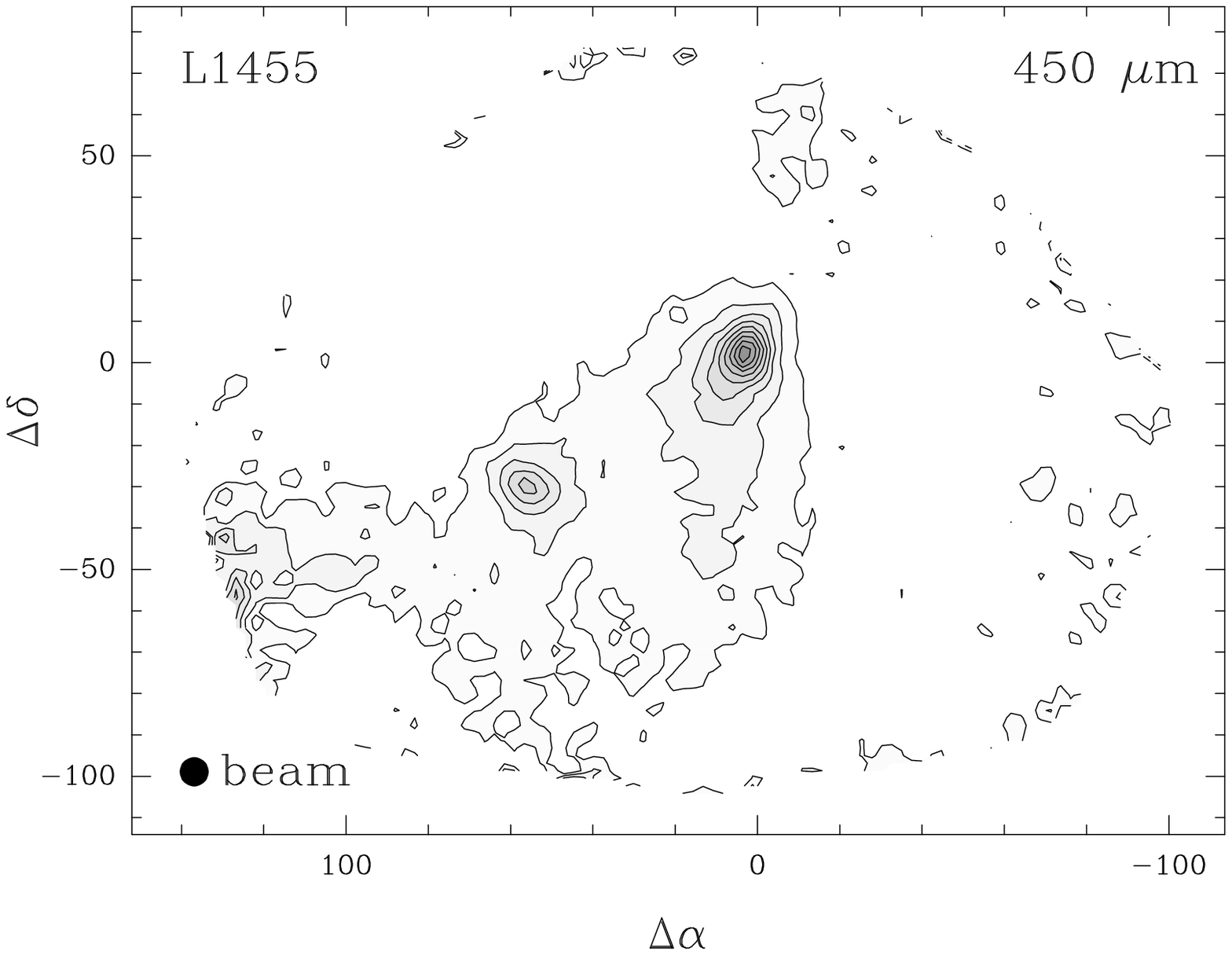}
   \includegraphics{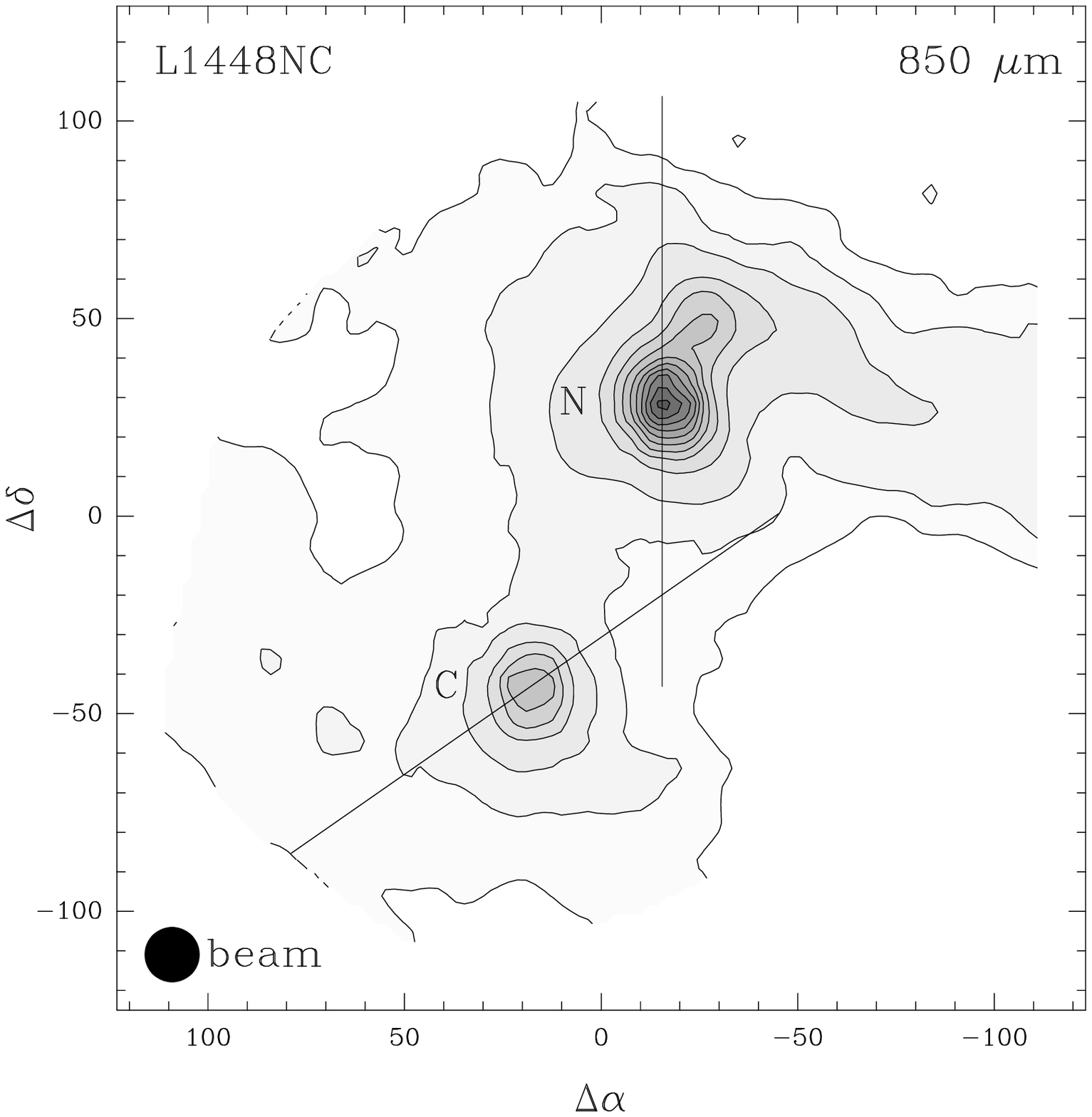}
   \includegraphics{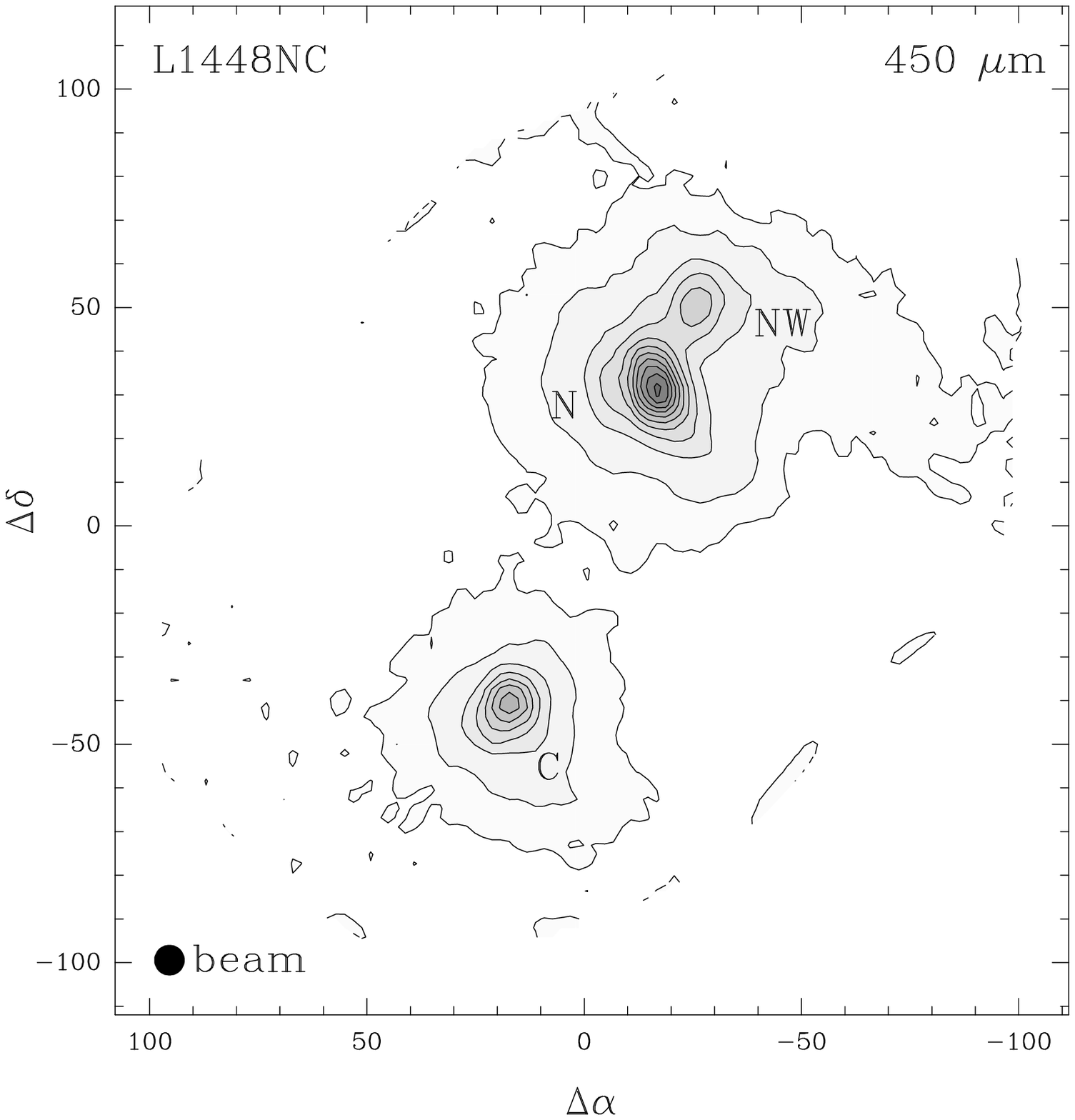}
   \includegraphics{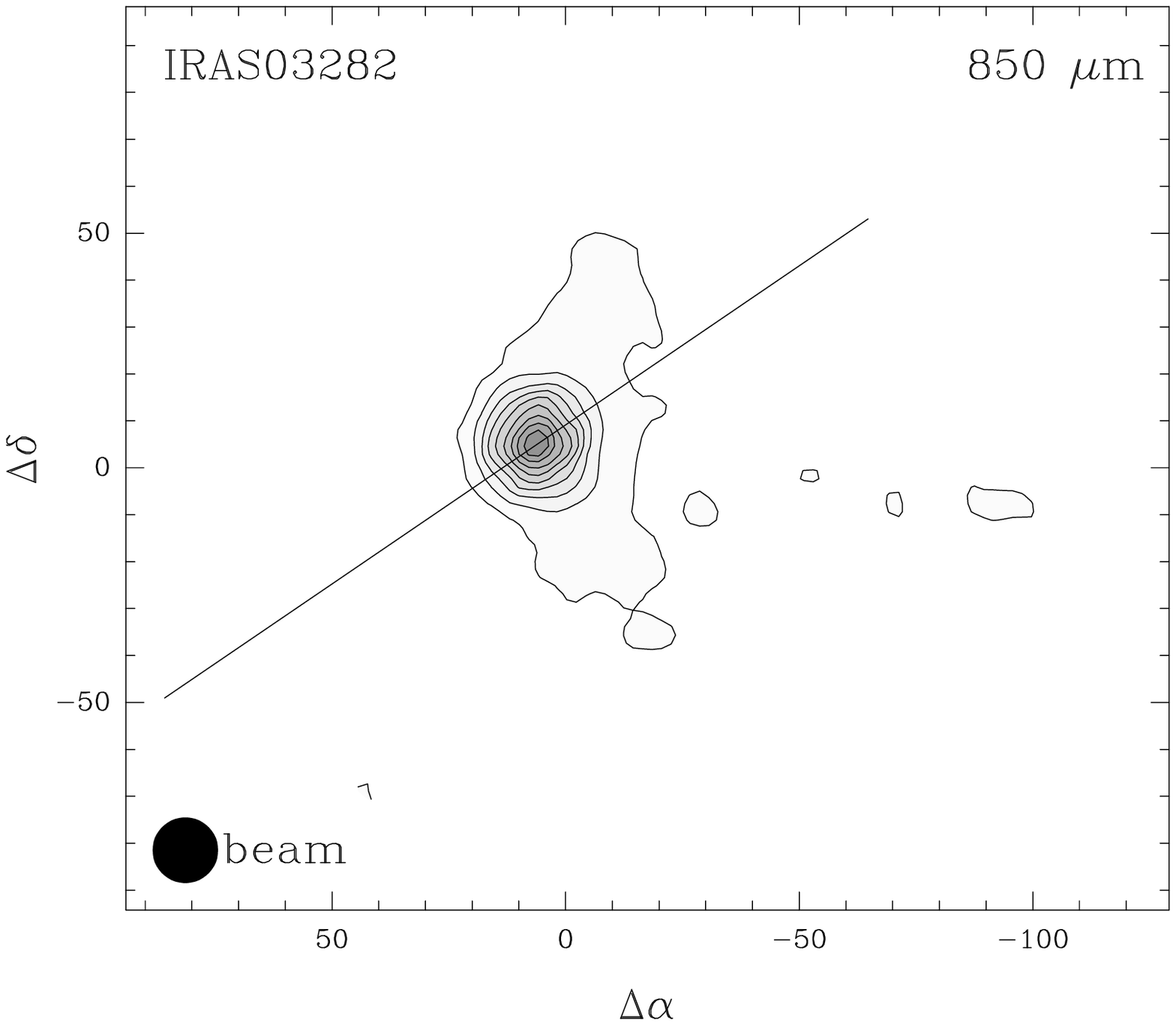}
   \includegraphics{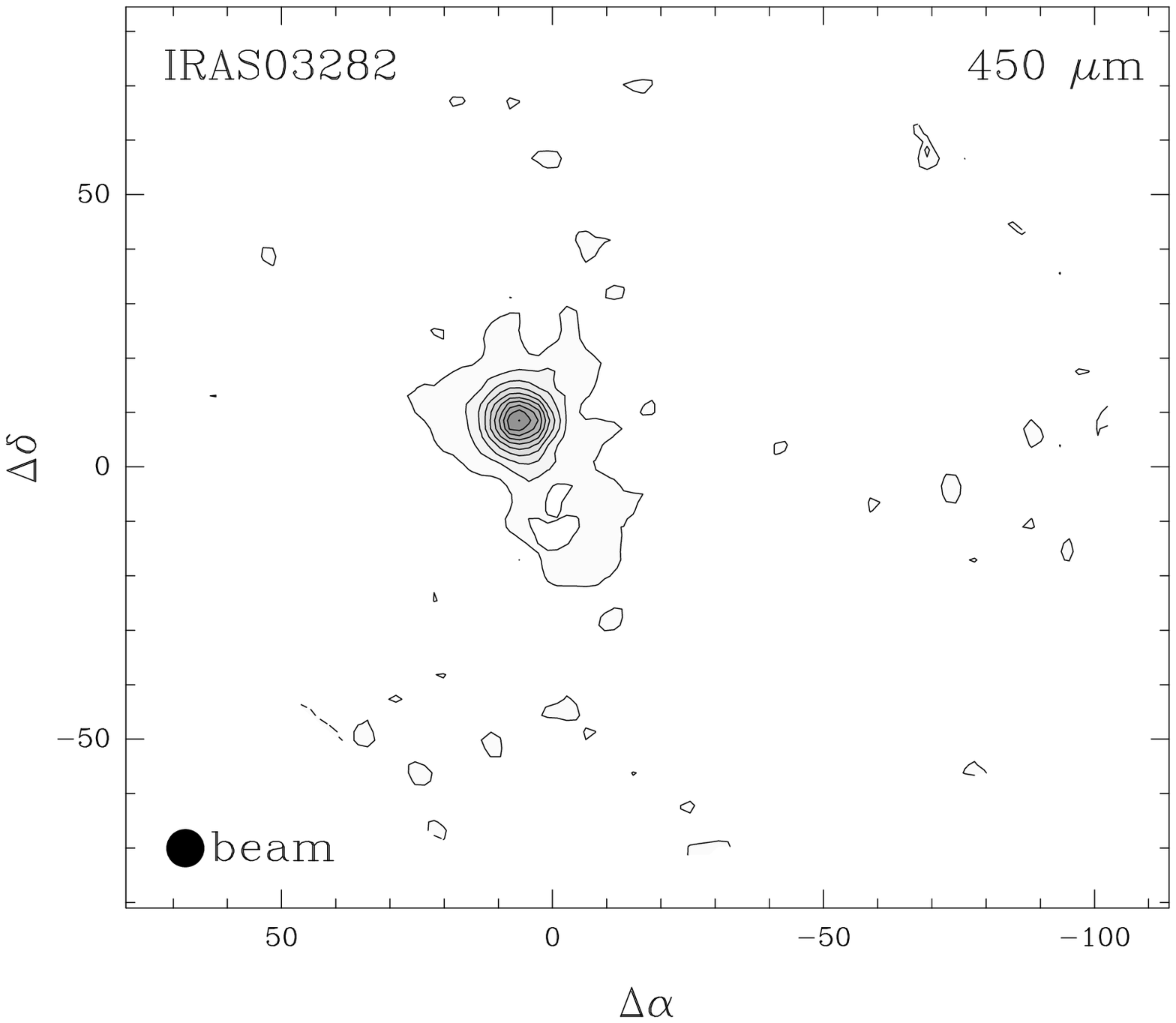}
\vskip 4.25in
\figcaption{
Contour maps of Class 0 sources.  
The left column is 850 \micron\ and the right column is 450 \micron.
The solid line indicates the outflow direction.
The contour levels are as follows (lowest contour and contour increment
in percentage of the peak flux).  
L1455 (850\micron) 10\%(7$\sigma$) increasing by 10\%.
L1455 (450\micron) 10\%(4$\sigma$) increasing by 10\%.
L1448N (850\micron) 2\%(7$\sigma$), 5\%(19$\sigma$), 10\%(37$\sigma$)
increasing by 10\%.
L1448N (450\micron) 5\%(5$\sigma$), 10\%(10$\sigma$) increasing by 10\%.
IRAS03282+3035 (850\micron) 10\%(6$\sigma$) increasing by 10\%.
IRAS03282+3035 (450\micron) 10\%(5$\sigma$) increasing by 10\%. 
Contours near the edge of the maps should be ignored due to 
noisy pixels, less integration time, and inability of the plotting
package to handle irregular edges.
}
\end{figure}

\begin{figure}
\figurenum{5}
\centering
 \vspace*{7.8cm}
   \leavevmode
   \includegraphics{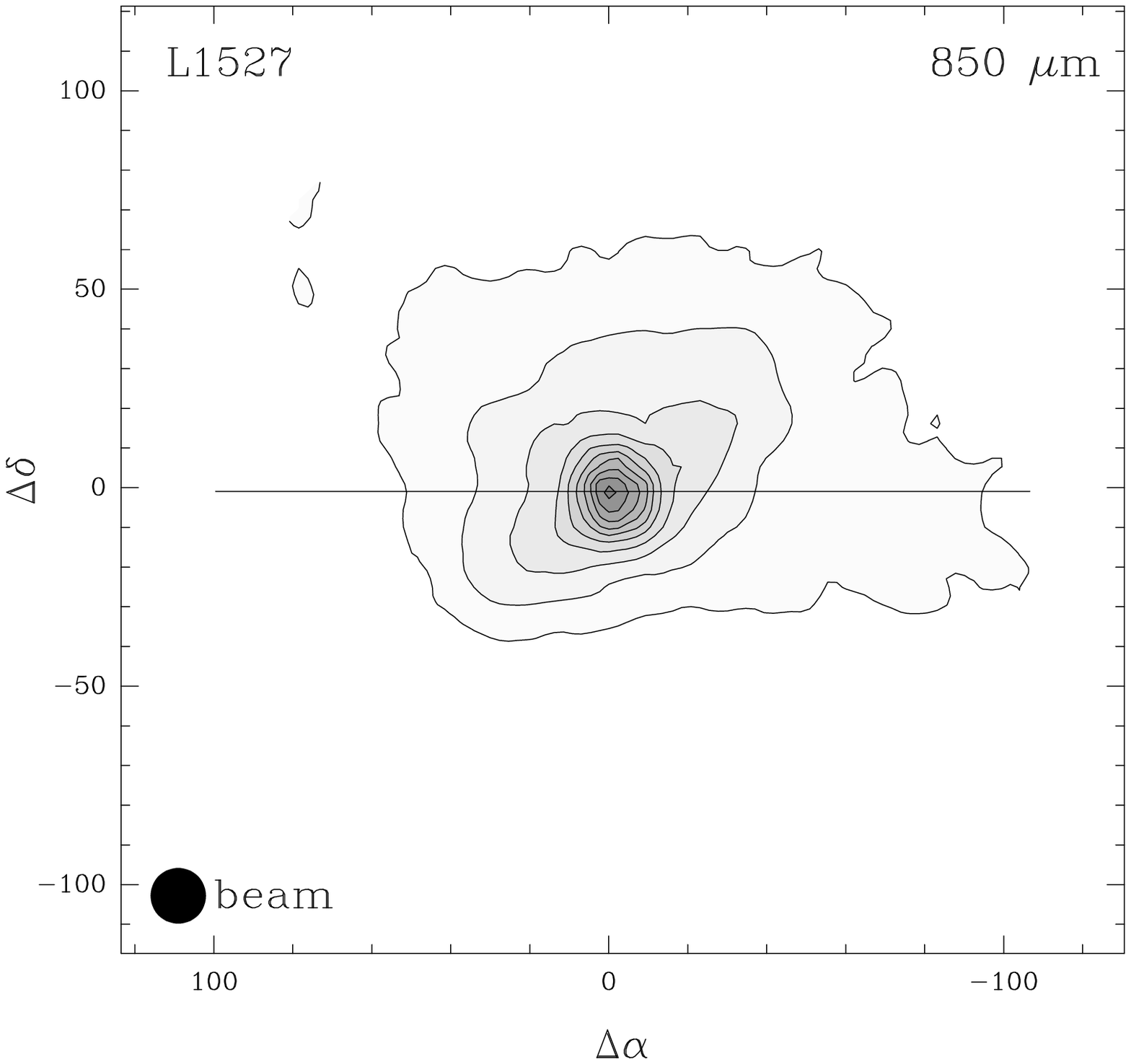}
   \includegraphics{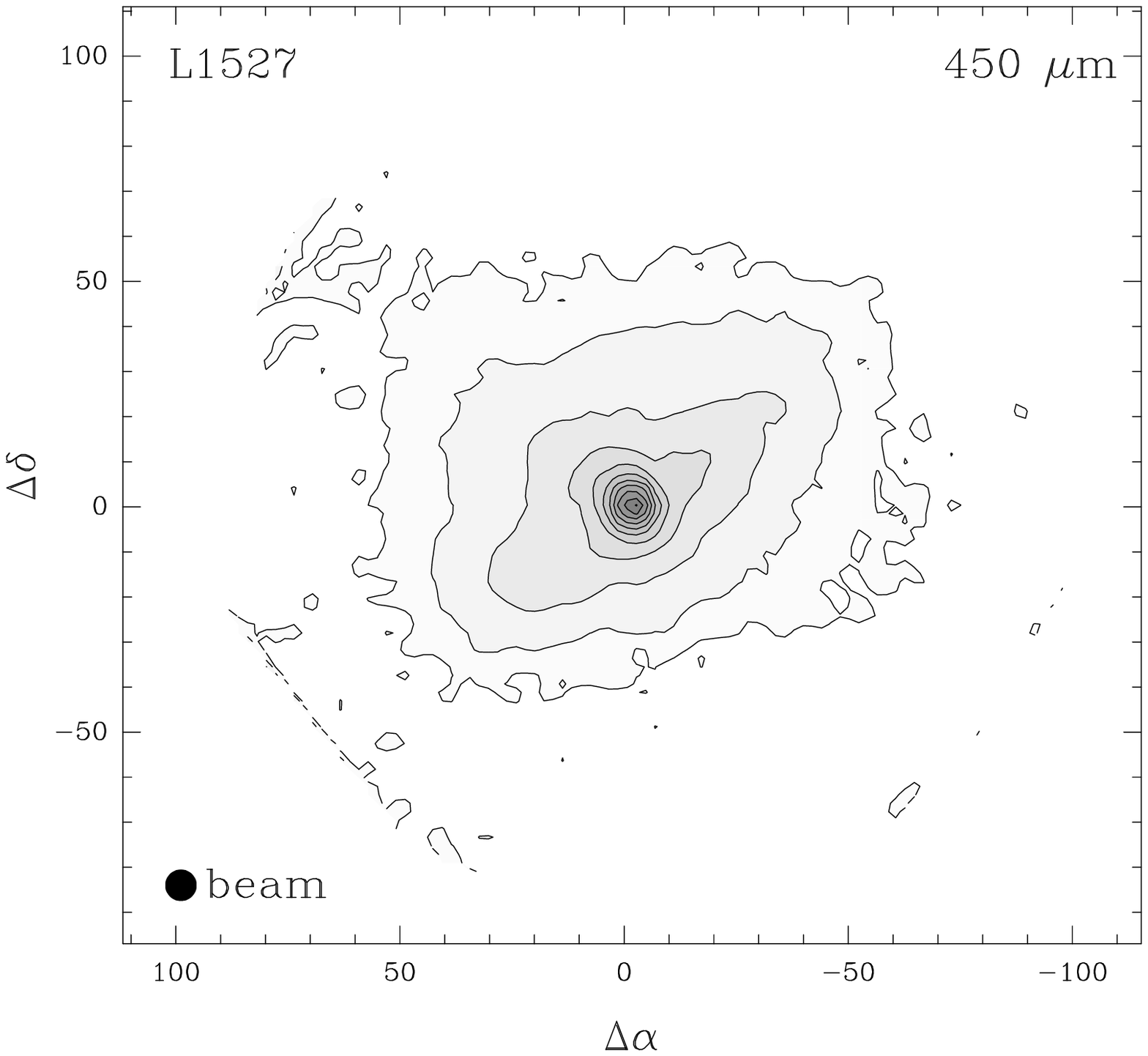}
   \includegraphics{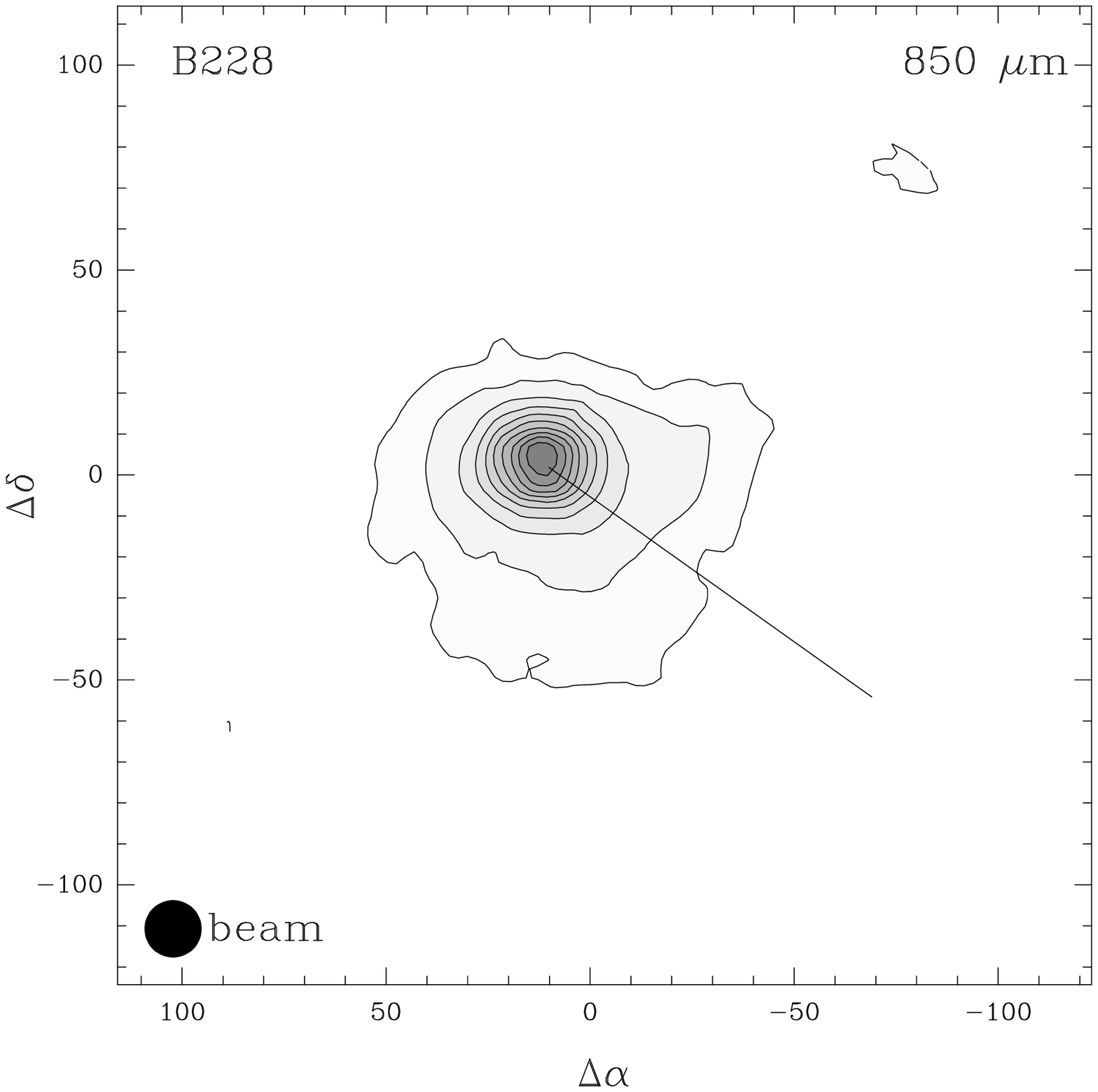}
   \includegraphics{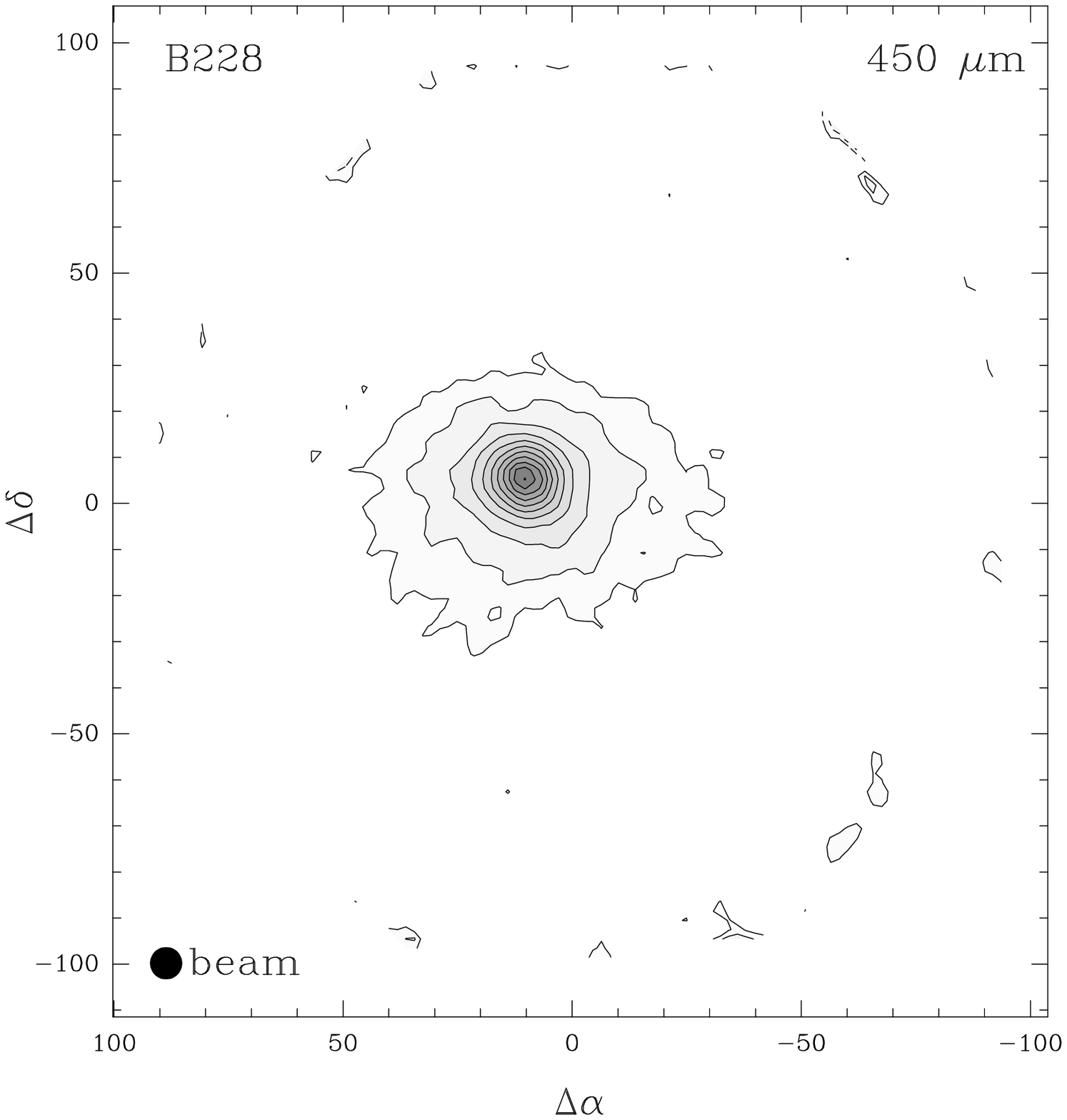}
   \includegraphics{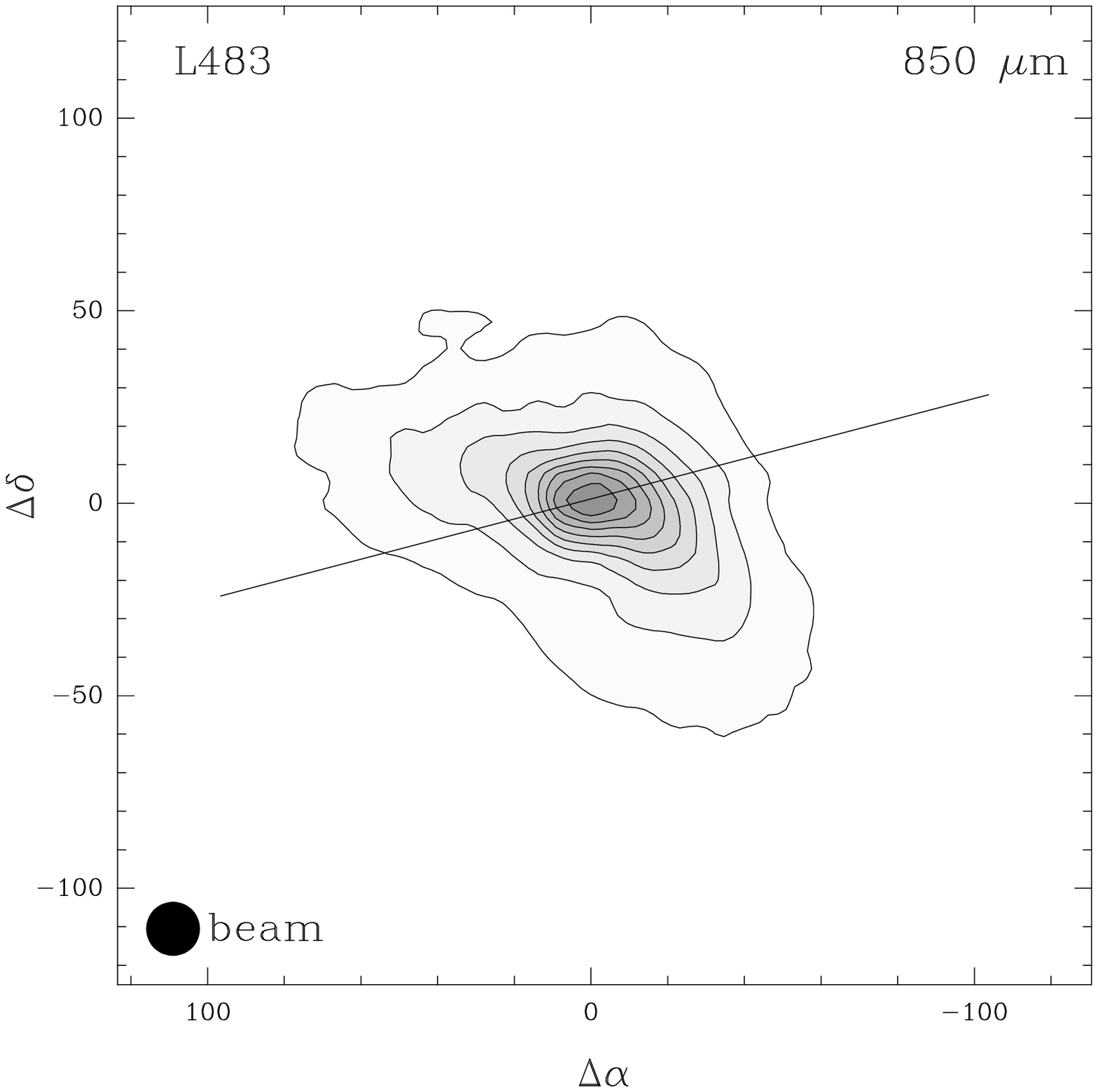}
   \includegraphics{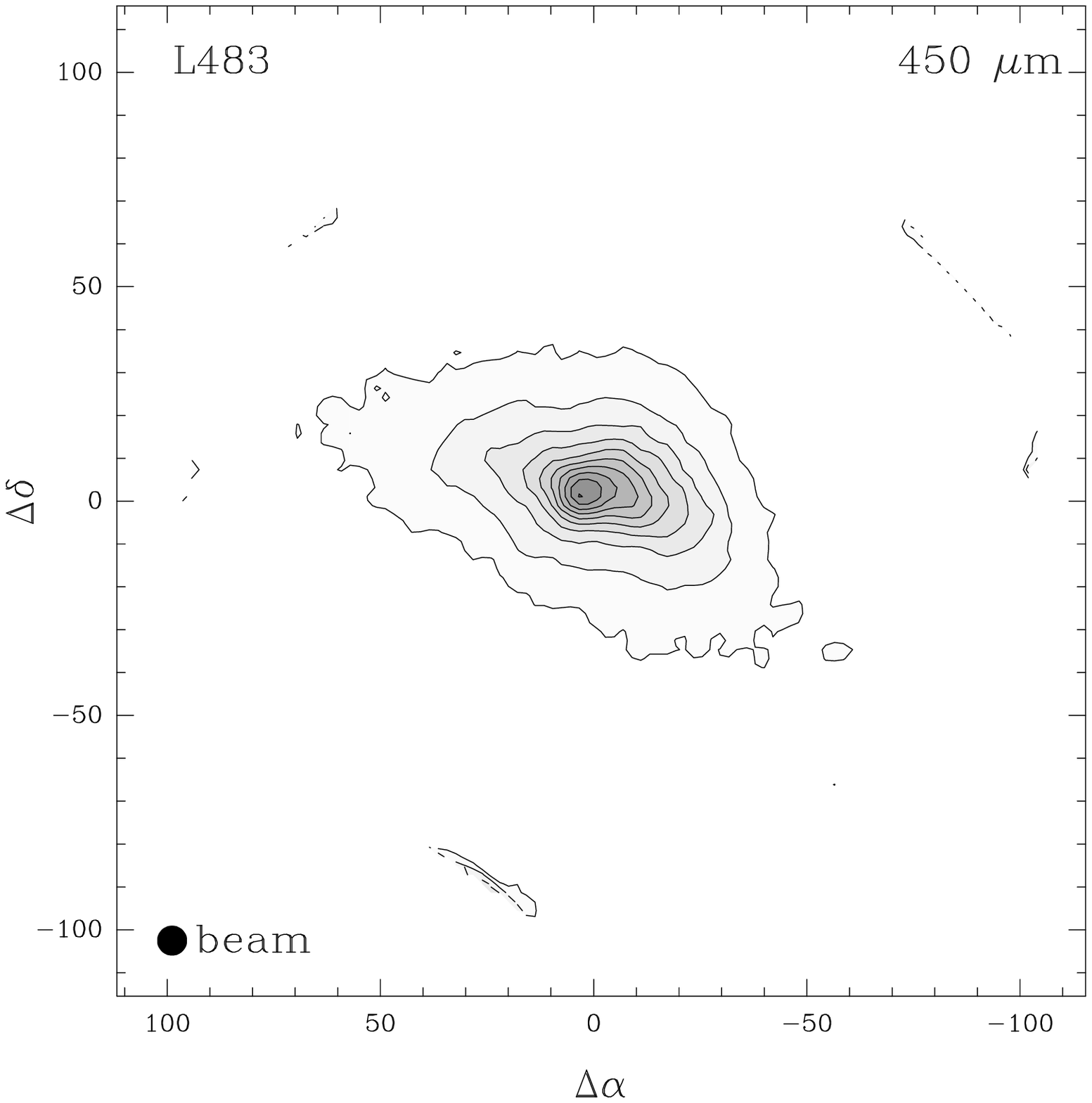}
\vskip 4.25in
\figcaption{
Contour maps of Class 0 sources.  
The left column is 850 \micron\ and the right column is 450 \micron.
The solid line indicates the outflow direction.
The contour levels are as follows (lowest contour and contour increment
in percentage of the peak flux).  
L1527 (850\micron) 10\%(4$\sigma$) increasing by 10\%.
L1527 (450\micron) 5\%(5$\sigma$), 10\%(8$\sigma$) increasing by 10\%.
B228 (850\micron) 5\%(4$\sigma$), 10\%(8$\sigma$) increasing by 10\%.
B228 (450\micron) 5\%(3$\sigma$), 10\%(6$\sigma$) increasing by 10\%.
L483 (850\micron) 10\%(5$\sigma$) increasing by 10\%.
L483 (450\micron) 10\%(6$\sigma$) increasing by 10\%.
Contours near the edge of the maps should be ignored due to 
noisy pixels, less integration time, and inability of the plotting
package to handle irregular edges.
}
\end{figure}

\begin{figure}
\figurenum{6}
\centering
 \vspace*{7.8cm}
   \leavevmode
   \includegraphics{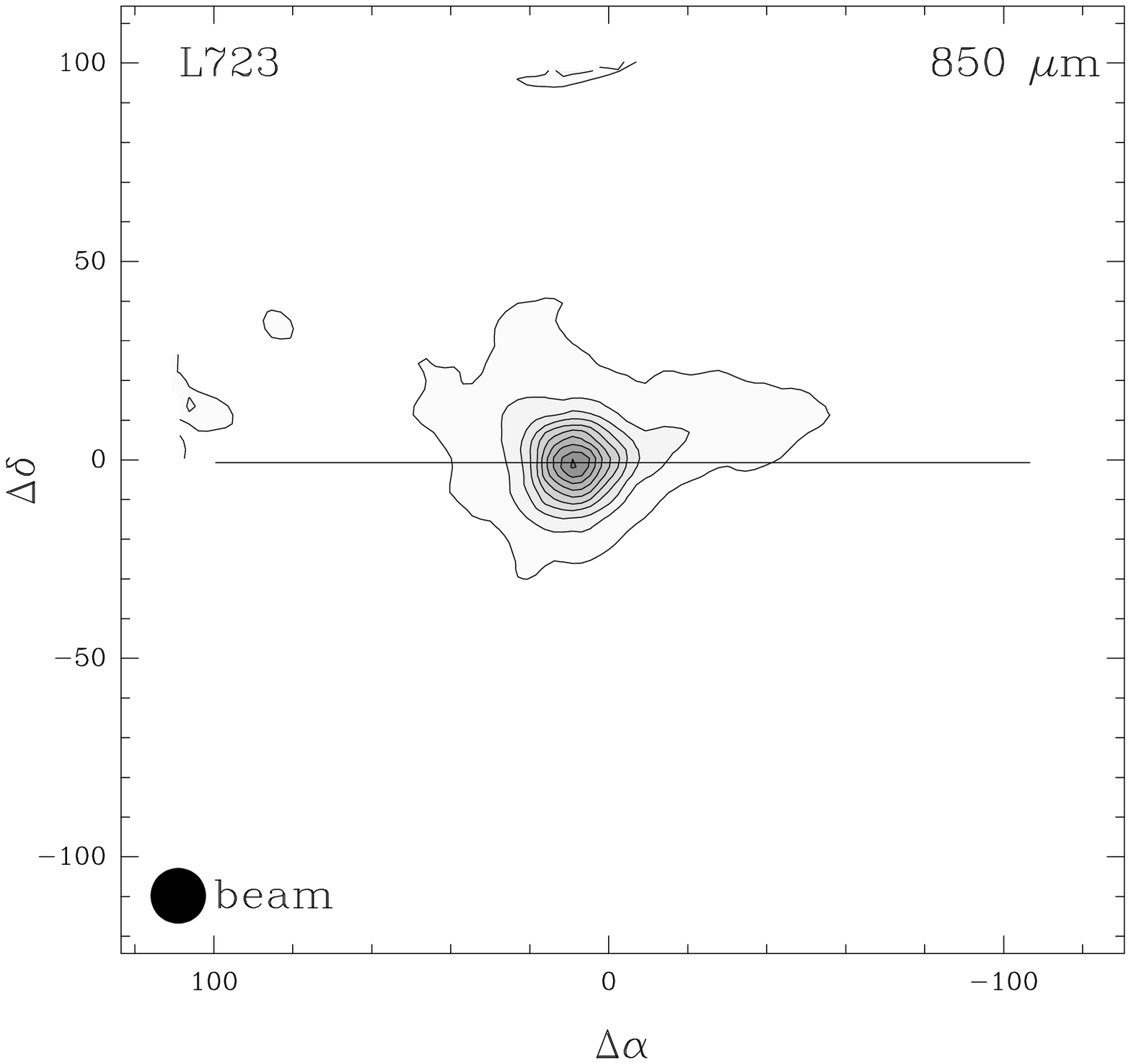}
   \includegraphics{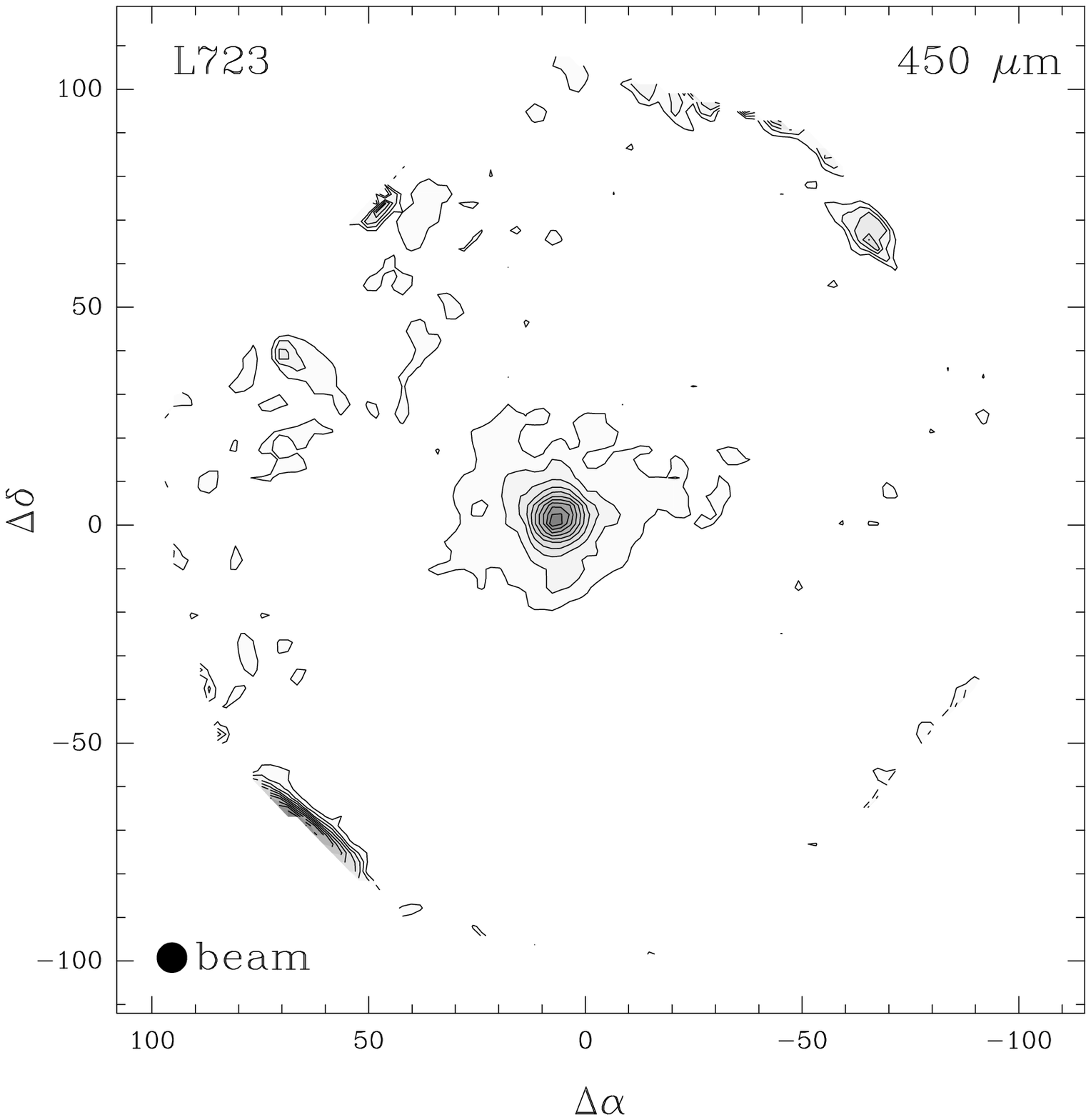}
   \includegraphics{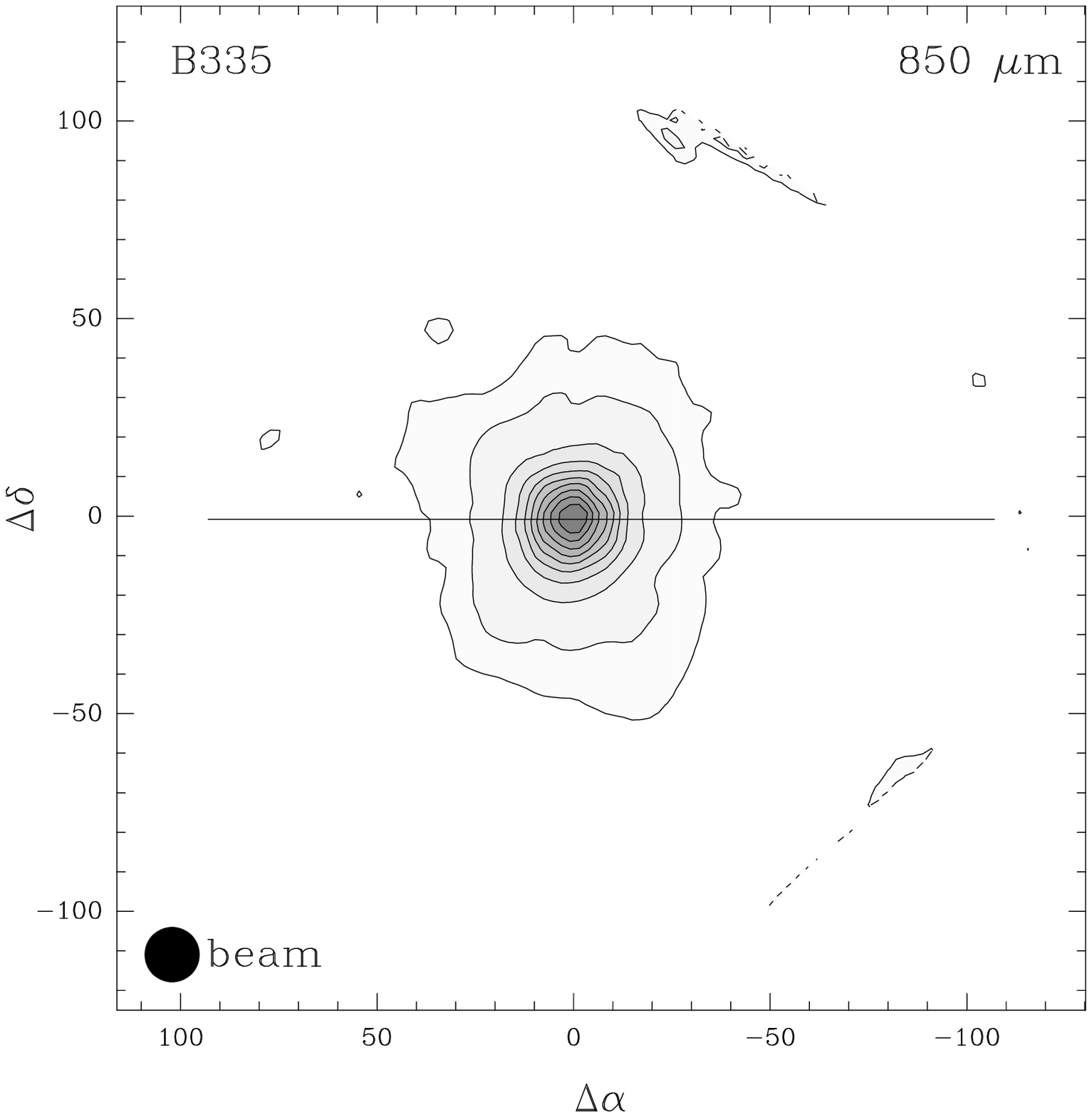}
   \includegraphics{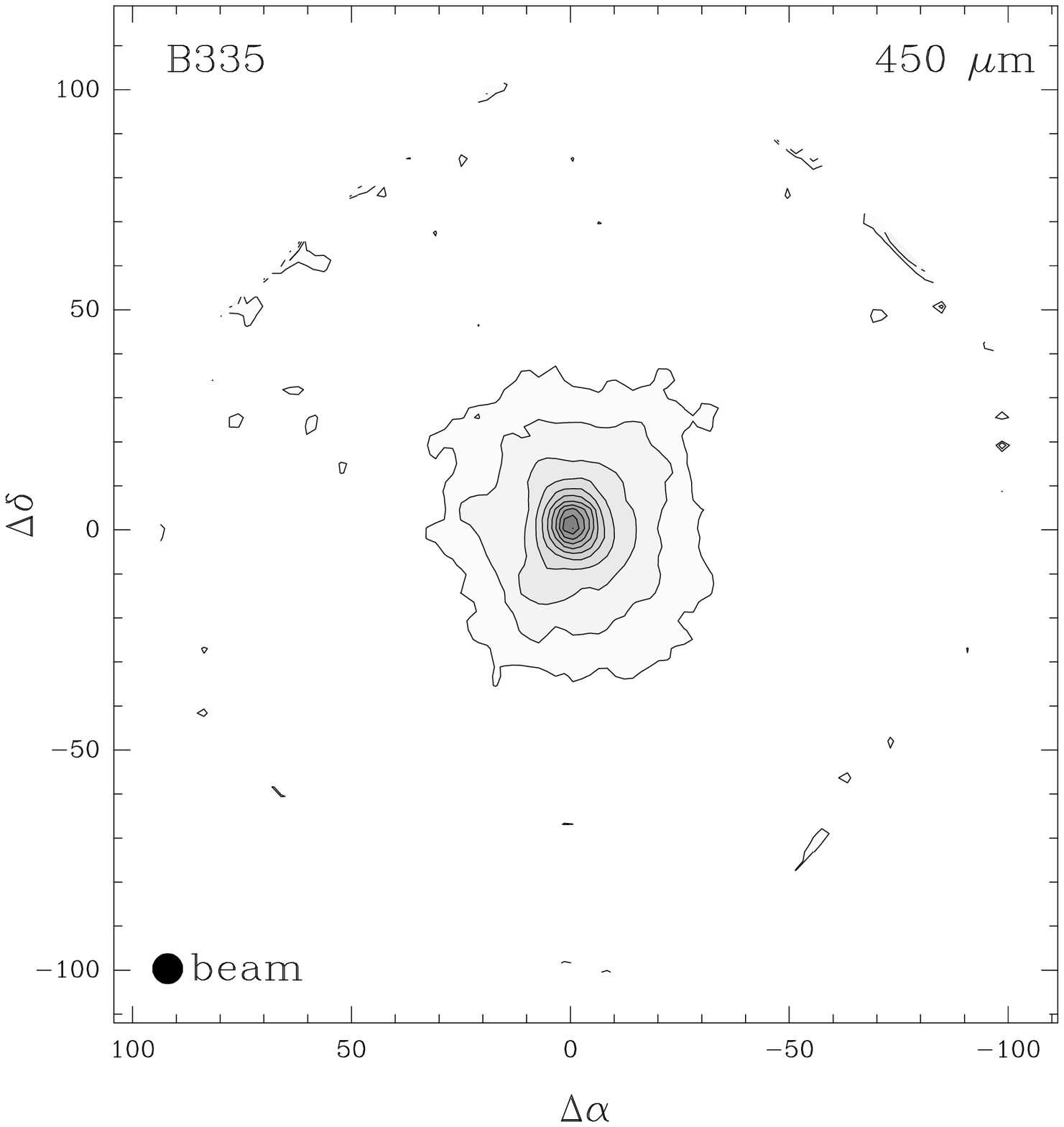}
   \includegraphics{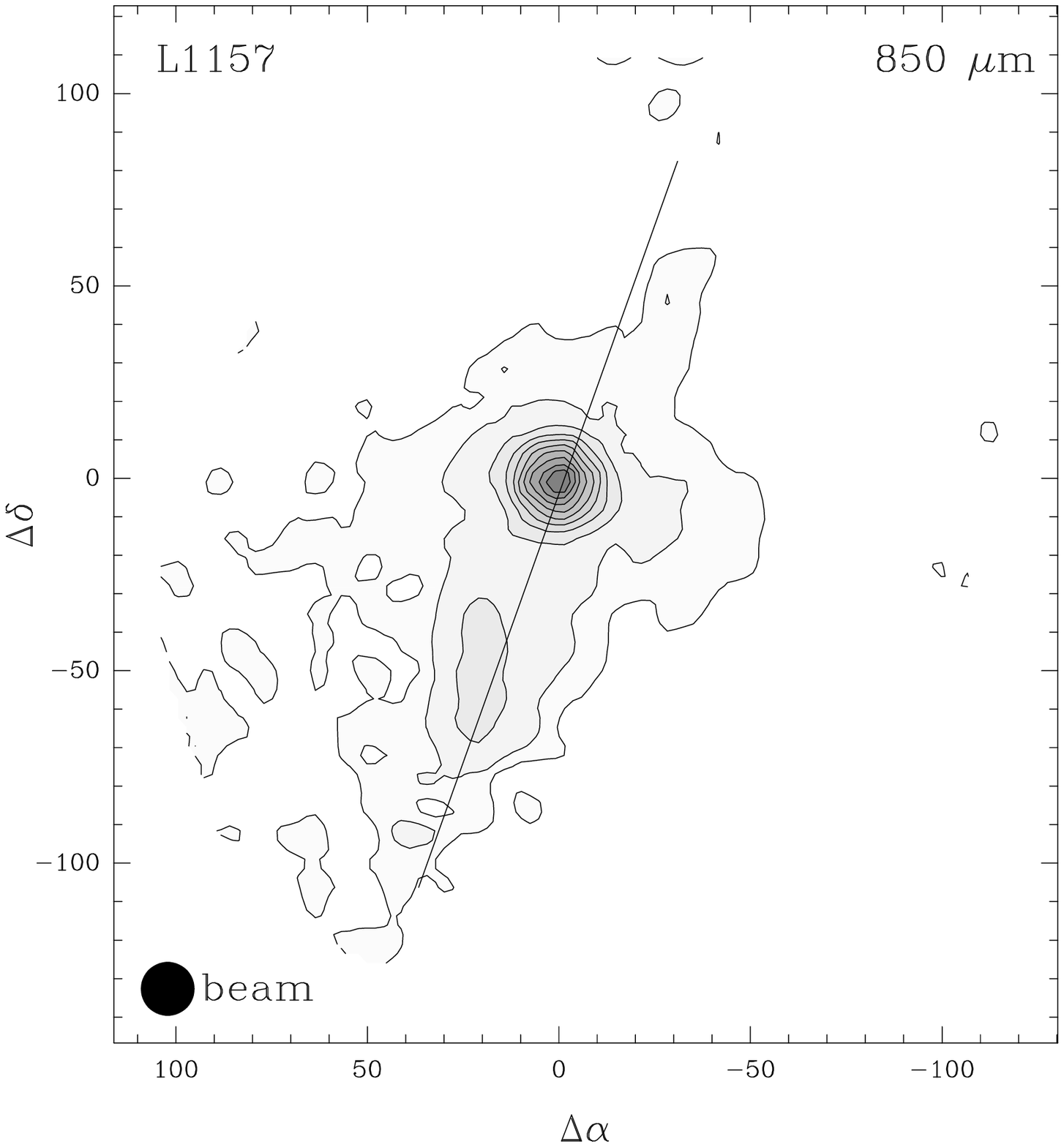}
\vskip 4.25in
\figcaption{
Contour maps of Class 0 sources.  
The left column is 850 \micron\ and the right column is 450 \micron.
The solid line indicates the outflow direction.
The contour levels are as follows (lowest contour and contour increment
in percentage of the peak flux).  
L723 (850\micron) 10\%(4$\sigma$) increasing by 10\%.
L723 (450\micron) 10\%(3$\sigma$) increasing by 10\%.
B335 (850\micron) 5\%(5$\sigma$), 10\%(9$\sigma$) increasing by 10\%.
B335 (450\micron) 5\%(3$\sigma$), 10\%(6$\sigma$) increasing by 10\%.
L1157 (850\micron) 5\%(3$\sigma$), 10\%(5$\sigma$) increasing by 10\%.
Contours near the edge of the maps should be ignored due to 
noisy pixels, less integration time, and inability of the plotting
package to handle irregular edges.
}
\end{figure}

\begin{figure}
\figurenum{7}
\centering
 \vspace*{7.8cm}
   \leavevmode
   \includegraphics{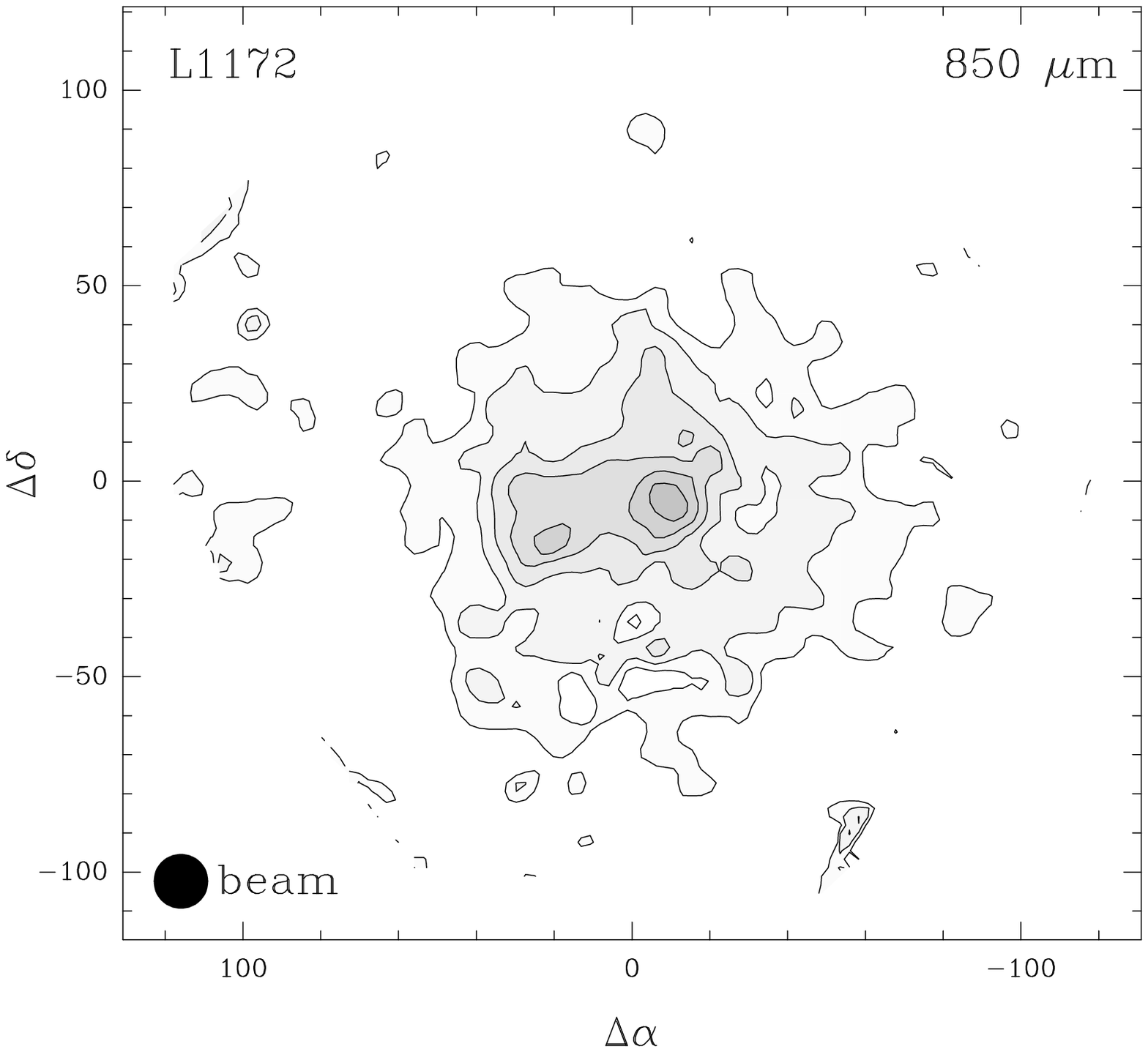}
   \includegraphics{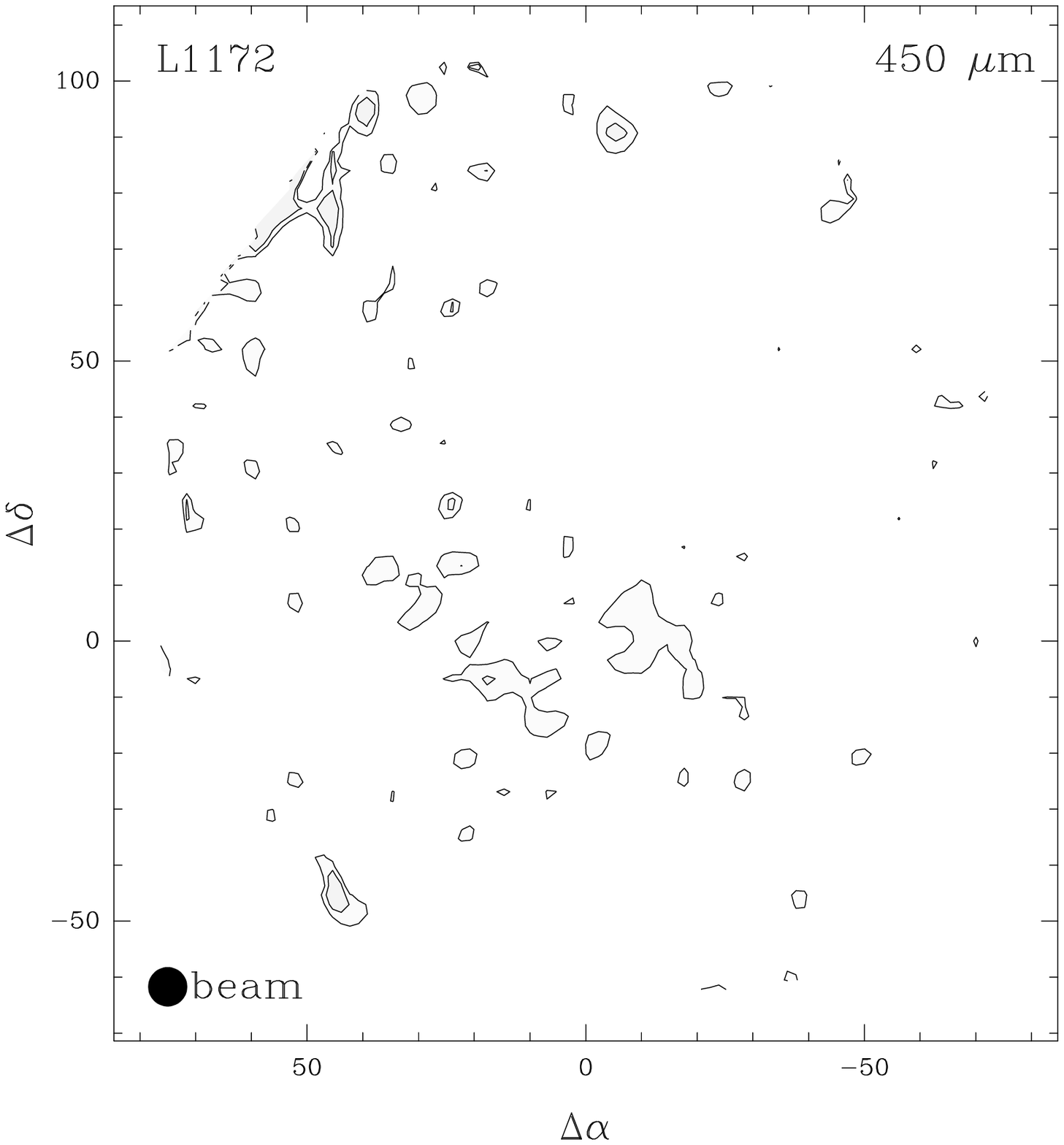}
   \includegraphics{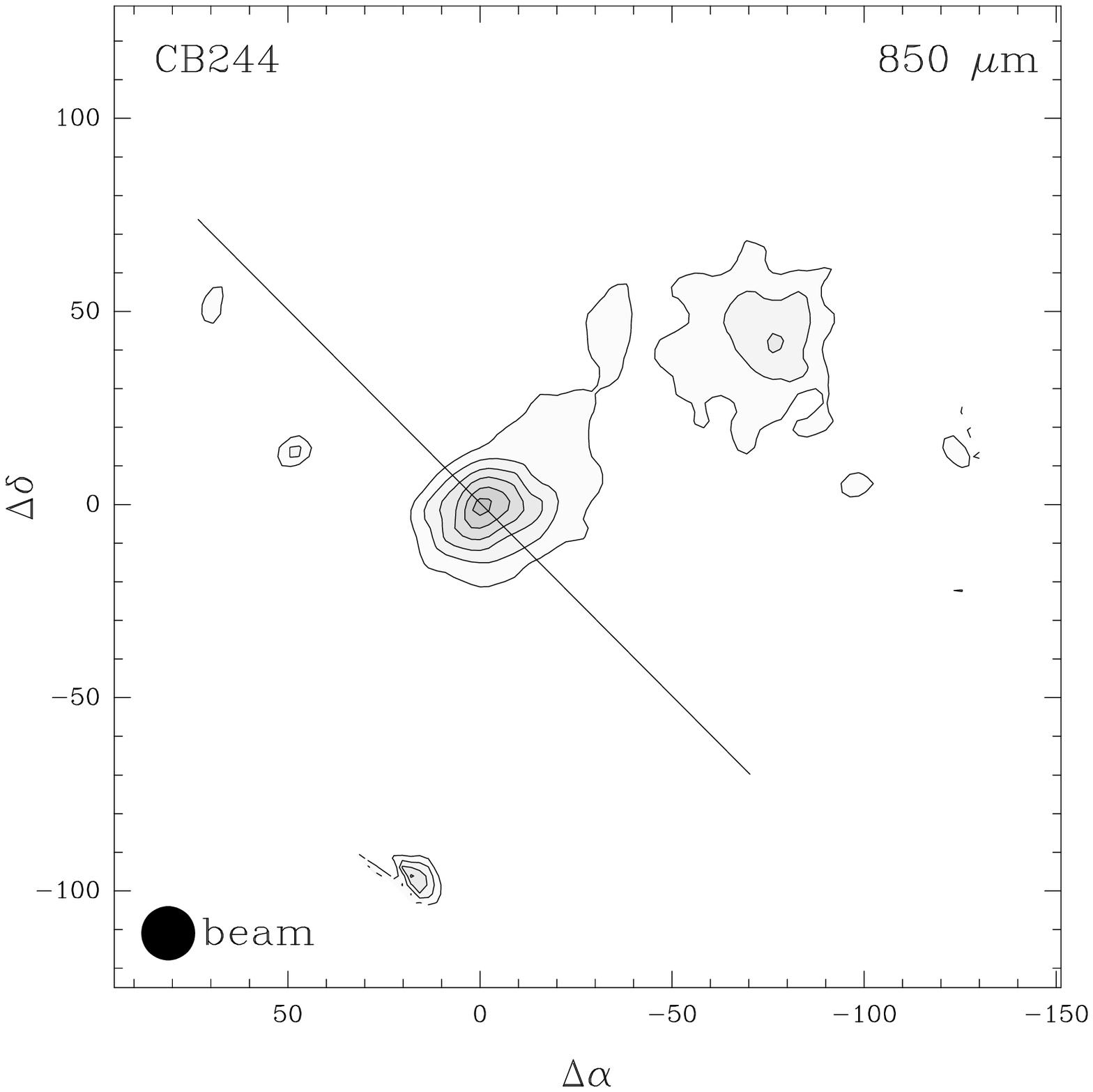}
   \includegraphics{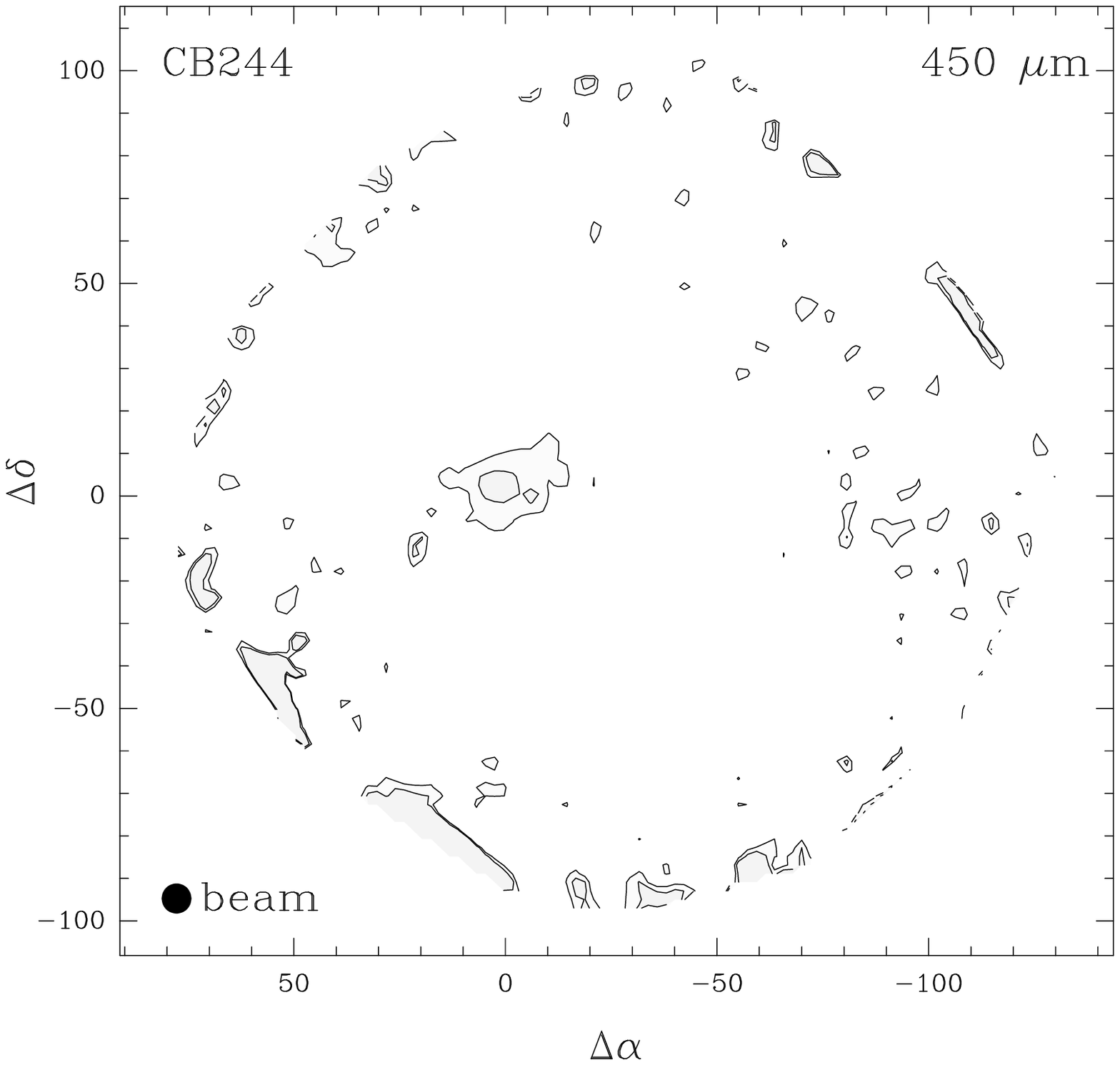}
   \includegraphics{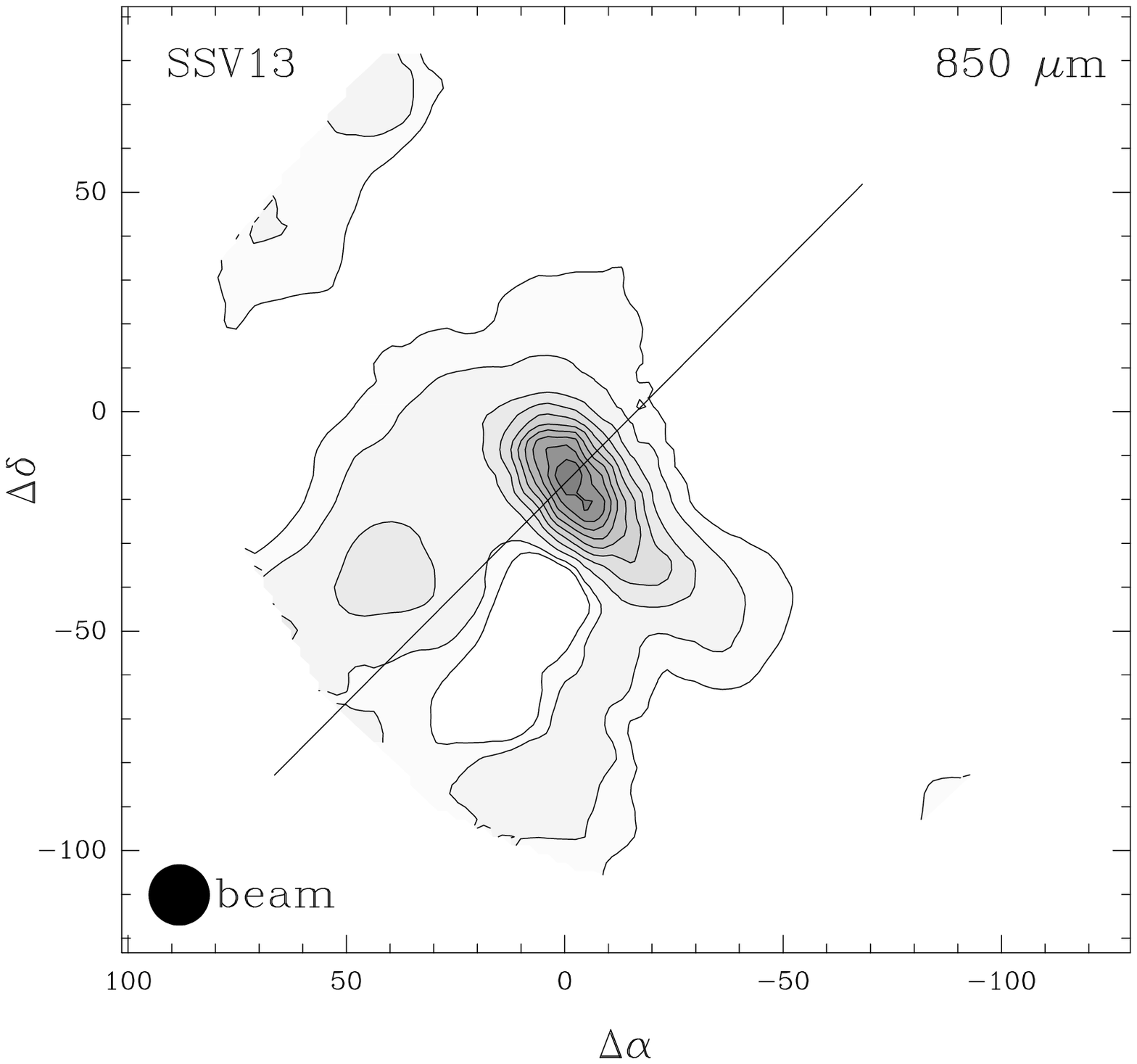}
   \includegraphics{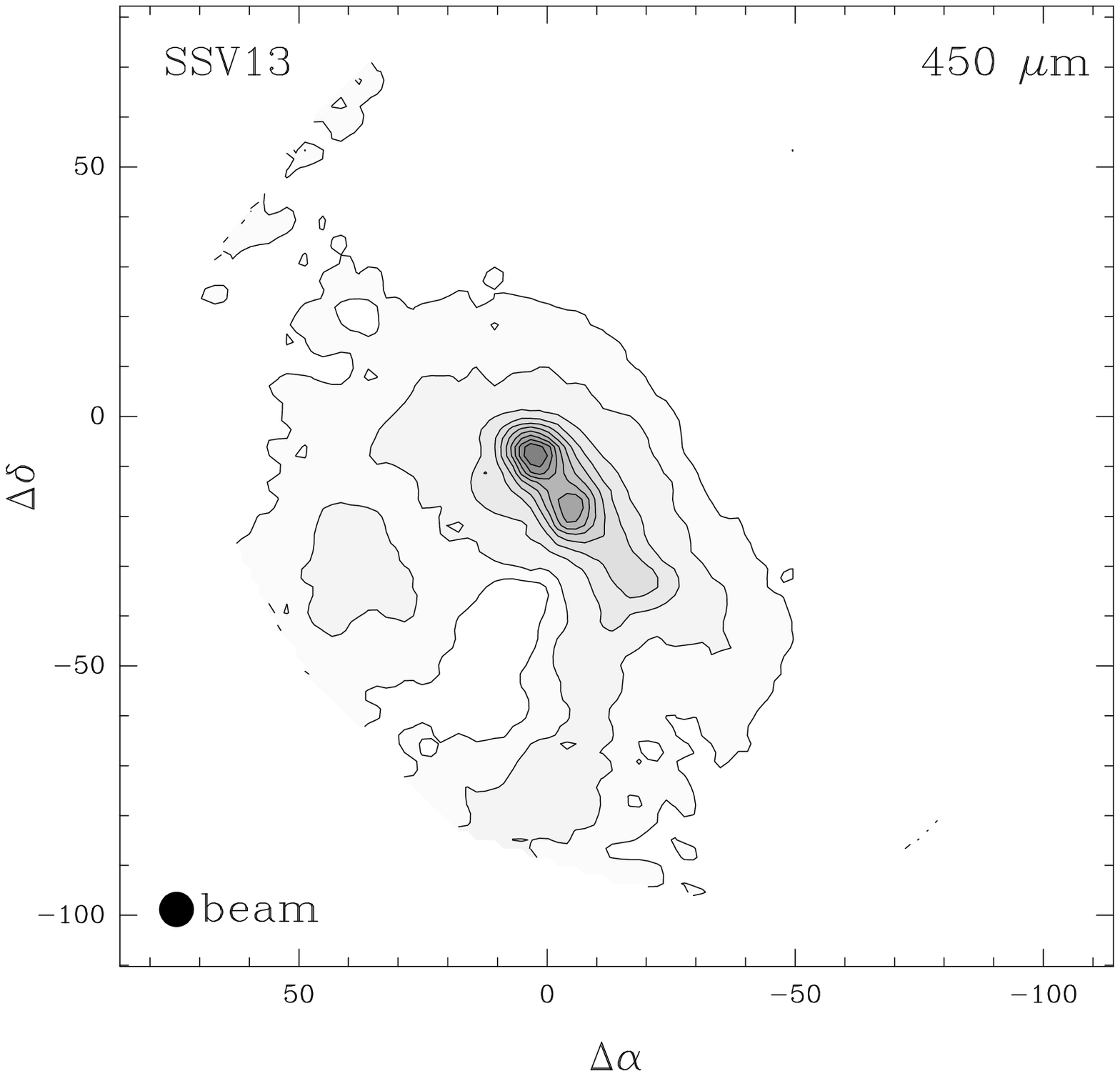}
\vskip 4.25in
\figcaption{
Contour maps of Class 0 sources.  
The left column is 850 \micron\ and the right column is 450 \micron.
The solid line indicates the outflow direction.
The contour levels are as follows (lowest contour and contour increment
in percentage of the peak flux).  
L1172 (850\micron) 22\%(3$\sigma$) increasing by 14\%(2$\sigma$).
L1172 (450\micron) 53\%(3$\sigma$) increasing by 36\%(2$\sigma$).
CB244 (850\micron) 22\%(3$\sigma$) increasing by 15\%(2$\sigma$).
CB244 (450\micron) 47\%(3$\sigma$) increasing by 32\%(2$\sigma$).
SSV13 (850\micron) 5\%(4$\sigma$), 10\%(8$\sigma$) increasing by 10\%.
SSV13 (450\micron) 5\%(6$\sigma$), 10\%(12$\sigma$) increasing by 10\%.
Contours near the edge of the maps should be ignored due to 
noisy pixels, less integration time, and inability of the plotting
package to handle irregular edges.
}
\end{figure}

\begin{figure}
\figurenum{8}
\centering
 \vspace*{7.8cm}
   \leavevmode
   \includegraphics{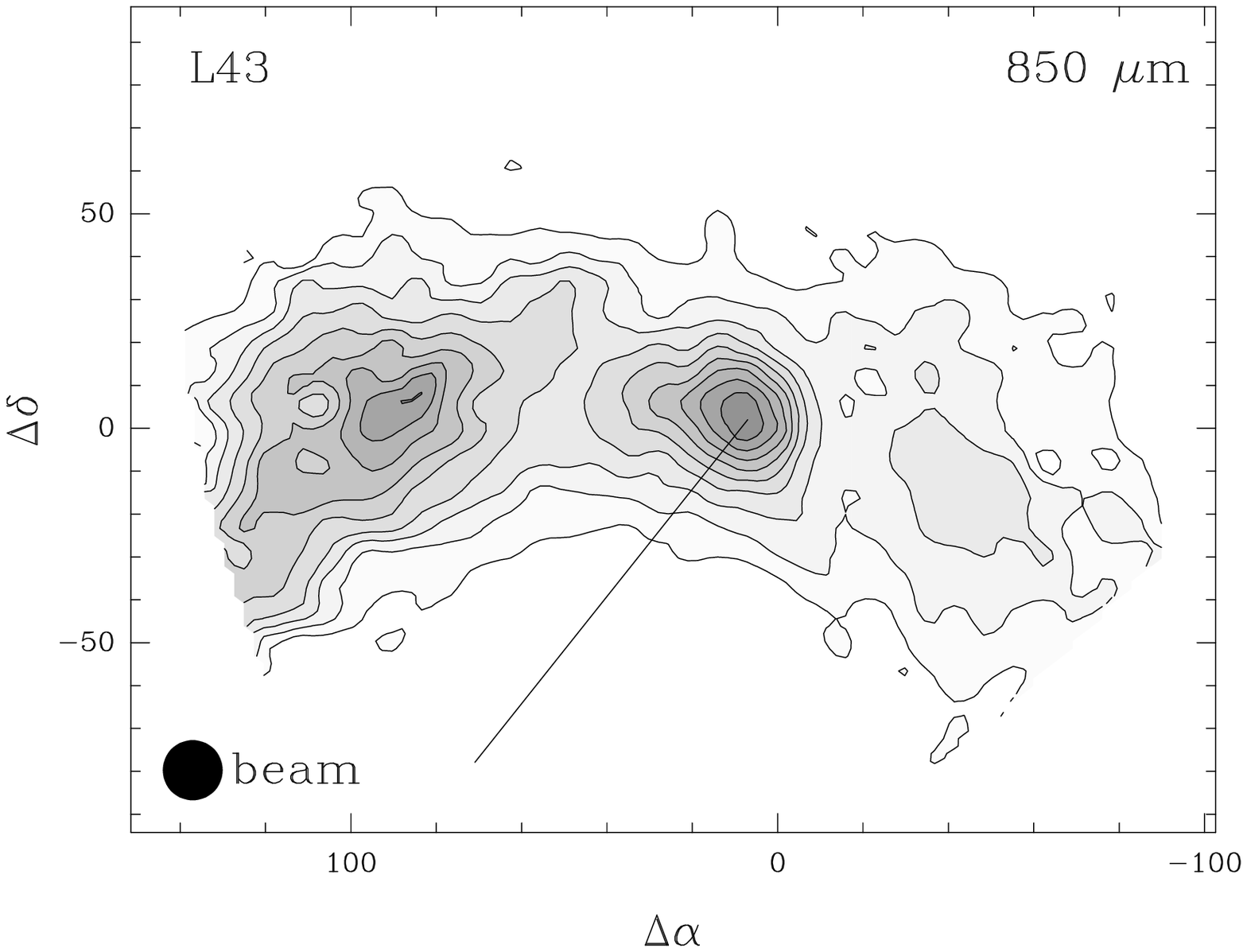}
   \includegraphics{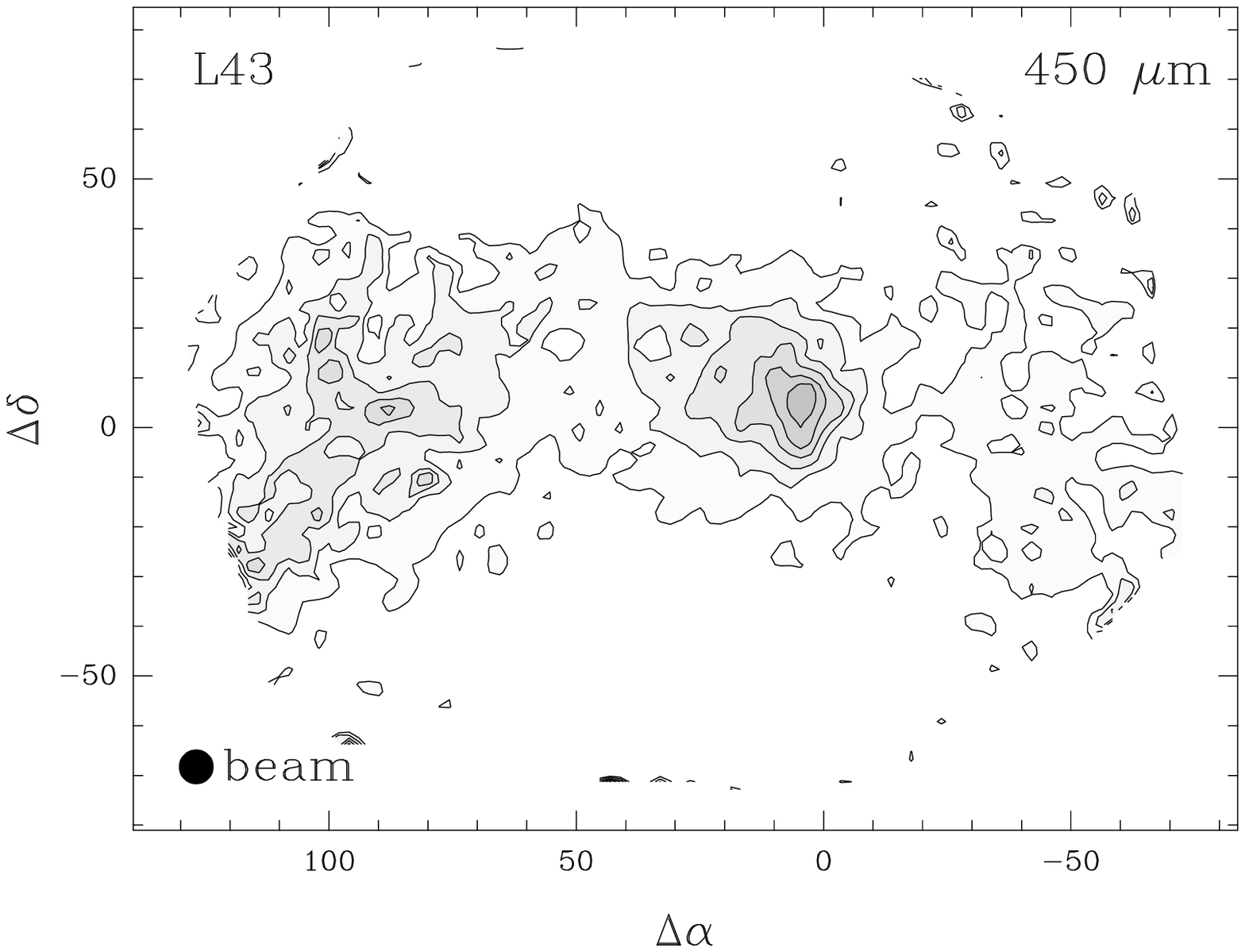}
   \includegraphics{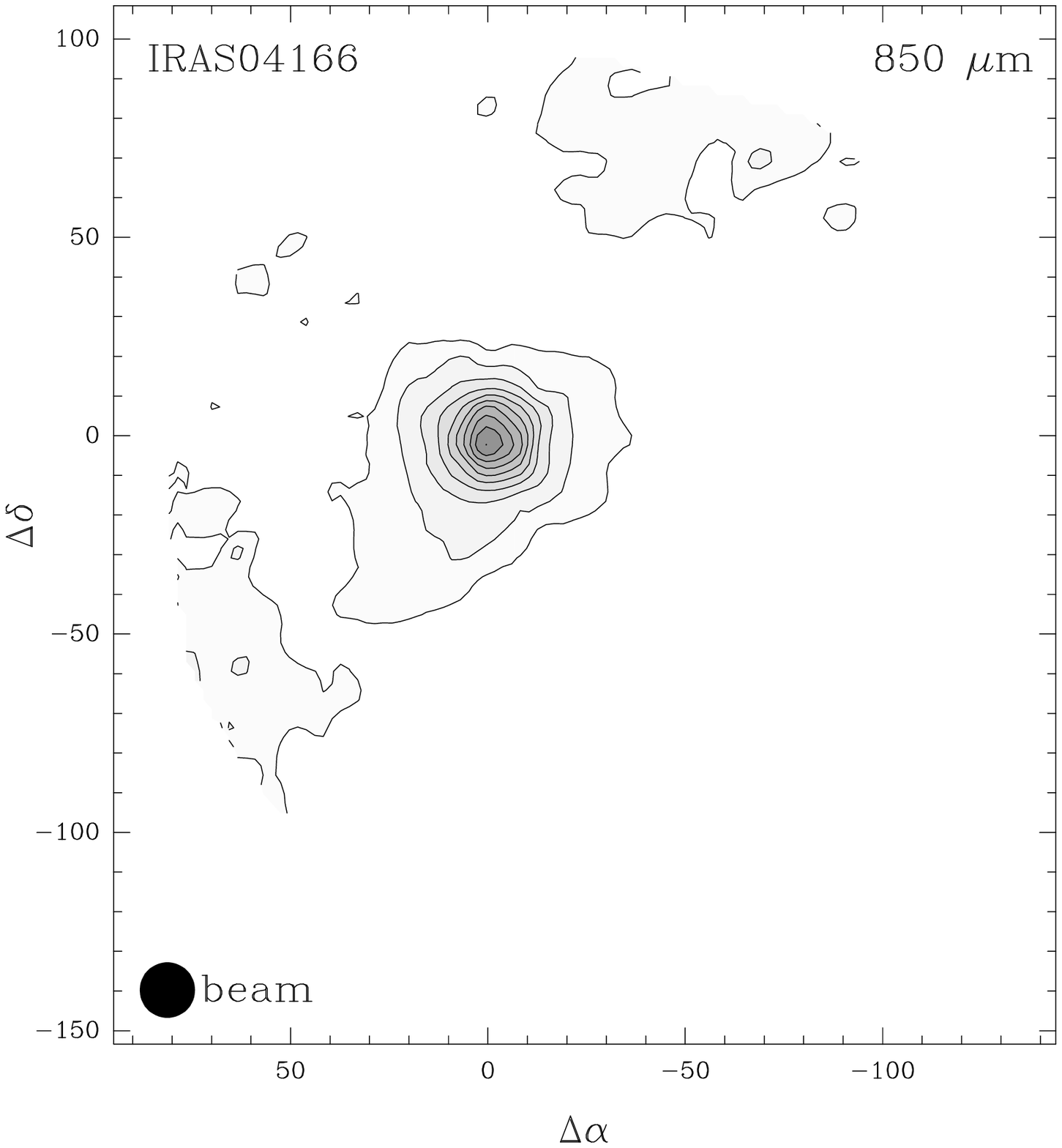}
   \includegraphics{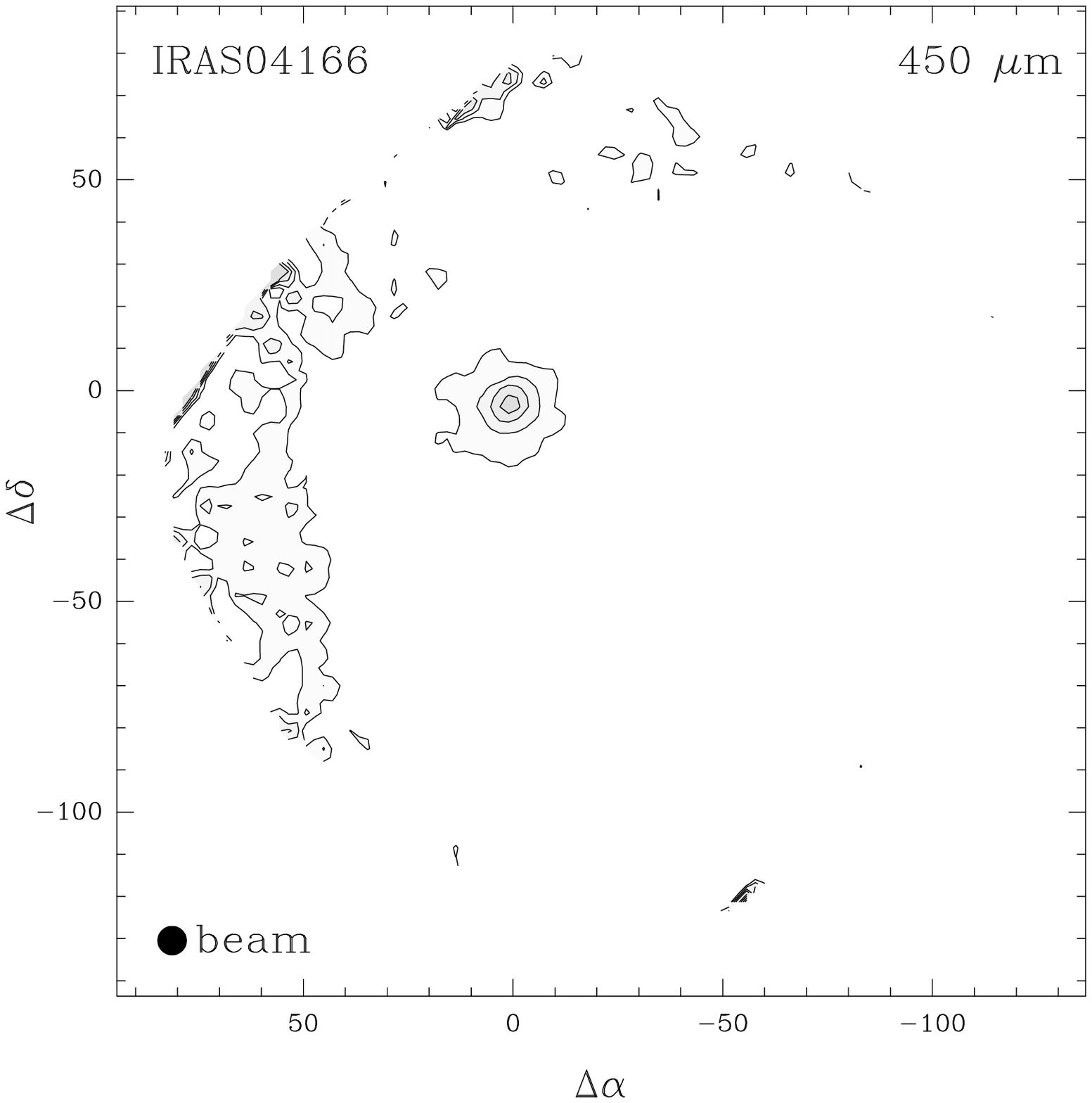}
\vskip 4.25in
\figcaption{
Contour maps of Class I sources.  
The left column is 850 \micron\ and the right column is 450 \micron.
The solid line indicates the outflow direction.
The contour levels are as follows (lowest contour and contour increment
in percentage of the peak flux).
L43 (850\micron) 15\%(3$\sigma$) increasing by 10\%(2$\sigma$).
L43 (450\micron) 20\%(3$\sigma$) increasing by 14\%(2$\sigma$).
IRAS04166+2706 (850\micron) 10\%(3$\sigma$) increasing by 10\%.
IRAS04166+2706 (450\micron) 29\%(3$\sigma$) increasing by 19\%(2$\sigma$).
Contours near the edge of the maps should be ignored due to 
noisy pixels, less integration time, and inability of the plotting
package to handle irregular edges.
}
\end{figure}


\begin{figure}
\figurenum{9}
\plotone{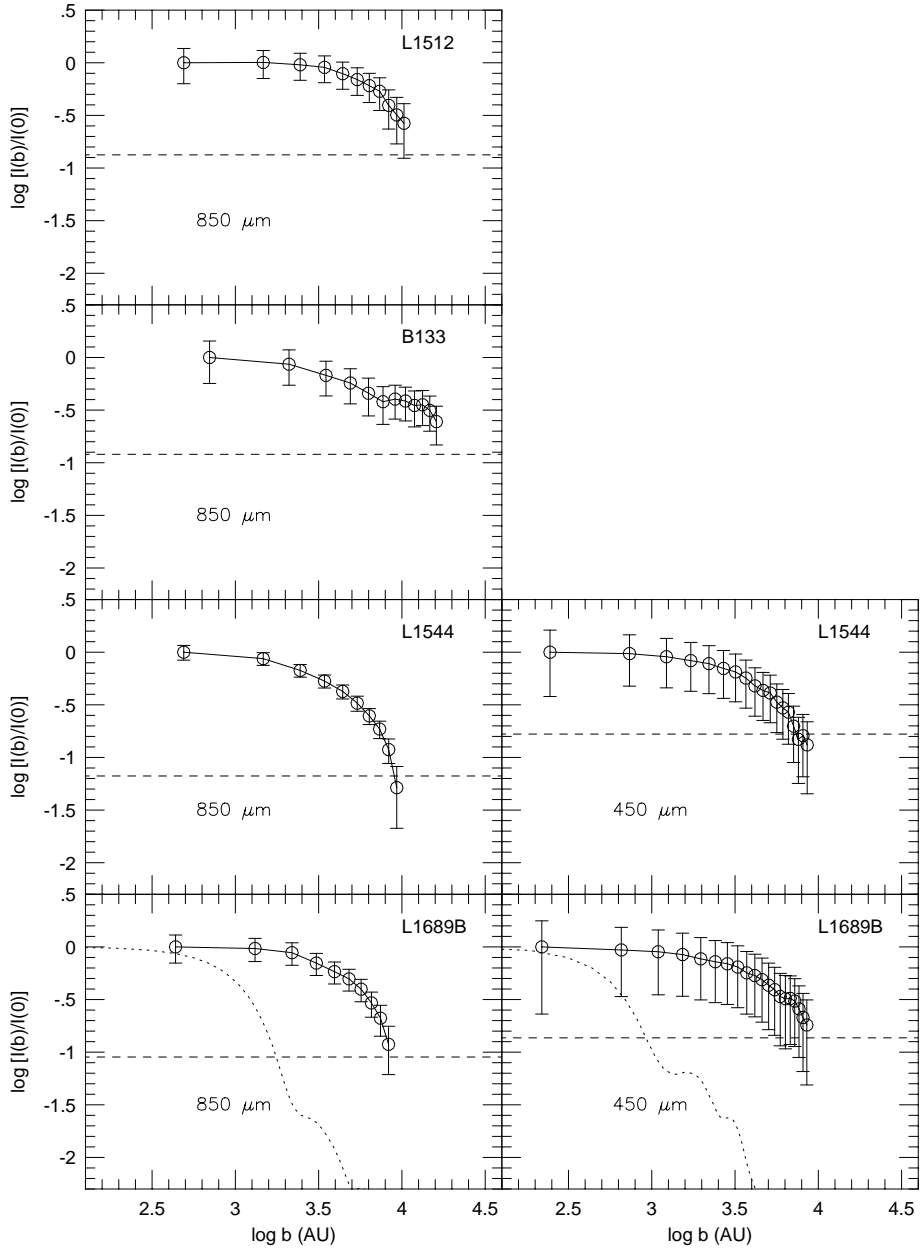}
\figcaption{
Radial profiles of pre-protostellar cores.  The normalized intensity is
plotted as a function of impact parameter, $b$(AU). 
The horizontal dashed lines represent the $1\sigma$ noise level at the
edge of the map. 
Note that no section of the profile is fit by a power law.
The beam profiles are shown as dashed lines in the bottom panel.
}
\end{figure}


\begin{figure}
\figurenum{10}
\plotone{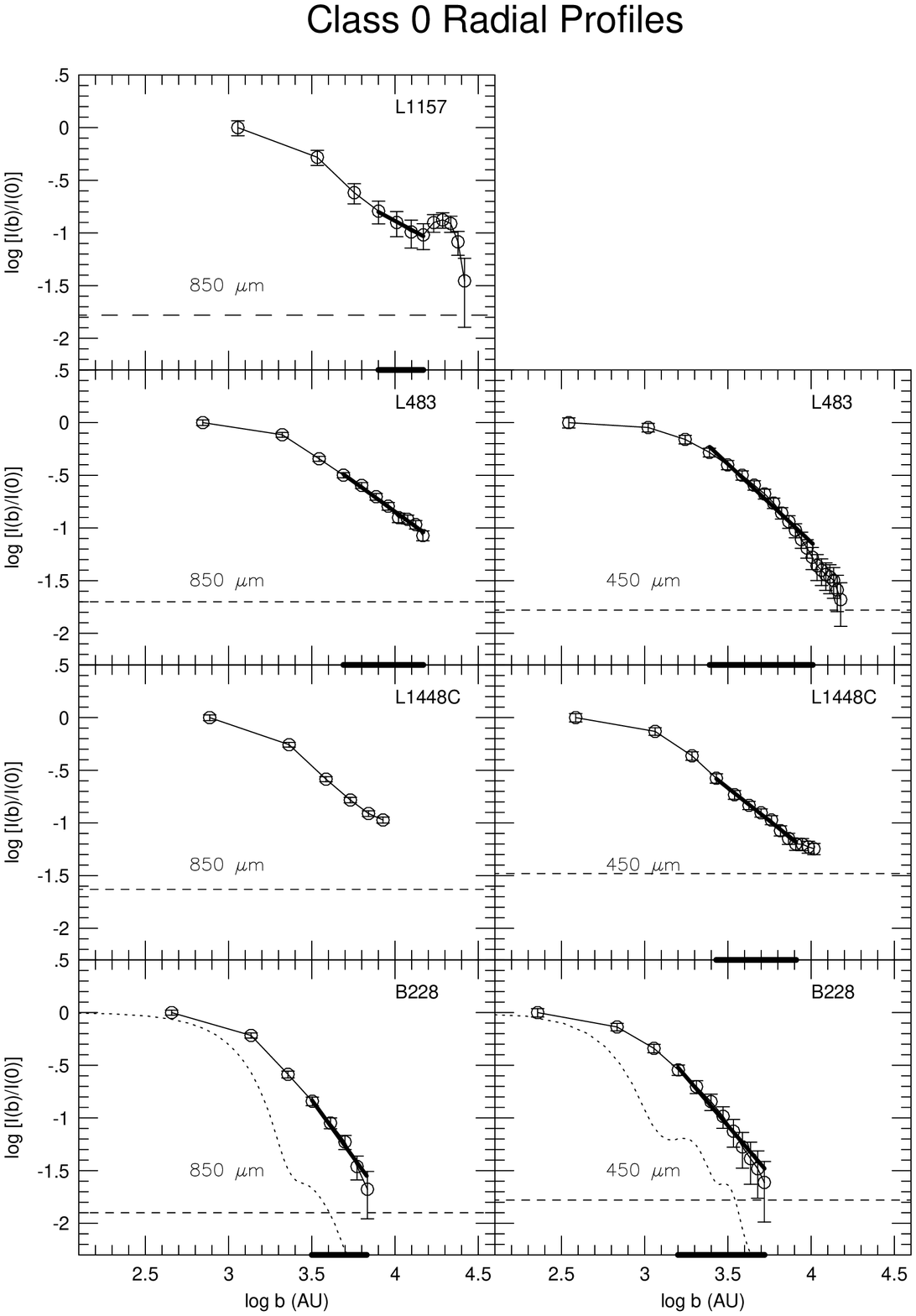}
\figcaption{
Radial profiles of Class 0 sources.  The normalized intensity is
plotted as a function of impact parameter, $b$(AU).  Power law fits are shown
as bold lines.  The range of the fits are indicated by the bold 
lines on the x axis.
The horizontal dashed lines represent the $1\sigma$ noise level at the
edge of the map.
The beam profiles are shown as dashed lines in the bottom panel.
}
\end{figure}


\begin{figure}
\figurenum{11}
\plotone{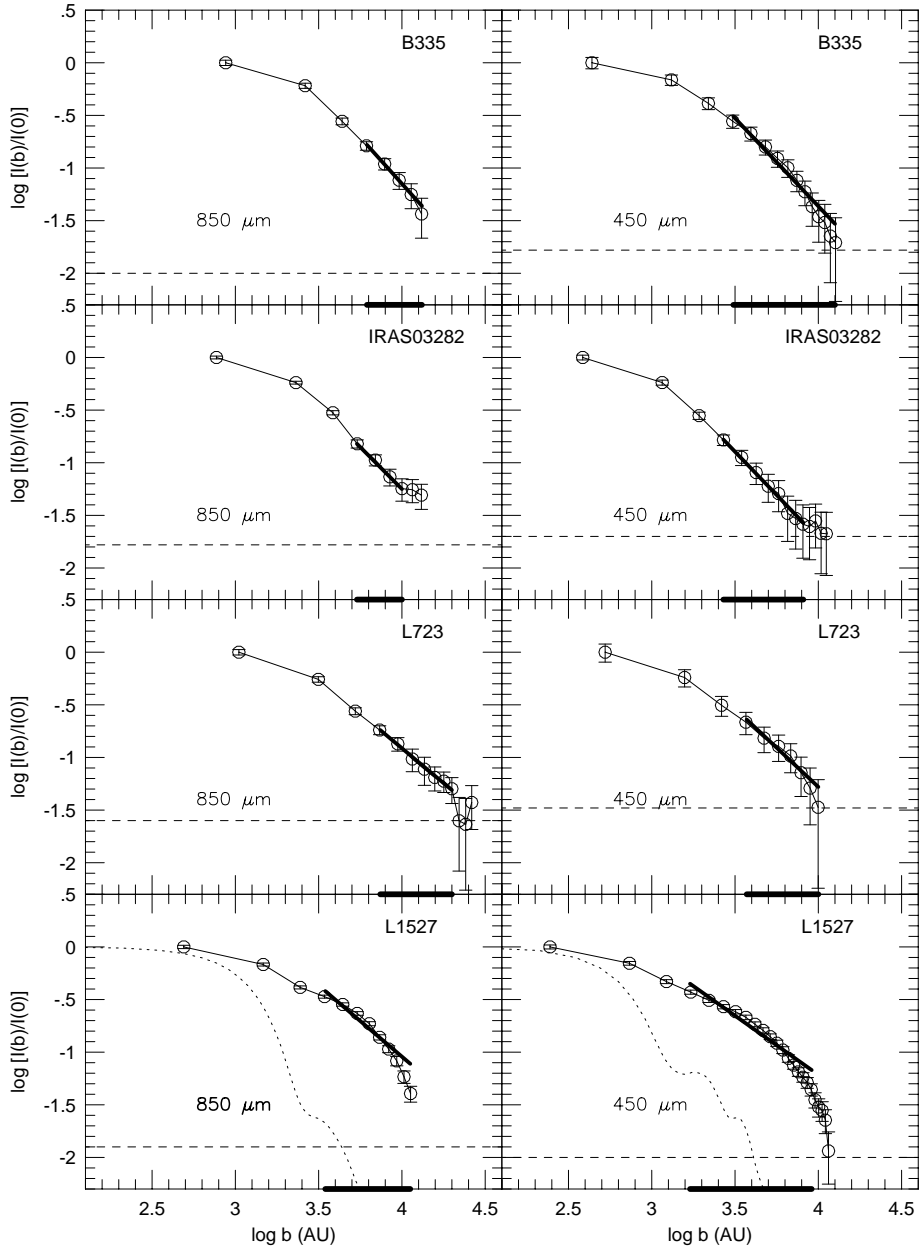}
\figcaption{
Radial profiles of Class 0 sources.  The normalized intensity is
plotted as a function of impact parameter, $b$(AU).  Power law fits are shown
as bold lines.  The range of the fits are indicated by the bold lines
on the x axis.
The horizontal dashed lines represent the $1\sigma$ noise level at the
edge of the map.
The beam profiles are shown as dashed lines in the bottom panel.
}
\end{figure}


\begin{figure}
\figurenum{12}
\plotone{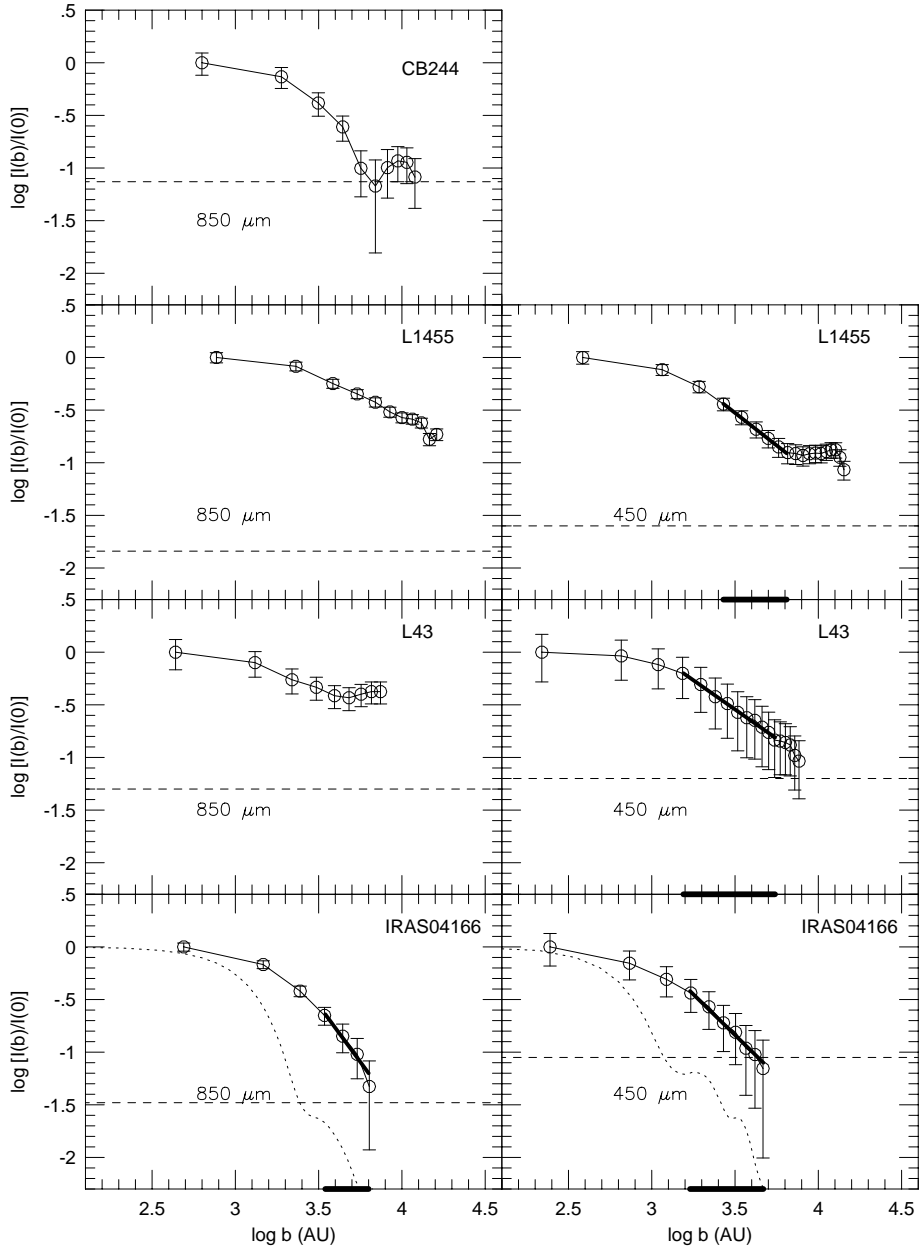}
\figcaption{
Radial profiles of Class 0/I sources.  The normalized intensity is
plotted as a function of impact parameter, $b$(AU).  Power law fits are shown
as bold lines.  The range of the fits are indicated by the bold lines
on the x axis.  
The horizontal dashed lines represent the $1\sigma$ noise level at the
edge of the map.
Emission from secondary sources contaminates all of the
profiles.
The beam profiles are shown as dashed lines in the bottom panel.
}
\end{figure}


\begin{figure}
\figurenum{13}
\epsscale{0.7}
\plotone{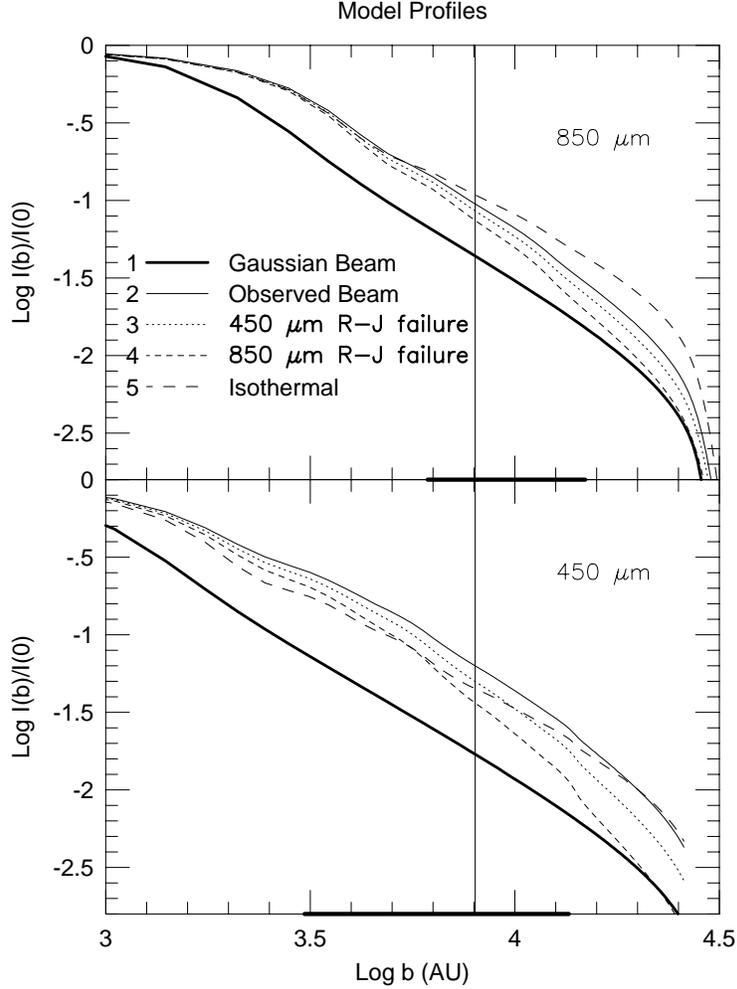}
\figcaption{
Five models of normalized intensity profiles are shown.  Power laws in the
density ($p = 2$) and temperature ($q = 0.4$) were assumed, and the resulting
intensity was convolved with a beam profile.  
Model 1 (bold line) assumes a gaussian beam of FWHM 14\arcsec\ and 7\arcsec\ 
respectively.
The other models use the observed beam profiles shown in Fig. 1.
Model 1 and 2 assumed no R-J failure
($\Tdr \geq 64 K , 60 {\rm AU} \leq r \leq 30000$ AU) throughout the envelope.  
Model 3 allows R-J failure to occur at 450 \micron\ at 8000 AU
($T(8000 {\rm AU}) = 32 $K).  Model 4 allows R-J failure at 850 \micron\
at 8000 AU ($T(8000 {\rm AU}) = 17 K$).
Model 5 is isothermal ($\Tdr = 10$ K)  in the outer envelope
(3000 AU $\leq r \leq 30000$ AU).  The vertical line
corresponds to $b = 8000$ AU.  The bold line on the x-axis represents the
range over which a fit would be made.
}
\end{figure}

\begin{deluxetable}{llrccrccc}
\tablecolumns{9}
\footnotesize
\tablecaption{Observed Sources \label{tab1}}
\tablewidth{0pt} 
\tablehead{
\colhead{Source}                &
\colhead{$\alpha$ (1950.0)}    &
\colhead{$\delta$ (1950.0)}    &
\colhead{Observed}              &
\colhead{Class\tablenotemark{a}}           &
\colhead{Dist.}              &
\colhead{Dist.}              &
\colhead{Outflow}          &
\colhead{Collapse\tablenotemark{b}}                \\
\colhead{}                      &
\colhead{($^h$~~$^m$~~$^s$~)~}        &  
\colhead{($\degree$ ~\arcmin\ ~\arcsec)}          &  
\colhead{}                      &
\colhead{}                      &
\colhead{(pc)}                  &
\colhead{Ref.}                  &
\colhead{Ref.}                      &
\colhead{Candidate?}                
}
\startdata 
 L1512  & 05 00 54.4    & 32 39 37    & 1/25/98 & PPC & 140 & 1 & ...   & N \\
 L1544  & 05 01 13.1    & 25 06 36    & 1/25/98 & PPC & 140 & 1 & ...   & Y \\
 L1689A & 16 29 10.5    &$-$24 57 22  & 4/18/98 & PPC & 125 & 2 & ...   & Y \\ 
 L1689B & 16 31 47      &$-$24 31 45  & 4/18/98 & PPC & 125 & 2 & ...   & Y \\
 B133   & 19 03 27.3    &$-$06 57 00  & 4/14/98 & PPC & 200 & 3 & ...   & Y  \\
 L1448NW& 03 22 31.1    & 30 35 04    & 1/24/98 & 0   & 220 & 4 & 10,11 & N \\
 L1448N & 03 22 31.8    & 30 34 45    & 1/24/98 & 0   & 220 & 4 & 10,11 & N \\
 L1448C & 03 22 34.3    & 30 33 30    & 1/24/98 & 0   & 220 & 4 & 10,11 & N \\
 L1455  & 03 24 34.9    & 30 02 36    & 1/24/98 & 0   & 220 & 4 & 6     & N \\
 IRAS03282+3035 & 03 28 15.2 & 30 35 14    & 1/24/98 & 0   & 220 & 4 & 12    & Y \\
 L1527  & 04 36 49.6    & 25 57 21    & 1/25/98 & 0   & 140 & 1 & 13    & Y \\
 B228   & 15 39 50.4    &$-$33 59 42  & 4/15/98 & 0   & 130 & 5 & 14    & Y \\
 L483   & 18 14 50.6    &$-$04 40 49  & 4/17/98 & 0   & 200 & 3 & 15    &N\tablenotemark{c} \\
 L723   & 19 15 41.3    & 19 06 47    & 4/20/98 & 0   & 300 & 6 & 16    & N \\
 B335   & 19 34 35.4    & 07 27 24    & 4/17/98 & 0   & 250 & 7 & 16    & Y \\ 
 L1157  & 20 38 39.6    & 67 51 33    & 4/19/98 & 0   & 325 & 8 & 17    & Y \\ 
 L1172  & 21 01 44.2    & 67 42 24    & 4/18/98 & 0   & 288 & 8 & ...   & N \\ 
 CB244  & 23 23 48.7    & 74 01 08    & 4/20/98 & 0   & 180 & 9 & 18    & Y \\
 SSV13  & 03 25 57.9    & 31 05 50    & 1/24/98 & I   & 220 & 4 & 19    & N \\
 IRAS04166+2706 & 04 16 37.8 & 27 06 29    & 8/30/98 & I   & 140 & 1 & ...   & Y \\ 
 L43    & 16 31 37.7    &$-$15 40 52  & 4/17/98 & I   & 125 & 2 & 20 & N \\
\enddata

\tablenotetext{a}{PPC=Pre-protostellar core}
\tablenotetext{b}{As indicated by studies of \hcop\ (Gregersen 1998) }
\tablenotetext{c}{Red asymmetry in \hcop, but blue in other lines}
\tablerefs{1. Taurus -- Elias 1978;  2. Ophiuchus -- de Geus et al. 1990; 
           3. Aquila Rift -- Dame \& Thaddeus 1985; 4. NGC1333 region -- \v{C}ernis 1990, 
            (but see Herbig \& Jones 1983 who get 350 pc);
           5. Lupus -- Murphy et al. 1986; 6. Goldsmith et al. 1984; 
           7. Tomita et al. 1979; 8. Strai\v{z}ys et al. 1992; 
           9. Lindblad Ring -- Launhardt \& Henning 1997; 
	   10. Bachiller et al. 1990;
	   11. Barsony et al. 1998; 12. Bachiller et al. 1991a; 
	   13. Zhou et al. 1996; 14. Heyer \& Graham 1989; 15. Fuller et al. 1995;
	   16. Hirano et al. 1998; 17. Bachiller \& Perez Gutierrez 1997; 
	   18. Yun \& Clemens 1994; 19. Liseau et al. 1988; 20. Bence et al. 1998  }

\end{deluxetable}

\begin{deluxetable}{lcccccccc}
\footnotesize
\tablecolumns{9}
\tablecaption{Opacity \& Calibration Summary \label{tab2}}
\tablewidth{0pt} 
\tablehead{
\colhead{Date}                  &
\colhead{$\tau_{850}$}          &
\colhead{$\tau_{450}$}          &
\colhead{$\tau_{1.3}$}          &
\colhead{$C_{40}^{850}$}        &
\colhead{$C_{120}^{850}$}       & 
\colhead{$C_{40}^{450}$}        & 
\colhead{$C_{120}^{450}$}       &
\colhead{$C_{40}^{1.3}$}       \\
\colhead{}                      &
\colhead{}                      &
\colhead{}                      &
\colhead{}                      &
\colhead{Jy/V\tablenotemark{a}}           &
\colhead{Jy/V\tablenotemark{b}}           &
\colhead{Jy/V\tablenotemark{a}}           &
\colhead{Jy/V\tablenotemark{b}}           & 
\colhead{Jy/V\tablenotemark{a}}
}
\startdata
        January  & & & & 1.02 (0.06) & 0.84 (0.04) & 5.72 (0.99) & 5.24 (1.97) & ... \\ 
        01/24/98 & 0.17 (0.01) & 0.84 (0.01) & ...  \\
        01/25/98 & 0.12 (0.01) & 0.48 (0.06) & ... \\
        April    & & & & 0.94 (0.05) & 0.74 (0.04) & 5.76 (0.76) & 4.14 (0.57) & ... \\
        04/14/98 & 0.12 (0.01) & 0.51 (0.01) & ... \\ 
        04/15/98 & 0.14 (0.01) & 0.60 (0.02) & ... \\
        04/17/98 & 0.14 (0.01) & 0.66 (0.06) & 0.06 (0.01) \\ 
        04/18/98 & 0.15 (0.01) & 0.69 (0.05) & ... \\ 
        04/19/98 & 0.34 (0.03) & 1.7  (0.1) & ... \\ 
        04/20/98 & 0.15 (0.03) & 0.7  (0.2) & ... \\ 
        August   & & & & 1.00 (0.05) & ... & 4.63 (0.99) & ... & 0.26(0.04) \\
        08/28/98 & 0.39 (0.01) & 2.7  (0.2) & 0.15 (0.01) \\ 
        08/29/98 & 0.20 (0.01) & 1.2  (0.1) & ... \\ 
        08/30/98 & 0.19 (0.01) & 1.0  (0.1) & ... \\  

\enddata

\tablenotetext{a}{Calibration Factor for a 40\as\ diameter aperture}
\tablenotetext{b}{Calibration Factor for a 120\as\ diameter aperture}

\end{deluxetable}

\begin{deluxetable}{llccccc}
\footnotesize
\tablecolumns{7}
\tablecaption{Observed Flux Densities of Sources \label{tab3}}
\tablewidth{0pt} 
\tablehead{
\colhead{Source}                &
\colhead{Centroid}              &
\colhead{}                      &
\colhead{}                      &
\colhead{$S_{\nu}$ (Jy)}        &
\colhead{}                      &
\colhead{}                      \\
\colhead{}                      &
\colhead{($\Delta\alpha$\as,$\Delta\delta$\as)}                  &
\colhead{1.3 mm\tablenotemark{a}} &
\colhead{850 $\mu$m\tablenotemark{a}}           &
\colhead{850 $\mu$m\tablenotemark{b}}           &
\colhead{450 $\mu$m\tablenotemark{a}}           &
\colhead{450 $\mu$m\tablenotemark{b}}
}
\startdata 
 L1512     & ($-$19,$-$26) &...        &0.35(0.02) &1.81(0.09) &1.6(0.3)   &8.5(3.2)  \\
 L1544     & ($-$1,4)      &0.27(0.04) &1.12(0.07) &3.64(0.18) &4.2(0.8)   &17.4(6.7) \\
 L1689A    & ($-$4,$-$9)   &...        &0.54(0.03) &...        &3.9(0.6)   &...       \\
           & (51,24)       &...        &0.55(0.03) &...        &3.4(0.5)   &...        \\
 L1689B    & ($-$1,$-$12)  &...        &0.90(0.05) &3.18(0.18) &3.2(0.5) &10.8(1.7)   \\
 B133      & ($-$13,$-$32) &...        &0.60(0.03) &2.06(0.12) &3.4(0.5)   &11.8(1.6)  \\
 L1448NW   & (0,0)         &...        &6.51(0.38) &...        &43.9(7.6)  &...        \\
 L1448N    & ($-1$,1)      &...        &8.19(0.49) &...        &56.4(9.8)  &...        \\
 L1448C    & (0,3)         &0.74(0.11) &3.95(0.24) &...        &31.8(5.5)  &...        \\
 L1455     & (3,1)         &...        &1.08(0.06) &...        &10.2(1.8)  &...        \\
           & (55,$-$30)    &...        &0.94(0.06) &...        &7.7(1.3)   &...        \\
           & ($-$9,59)     &...        &0.38(0.02) &...        &2.5(0.5) &...        \\
           & (10,$-$47)    &...        &0.91(0.05) &...        &6.9(1.2)   &...        \\
 IRAS03282+3035 & (6,8)         &...        &1.74(0.10) &3.59(0.17) &9.9(1.7)   &25.0(9.4)  \\
 L1527     & ($-$2,0)      &0.72(0.11) &3.19(0.19) &9.41(0.46) &18.2(3.2)  &55.5(20.9) \\
 B228      & (11,5)        &...        &2.63(0.15) &4.23(0.24) &19.6(2.7)  &25.9(3.7)  \\
 L483      & (1,2)         &...        &3.74(0.20) &9.25(0.51) &30.1(4.6)  &59.2(1.7)  \\
 L723      & (8,1)         &...        &1.79(0.11) &3.60(0.23) &8.5(2.1)   &12.4(3.1)  \\
 B335      & (0,1)         &0.57(0.09) &2.28(0.12) &3.91(0.22) &14.6(2.2)  &21.1(3.3)  \\ 
 L1157     & (0,1)         &0.58(0.09) &2.41(0.19) &5.03(0.40) &...        &...        \\ 
 L1172     & ($-$10,$-$2)  &...        &0.66(0.04) &2.69(0.15) &5.2(0.8)   &16(3)      \\ 
 CB244     & (0,1)         &0.25(0.04) &1.04(0.08) &1.86(0.14) &5.1(1.9)   &9.0(3.4)   \\
           & ($-$75,45)    &...        &0.78(0.06) &...        &3.1(1.2)   &...        \\
 SSV13     & (1,$-$10)     &1.83(0.28) &6.95(0.41) &...        &52.4(9.1)  &...        \\
 IRAS04166+2706 & (1,$-4$)      &...        &1.08(0.06) &...        &4.2(1.0)   &...        \\ 
 L43       & (7,5)         &...        &1.60(0.09) &...        &11.8(2.0)  &...        \\ 
           & (89,6)        &...        &1.80(0.10) &...        &11.4(1.9)  &...        \\
\enddata

\tablenotetext{a}{In a 40\as\ aperture}
\tablenotetext{b}{In a 120\as\ aperture}

\end{deluxetable}

\begin{deluxetable}{llccc}
\footnotesize
\tablecolumns{5}
\tablecaption{Observed Spectral Indices of Sources \label{tab4}}
\tablewidth{0pt} 
\tablehead{
\colhead{Source}                &
\colhead{Centroid}              &
\colhead{$\alpha_{850/1.3}$\tablenotemark{a}}    &
\colhead{$\alpha_{450/850}$\tablenotemark{a}}    &
\colhead{$\alpha_{450/850}$\tablenotemark{b}}    
}
\startdata
L1512     & ($-$19,$-$26) & ...      &2.4(0.7) &2.4(1.4)\\
L1544     & ($-$1,4)      & 3.3(0.9) &2.1(0.7) &2.5(1.4)\\
L1689A    & ($-$4,$-$9)   & ...      &3.1(0.6) &...     \\
          & (51,24)       & ...      &2.9(0.6) &...     \\
L1689B    & ($-$1,$-$12)  & ...      &2.0(0.6) &1.9(0.6)\\
B133      & ($-$13,$-$32) & ...      &2.7(0.5) &2.7(0.5)\\ 
L1448NW   & (0,0)         & ...      &3.0(0.7) &...     \\ 
L1448N    & ($-1$,1)      & ...      &3.0(0.7) &...     \\ 
L1448C    & (0,3)         & 3.9(0.9) &3.3(0.7) &...     \\ 
L1455     & (3,1)         & ...      &3.5(0.7) &...     \\
          & (55,$-$30)    & ...      &3.3(0.7) &...     \\
          & ($-$9,59)     & ...      &3.0(0.7) &...     \\
          & (10,$-$47)    & ...      &3.2(0.7) &...     \\ 
IRAS03282+3035 & (6,8)         & ...      &2.7(0.7) &3.1(1.4)\\ 
L1527     & ($-$2,0)      & 3.5(0.9) &2.7(0.7) &2.8(1.4)\\ 
B228      & (11,5)        & ...      &3.2(0.5) &2.8(0.6)\\
L483      & (1,2)         & ...      &3.3(0.6) &2.9(0.6)\\ 
L723      & (8,1)         & ...      &2.4(0.9) &1.9(0.9)\\ 
B335      & (0,1)         & 3.3(0.9) &2.9(0.6) &2.6(0.6)\\
L1157     & (0,1)         & 3.3(0.9) &...      &...     \\ 
L1172     & ($-$10,$-$2)  & ...      &3.3(0.6) &2.8(0.6)\\
CB244     & (0,1)         & 3.4(1.0) &2.5(1.4) &2.5(1.4)\\
          & ($-$75,45)    & ...      &2.2(1.4) &...     \\
SSV13     & (1,$-$10)     & 3.1(0.9) &3.2(0.7) &...     \\
IRAS04166+2706 & (1,$-4$)      & ...      &2.1(0.9) &...     \\
L43       & (7,5)         & ...      &3.1(0.6) &...     \\
          & (89,6)        & ...      &2.9(0.6) &...     \\ 
\enddata

\tablenotetext{a}{In a 40\as\ aperture}
\tablenotetext{b}{In a 120\as\ aperture}

\end{deluxetable}

\begin{deluxetable}{lcccclcccc}
\footnotesize
\tablecolumns{10}
\tablecaption{Spectral Energy Distributions of Sources \label{tab5}}
\tablewidth{0pt} 
\tablehead{
\colhead{Source}                  &
\colhead{$\lambda$ (\micron)}     &
\colhead{$S_{\nu}$ (Jy)}          &
\colhead{$\theta$ (\arcsec)} &
\colhead{Ref.}                    &
\colhead{Source}                  &
\colhead{$\lambda$ (\micron)}     &
\colhead{$S_{\nu}$ (Jy)}          &
\colhead{$\theta$ (\arcsec)} &
\colhead{Ref.}              
}
\startdata 
L1512 	& 450	& $<$6.0($3\sigma$)	& 18  & 1  & 	L1544	&170\tablenotemark{*} & 220(80) & (?) & 9 \\
        & 800   & 0.11(0.02) 		& 18  & 1  & 	        &200\tablenotemark{*} & 280(100) & (?) & 9 \\
	& 1100  & 0.045(0.009)		& 18  & 1  &            &  450  &  1.3(0.24)  		&  18  & 2 \\
	& 1300  & $<$0.016($3\sigma$)   & 12  & 1  &            &  800  &  0.45(0.06) 		&  18  & 2 \\
L1689A  & 450   & 2.20(0.30) 		& 18  & 2  &	        & 1100  &  0.19(0.03) 		&  18  & 2 \\ 
	& 850   & 0.29(0.045)		& 18  & 2  &		& 1300\tablenotemark{*}  	&  2.3(0.5)& 260$\times$140 & 1 \\
	& 1100  & $<$0.10($3\sigma$)	& 18  & 2  &  		L1689B	& 12	& $<$0.25($3\sigma$)	& 300$\times$45  & 3 \\
	& 1300  & 0.054(0.015)		& 24  & 2  &		& 25    & $<$0.50($3\sigma$)	& 300$\times$45  & 3 \\
B133	& 450	& $<$1.8($3\sigma$)	& 18  & 2  &		& 60    & $<$0.63($3\sigma$)	& 90$\times$300  & 3 \\
	& 800	& 0.34(0.06)  		& 18  & 2  &		& 90    & $<$12.6($3\sigma$)	& 72  & 4 \\
	& 1100  & $<$0.12($3\sigma$)    & 18  & 2  &		& 100   & $<$32($3\sigma$)  	& 180$\times$300 & 3 \\
	& 1300  & 0.65(0.13)   & 164$\times$102 & 2  &          & 160\tablenotemark{*}   	&  43(15)& 72  & 4 \\
L1448C  & 12\tablenotemark{*}    & 0.33(0.07)   & 35$\times$28   & 5  &	& 190\tablenotemark{*}   	&  46(13)& 72  & 4 \\
	& 25\tablenotemark{*}    & 2.9(0.6)     & 35$\times$28   & 5  &		& 800   & 0.36(0.04)     & 18   & 2 \\
	& 60\tablenotemark{*}    & 31.2(6.5)    & 36  & 5  	 &		& 850   & 4.2(0.9)	& 120  & 2 \\
	& 100\tablenotemark{*}   & 70.3(14.8)   & 45$\times$40  & 5& & 1100  & 0.14(0.03)& 18   & 2 \\
	& 350    & 30(3.0)		& 19.5& 5  &		 &1100\tablenotemark{*} &  1.6(0.3)& 120 & 2 \\
	& 450   & 21(2.0)		& 18.5& 5  &		 & 1300  & 0.13(0.01)& 24   & 2 \\
	& 800   & 3.0(0.3)		& 16.5& 5  &	& 1300\tablenotemark{*} & 0.8(0.17) & 120 & 2 \\
	& 1100  & 1.0(0.1)		& 18.5& 5  &		L1448N	& 12\tablenotemark{*}& 0.67(0.15) & 35$\times$28& 5 \\
	& 1300  & 1.0(0.1)              & 12  & 6  & 		& 25\tablenotemark{*}& 5.7(1.2)	& 35$\times$28   & 5 \\
	& 2600  & 0.091(0.002)		& 2.7 & 7  &		& 60\tablenotemark{*}& 28.8(6.1) & 36   & 5 \\
	& 3500  & 0.026(0.002)		& 2.4 & 6  &		& 100\tablenotemark{*}& 89.0(18.7)& 45$\times$40 & 5 \\
L1448NW & 12    & $<$0.015($3\sigma$)   & 35$\times$28  & 5  &		& 350   & 45(3.0)           	& 19.5 & 5 \\
        & 25    & $<$0.05($3\sigma$)    & 35$\times$28  & 5  &		& 450   & 28(2)			& 18.5 & 5 \\
	& 60\tablenotemark{*}   & 3.2(0.5)              & 36  & 5&	& 800   & 5.8(0.4)          	& 16.5 & 5 \\
	& 100\tablenotemark{*}	& 23(7.5)               & 45$\times$40  & 5  & & 1100	& 12.3(0.2)	& 18.5 & 5 \\	
	& 800   & 2.0(0.5)      	& 16.5& 5  & & 1300  & 2.2(0.1)	& 12   & 5 \\	
	& 1300  & 0.4(0.1)		& 12  & 5  & & 2600  & 0.185(...) & 2.7  & 7 \\
	& 2720  & $<$0.021($3\sigma$)   & 7.0 & 8  &  & 2720  & $>$0.225(...)& 7.0  & 7,8\\
\enddata
\tablenotetext{*}{Flux value used in calculation of \lbol\ and \tbol}
\tablerefs{1. Ward-Thompson et al. 1999;  2. Ward-Thompson et al. 1994; 3. IRAS PSC; 
	   4. Ward-Thompson et al. 1998;  5. Barsony et al. 1998;  
	   6. Bachiller et al. 1991b
	   7. Bachiller et al. 1995;  8. Terebey et al. 1993;
	   9. Ward-Thompson \& Andr\'e 1999}
\end{deluxetable}
\newpage

\begin{deluxetable}{lcccclcccc}
\footnotesize
\tablecolumns{10}
\tablecaption{Spectral Energy Distributions of Sources \label{tab6}}
\tablewidth{0pt} 
\tablehead{
\colhead{Source}                  &
\colhead{$\lambda$ (\micron )}     &
\colhead{$S_{\nu}$ (Jy)}          &
\colhead{$\theta_{mb}$ (\arcsec )} &
\colhead{Ref.}                    &
\colhead{Source}                  &
\colhead{$\lambda$ (\micron )}     &
\colhead{$S_{\nu}$ (Jy)}          &
\colhead{$\theta_{MB}$ (\arcsec )} &
\colhead{Ref.}              
}
\startdata 
L1527	& 1.6	& 0.0056(0.0003)	& 10(?)  & 10 &  IRAS03282+3035 & 12	& $<$0.18($3\sigma$)	& 34$\times$29   & 5 \\
	& 2.2   & 0.0086(0.0002)	& 10(?)  & 10 &		& 25    & $<$0.29($3\sigma$)    & 36   & 5 \\
	& 3.4   & 0.020(0.002)          & 10(?)  & 10 &		& 60\tablenotemark{*}	& 2.32(0.5)		& 33$\times$36 & 5 \\
	& 12    & $<$0.25($3\sigma$)    & 300$\times$45& 3&            & 100\tablenotemark{*}	& 11.05(2.4)		& 40$\times$39 & 5 \\
	& 25\tablenotemark{*}    & 0.74(0.07)            & 300$\times$45& 3&		& 350   & 9.1(1.0)		& 19.5 & 5 \\
	& 60\tablenotemark{*}    & 17.8(1.6) 		& 90$\times$300& 3&		& 450   & 5.9(1.0)		& 18.5 & 5 \\
	& 100   & 89(36)		& 60  & 11 &		& 800	& 1.4(0.1)		& 16.5 & 5 \\
	& 100\tablenotemark{*}   & 73.3(11.7)		& 180$\times$300& 3&		& 1100\tablenotemark{*}  & 0.58(0.05)		& 18.5 & 5 \\
	& 160\tablenotemark{*}   & 94(38)		& 60  & 11 &		& 1300  & 0.3(...)		& 12   & 9 \\
	& 350	& 22(9)			& 60  & 11 &	B228	& 12\tablenotemark{*}	& 0.19(0.03)		& 300$\times$45  & 3 \\
	& 450   & 14(5.6)		& 60  & 11 & 		& 25\tablenotemark{*}    & 1.27(0.05)		& 300$\times$45  & 3 \\
	& 800   & 1.4(0.56)		& 60  & 11 &		& 60\tablenotemark{*}    & 14.7(0.59)		& 90$\times$300  & 3 \\
L483	& 12\tablenotemark{*}& $<$0.25(3$\sigma$)	& 330$\times$45& 3  &		& 100\tablenotemark{*}   & 41.1(2.46)		& 180$\times$300 & 3 \\
	& 25\tablenotemark{*}	& 6.91(0.48)		& 300$\times$45& 3  &	L723	& 12\tablenotemark{*}	& 0.28(0.06)		& 300$\times$45 & 13 \\ 
	& 60\tablenotemark{*}	& 89.1(11.6)		& 90$\times$300& 3  &		& 25\tablenotemark{*}	& 0.38(0.05)		& 300$\times$45 & 3  \\
	& 100\tablenotemark{*}   & 170(85)		& 60	& 11 &		& 60\tablenotemark{*}    & 6.93(0.62)		& 90$\times$300 & 3  \\ 
	& 100   & 165.5(20.0)		& 180$\times$300& 3 & 		& 95    & 27(6)			& 45	 & 14 \\
	& 160\tablenotemark{*}	& 290(145)		& 60 	& 11 &		& 100\tablenotemark{*}   & 20.7(1.7)		& 180$\times$300& 3  \\
	& 190\tablenotemark{*}	& 140(70)		& 60	& 11 &		& 130	& 32(11)		& 33	 & 14 \\
	& 450	& 15(2)			& 19 	& 12 &		& 140\tablenotemark{*}	& 23(8)			& 85	 & 14 \\
	& 800	& 1.98(0.02)		& 19	& 12 &		& 144   & 33(10)		& 33     & 14 \\
	& 1100  & 0.64(0.02)		& 19	& 12 &		& 166   & 40(12)		& 45     & 14 \\
	& 2700	& 0.0072(?)		& 5	& 12 &		& 195\tablenotemark{*}   & 35(7)			& 85     & 14 \\
B335	& 12	& 0.32(0.08)		& 300$\times$45& 13 &		& 400   & 13(3)			& 48     & 14 \\
	& 25	& 0.19(0.03)		& 300$\times$45& 13 &		& 1000\tablenotemark{*}  & 1.0(0.5)		& 102	 & 14 \\
	& 60    & 7(2)			& 33    & 16 &		& 1300  & 0.357(0.017)		& 23     & 15 \\
	& 60\tablenotemark{*}    & 8.3(0.8)		& 90$\times$300& 3  &	L1157	& 12\tablenotemark{*}	& 0.066(0.011)		& 300$\times$45 & 13 \\
	& 85\tablenotemark{*}	& 24(2.4)		& 80	& 17 &		& 25\tablenotemark{*}	& 0.226(0.016)		& 300$\times$45 & 13 \\
	& 100\tablenotemark{*}	& 31(3.1)		& 80	& 17 &		& 60\tablenotemark{*}	& 9.97(0.50)		& 90$\times$300 & 13 \\
	& 100   & 42.0(7.6)		& 180$\times$300& 3 &		& 100	& 42.0(1.7)		& 180$\times$300& 13 \\
	& 110   & 35(9)			& 42    & 16 &		& 1300  & 0.9(0.1)		& (?)	 & 19 \\
	& 115	& 40(4)			& 80	& 17 &		& 2700  & 0.04(?)		& 5	 & 19 \\
\enddata
\tablenotetext{*}{Flux value used in calculation of \lbol\ and \tbol}
\tablerefs{3. IRAS PSC; 5. Barsony et al. 1998;  6. Bachiller et al. 1991b; 
           9. Bachiller at al. 1994
	   10. Kenyon et al. 1990;  11. Ladd et al. 1991; 12.Fuller et al. 1995;
	   13. IRAS FSC;  14. Davidson 1987;  15. Reipurth et al. 1993;
           16. Keene et al. 1983;  17. Larsson 1998;  19. Gueth et al. 1997}
\end{deluxetable}
\newpage

\begin{deluxetable}{lcccclcccc}
\footnotesize
\tablecolumns{10}
\tablecaption{Spectral Energy Distributions of Sources \label{tab7}}
\tablewidth{0pt} 
\tablehead{
\colhead{Source}                  &
\colhead{$\lambda$ (\micron)}     &
\colhead{$S_{\nu}$ (Jy)}          &
\colhead{$\theta_{mb}$ (\arcsec)} &
\colhead{Ref.}                    &
\colhead{Source}                  &
\colhead{$\lambda$ (\micron)}     &
\colhead{$S_{\nu}$ (Jy)}          &
\colhead{$\theta_{MB}$ (\arcsec)} &
\colhead{Ref.}              
}
\startdata 
B335    & 140	& 38(9)			& 42	& 16 &		CB244	& 12\tablenotemark{*}	& 0.055(0.12)		& 300$\times$45 & 13 \\
	& 150\tablenotemark{*}   & 56(5.6)& 80  & 17 &			& 25\tablenotemark{*}	& 0.775(0.039)		& 300$\times$45 & 13 \\
	& 170\tablenotemark{*}	 & 60(6.0)& 80	& 17 &			& 60\tablenotemark{*} 	& 9.06(0.45)		& 90$\times$300 & 13 \\
	& 180   & 80(18)		& 90    & 16 &			& 100\tablenotemark{*}	& 15.0(0.9)		& 180$\times$300& 13 \\
	& 190	& 84(24)		& 102   & 16 &			& 350\tablenotemark{*}	& 9.3(2.8)		& 19.5	 & 20 \\
	& 200\tablenotemark{*}   & 67(14)& 90   & 16 &			& 450 	& 3.5(1.1)		& 18.5   & 20 \\
	& 235\tablenotemark{*}   & 61(14)& 102  & 16 &			& 800	& 0.65(0.13)		& 16.5   & 20 \\
	& 360\tablenotemark{*}   & 41(8)& 55    & 18 &			& 1100\tablenotemark{*}  & 0.27(0.05)		& 18.5   & 20 \\
	& 750\tablenotemark{*}   & 5.3(1.0)& 58 & 18 &			& 1300  & 0.12(0.02)		& 16.5   & 20 \\
SSV13   & 1.6\tablenotemark{*}   & 0.033(0.003) & 3  & 22 &	L1455	& 12\tablenotemark{*}	& 0.18(0.05)		& 300$\times$45 & 13 \\
        & 2.2\tablenotemark{*}   & 0.098(0.010) & 3  & 22 &		& 25\tablenotemark{*}	& 4.24(0.21)		& 300$\times$45 & 13 \\
        & 3.4\tablenotemark{*}   & 0.34(0.03)   & 3  & 22 &		& 60\tablenotemark{*} 	& 48.8(2.4)		& 90$\times$300 & 13 \\
	& 12\tablenotemark{*}	& 13.6(3.7)& 300$\times$45& 13 &	& 100\tablenotemark{*}	& 82.2(4.9)		& 180$\times$300& 13 \\
	& 25\tablenotemark{*}   & 46.5(2.8)& 300$\times$45& 13 &	& 160\tablenotemark{*}	& 55(25)		& 49	 & 21 \\
	& 60\tablenotemark{*}	& 204(20)& 90$\times$300& 13 &		& 190\tablenotemark{*}   & 40(15)		& 49	 & 21 \\
	& 100\tablenotemark{*}	& 381(23)& 180$\times$300& 13&		& 400	& 20(5)			& 49	 & 21 \\
	& 870	& 3.85(0.09)		& 18	& 15 &		IRAS04166+2706& 1.6\tablenotemark{*}   & 0.00010(0.00002)      & 10(?) & 10 \\
	& 1300	& 1.23(0.04)		& 23	& 15 &  		& 2.2\tablenotemark{*}   & 0.00019(0.00009)      & 10(?) & 10 \\
L43     & 0.45\tablenotemark{*}  & 0.00064(0.00006)& 12    & 23 &	& 12\tablenotemark{*}    & 0.07(0.007)           & 300$\times$45 & 3  \\
	& 0.55\tablenotemark{*}  & 0.0031(0.0003) & 12& 23 &		& 25\tablenotemark{*}    & 0.58(0.058)		& 300$\times$45 & 3  \\
	& 0.7\tablenotemark{*}0  & 0.0086(0.0009) & 12& 23 &		& 60\tablenotemark{*}    & 5.9(0.59)		& 90$\times$300 & 3  \\
        & 0.90\tablenotemark{*}  & 0.024(0.002)   & 12& 23 &		& 100\tablenotemark{*}   & 9.5(0.95)             & 180$\times$300& 3  \\
	& 1.25\tablenotemark{*}  & 0.096(0.005)   & 12& 23 &	L1172   & 12\tablenotemark{*}	& 0.14(0.03)		& 300$\times$45 & 13 \\
	& 1.6\tablenotemark{*}   & 0.25(0.01)     & 12& 23 &		& 25\tablenotemark{*}    & 0.30(0.02)		& 300$\times$45 & 13 \\
        & 2.2\tablenotemark{*}   & 0.48(0.02)     & 12& 23 &		& 60\tablenotemark{*}    & 1.31(0.09)		& 90$\times$300 & 13 \\
	& 3.4\tablenotemark{*}   & 0.51(0.02)     & 12& 23 &		& 100   & 11(4.4)		& 60     & 11 \\
	& 12\tablenotemark{*}	& 1.47(0.12)& 300$\times$45& 13 &	& 100\tablenotemark{*}   & 4.76(1.33)		& 180$\times$300& 13 \\
	& 25\tablenotemark{*}    & 6.00(0.36)& 300$\times$45& 13 &	& 160\tablenotemark{*}   & 10(4.0)		& 60     & 11 \\
	& 60\tablenotemark{*}    & 34.0(2.7)             & 90$\times$300& 13 &\\
	& 100\tablenotemark{*}   & 68.0(3.4)             & 180$\times$300& 13&\\ 
	& 160\tablenotemark{*}   & 79(32)                & 60    & 11 &\\
	& 190\tablenotemark{*}   & 38(15.2)              & 60    & 11 &\\
\enddata

\tablenotetext{*}{Flux value used in calculation of \lbol\ and \tbol}

\tablerefs{3. IRAS PSC; 10. Kenyon et al. 1990;  11. Ladd et al. 1991; 13. IRAS FSC;  
	   15. Reipurth et al. 1993; 16. Keene et al. 1983;  17. Larsson 1998;  18. Gee et al. 1988; 
           20. Launhardt \& Henning 1997; 21. Davidson \& Jaffe 1984;  22. Aspin \& Sandell 1994;  23. Myers et al. 1987}
\end{deluxetable}
\newpage

\begin{deluxetable}{lcrrrccc}
\footnotesize
\tablecolumns{8}
\tablecaption{Source Properties \label{tab8}}
\tablewidth{0pt} 
\tablehead{
\colhead{Source}                &
\colhead{Class}                 &
\colhead{\lbol}                 &
\colhead{\tbol}                 &
\colhead{\fsmm}                 &
\colhead{$M_{D}$(20K)}          &
\colhead{$M_{V}$}               &
\colhead{Ref.\tablenotemark{a}}  \\
\colhead{}                      &
\colhead{}                      &
\colhead{\lsun}                 &
\colhead{(K)}                   &
\colhead{}                      &  
\colhead{\msun}                 &  
\colhead{\msun}                 &
\colhead{}                           
}
\startdata 
        L1512   & PPC & ... & ... & ... & 0.2 & 0.3 & 1 \\
        L1544   & PPC & 1.0(0.3) & 18(6) & 0.03(0.01) & 0.4 & 0.4 & 1 \\
        L1689A  & PPC & ... & ... & ... & ... & 13   & 2 \\ 
        L1689B  & PPC & 0.2(0.03) & 18(4) & 0.09(0.01) & 0.24 & 2.0 & 3 \\
        B133    & PPC & ... & ... & ... & ... & 4.7  & 1 \\ 
        L1448NW & $0$   & 2.2(0.5) & 24(5) & 0.09(0.02) & ... & ... & ...\\
        L1448N  & $0$   & 8.0(1.0) & 55(7) & 0.028(0.007) & ... & ... &... \\ 
        L1448C  & $0$   & 6.0(0.5) & 54(7) & 0.020(0.005) & ... & 9.4 & 1 \\ 
        L1455   & $0$   & 6.9(0.3) & 67(3) & 0.0053(0.0007) & ... & 6.4 & 4\\ 
        IRAS03282+3035 & $0$ & 1.2(0.3) & 23(5) & 0.09(0.03) & 2.2 & 2.9 & 4 \\ 
        L1527   & $0$   & 2.2(0.2) & 36(5) & 0.04(0.02) & 0.9 & 0.9 & 4 \\ 
        B228    & $0$   & 1.2(0.2) & 48(2) & 0.03(0.01) & 0.4 & 2.6  & 5 \\ 
        L483    & $0$   & 13(2)    & 52(8) & 0.015(0.003) & 1.8 & 2.7 & 4 \\ 
        L723    & $0$  & 3.3(0.2) & 47(3) & 0.035(0.008) & 1.6 & 7.3 & 4 \\ 
        B335    & $0$   & 3.1(0.1) & 28(1) & 0.060(0.007) & 1.2 & 3.5 & 4 \\ 
        L1157   & $0$   & 5.8(0.8) & 42(4) & 0.009(0.003) & 2.6 & 10 & 4 \\ 
        L1172   & $0$   & 1.1(0.1) & 44(4) & 0.010(0.008) & ... & 4.9 & 4 \\ 
        CB244   & $0$   & 1.0(0.1) & 56(3) & 0.024(0.004) & 0.3 & 2.2 & 4 \\ 
        SSV13   & I   & 43(2)    & 136(15) & 0.0047(0.0012) & ... & 6.9 & 4\\ 
        IRAS04166+2706 & I & 0.42(0.03) & 91(12) & 0.019(0.002) & ... & 1.0 & 4 \\ 
        L43     & I   & 2.7(0.1) & 370(20) & 0.0054(0.0010 & ... & 1.7 & 4\\  
\enddata

\tablenotetext{a}{Reference for linewidth used to calculate $M_V$}
\tablerefs{1. Benson et al. 1998(\nthp); 2. Benson \& Myers 1989 (\ammonia); 
3. Gregersen \& Evans 2000 (H$^{13}$CO$^+$); 4. Mardones et. al. 1997 (\nthp);
5. Gregersen et al. 2000 (H$^{13}$CO$^+$)}

\end{deluxetable}

 
\begin{deluxetable}{lclcclcc}
\footnotesize
\tablecolumns{8}
\tablecaption{Radial Profile Power Law Fits \label{tab9}}
\tablewidth{0pt} 
\tablehead{
\colhead{Source}           &
\colhead{Class}            &
\colhead{T$_{bol}$ }       &
\colhead{$\lambda$}        &
\colhead{$m$ }             &
\colhead{Range}            &
\colhead{Num\tablenotemark{a}}   &
\colhead{S/N}              \\
\colhead{}                 &
\colhead{}                 &
\colhead{(K)}              &
\colhead{($\mu$m)}         &
\colhead{}                 &
\colhead{(AU)\tablenotemark{b}}   &
\colhead{}                 &
\colhead{}
}
\startdata
 L1448C    & $0$  & 54(7) & 450 & 1.27 (0.11) & 2700 - 8100  & 8  & 30  \\
 L1455     & $0$  & 67(3) & 450 & 1.22 (0.23) & 2700 - 6550  & 6  & 40  \\ 
 IRAS03282+3035 & $0$  & 23(5) & 850 & 1.57 (0.31) & 5400 - 10000 & 4  & 60  \\
           &      &       & 450 & 1.66 (0.25) & 2700 - 8100  & 8  & 50  \\
 L1527     & $0$  & 36(5) & 850 & 1.32 (0.05) & 3450 - 11250 & 9  & 50 \\
           &      &       & 450 & 1.13 (0.03) & 1700 - 9050  & 16 & 80  \\ 
 B228      & $0$  & 48(2) & 850 & 2.17 (0.29) & 3200 - 6800  & 5  & 80 \\ 
           &      &       & 450 & 1.86 (0.21) & 1600 - 5250  & 9  & 60  \\ 
 L483      & $0$  & 52(8) & 850 & 1.16 (0.08) & 4900 - 14700 & 8  & 50 \\ 
           &      &       & 450 & 1.49 (0.09) & 4550 - 10150 & 12 & 60 \\ 
 L723      & $0$  & 47(3) & 850 & 1.30 (0.19) & 7350 - 19950 & 7  & 40 \\
           &      &       & 450 & 1.47 (0.42) & 3700 - 10000 & 7  & 30 \\
 B335      & $0$  & 28(1) & 850 & 1.74 (0.30) & 6100 - 13100 & 5  & 100 \\
           &      &       & 450 & 1.65 (0.17) & 3050 - 12700 & 12 & 60 \\ 
 L1157     & $0$  & 42(4) & 850 & 0.87 (0.56) & 8000 - 14800 & 4  & 60 \\
 IRAS04166+2706 & I    & 91(12)& 850 & 2.07 (0.78) & 3450 - 6400  & 4  & 30 \\
           &      &       & 450 & 1.57 (0.58) & 1700 - 4650  & 7  & 10 \\ 
 L43       & I    & 370(20)&450 & 1.10 (0.39) & 1550 - 5450  & 10 & 15 \\ 
\enddata

\tablenotetext{a}{Number of points used in fit}
\tablenotetext{b}{Range (AU) over which fit was made}
\end{deluxetable}

\end{document}